\documentclass[twocolumn,floats,eqsecnum,pra]{revtex4}

\newcommand {\be}{\begin{equation}}
\newcommand {\ee} {\end{equation}}
\newcommand {\bea}{\begin{eqnarray}}
\newcommand {\eea} {\end{eqnarray}}
\newcommand{\non}{\nonumber}
\newcommand{\bk}{{\bf k}}
\newcommand{\br}{{\bf r}}

\begin{document}


\title{Non-Abelian adiabatic statistics and Hall viscosity\\in quantum Hall states and p$_x+i$p$_y$ paired superfluids}
\author{N. Read}
\affiliation{Department of Physics, Yale
University, P.O. Box 208120, New Haven, CT 06520-8120, USA}
\date{October 31, 2008}

\begin{abstract}
Many trial wavefunctions for fractional quantum Hall states in a
single Landau level are given by functions called conformal
blocks, taken from some conformal field theory. Also,
wavefunctions for certain paired states of fermions in two
dimensions, such as p$_x+i$p$_y$ states, reduce to such a form at
long distances. Here we investigate the adiabatic transport of
such many-particle trial wavefunctions using methods from
two-dimensional field theory. One context for this is to
calculate the statistics of widely-separated quasiholes, which has
been predicted to be non-Abelian in a variety of cases. The Berry phase or matrix (holonomy) resulting from adiabatic
transport around a closed loop in parameter space is the same as
the effect of analytic continuation around the same loop with the
particle coordinates held fixed (monodromy), provided the trial
functions are orthonormal and holomorphic in the parameters
so that the Berry vector potential (or connection) vanishes. We
show that this is the case (up to a simple area term) for paired states (including the Moore-Read quantum Hall state), and present general conditions for it to hold for other trial states (such as the Read-Rezayi
series). We argue that trial states based on a non-unitary conformal field theory do not describe a gapped topological phase, at least in many cases. By considering adiabatic variation of the aspect ratio of the torus, we calculate the {\em Hall viscosity}, a non-dissipative viscosity coefficient analogous to Hall conductivity, for paired states, Laughlin states, and more general quantum Hall states. Hall viscosity is an invariant within a topological phase, and is generally proportional to the ``conformal spin density'' in the ground state.
\end{abstract}

\pacs{PACS numbers: } 

\maketitle


\section{Introduction}
\label{intro}
\subsection{Background and motivation}

There has been renewed interest in the past few years in
non-Abelian quantum states of matter, both in the original setting
of quantum Hall states \cite{mr}, and also in other systems
including ones in which the symmetries under time reversal and
parity are unbroken (see e.g.\ Ref.\ \cite{other}). Briefly, a non-Abelian phase of matter is a gapped (topological) phase in which there are
quasiparticle excitations over the ground state, the adiabatic
exchange of which produces a matrix effect on the state of the
system, with the matrices corresponding to distinct exchanges not
all commuting (thus the term ``non-Abelian''). This requires that
there be a degenerate space of states when there are
quasiparticles at well-separated positions in the system. Given a
reasonably local Hamiltonian, such behavior can only occur when
the dimensionality of space is two, for topological reasons. The
basic example proposed by Moore and Read (MR) \cite{mr} is a
paired state, which can be viewed as a p-wave, or more precisely a
$p_x+i p_y$-wave, Bardeen-Cooper-Schrieffer (BCS) \cite{bcs} paired state of
spinless or spin-polarized composite fermions in zero or almost
zero net magnetic field. It turns out that much of the physics of
this state is also found in such paired states of ordinary
fermions \cite{rg}. In this paper we address both of these
situations together, not only quantum Hall systems. Much of the current interest in these systems is driven by their potential use for topological quantum computing \cite{freed}.

The major issue that we address in this paper is the derivation of
the non-Abelian statistics of the quasiholes (or vortices) when
they are exchanged adiabatically, where adiabatic transport is
calculated using trial forms of the wavefunctions. The trial
wavefunctions are taken to be ``conformal blocks'' obtained from
some conformal field theory (CFT), as in MR.

A central idea of MR is that the adiabatic effect of an exchange
of quasiholes in the trial wavefunctions given by conformal blocks
is the same as the effect inferred from simple analytic
continuation of the wavefunctions viewed as functions of the
quasihole coordinates, with the particle coordinates fixed. In
contrast to analytic continuation, adiabatic transport involves
integration over the particle coordinates for each infinitesimal
time step, since this defines the inner product in the Hilbert
space. In several examples with Abelian statistics, this was shown
in MR to give the same result, but the result was not demonstrated
for the trial states with non-Abelian statistics. Here we provide
a detailed explicit calculation for the cases of two or four
quasiholes in the MR paired state (with the charge part of the
wavefunctions removed); the four quasihole case is the first to
exhibit non-Abelian statistics. Finally, we also give a general
criterion for other states, and discuss when this may hold, with
examples.

An additional issue that we address is adiabatic variation of the
geometry in a closed finite system, for example a parallelogram
with periodic boundary conditions (i.e.\ a torus). There is a
discrete group of transformations that map the geometry to one
equivalent to the original (called the modular group). It is of
interest to perform these adiabatically also. It turns out that in
these cases the relevant Berry connection (vector potential on the
space of inequivalent geometries) has non-zero curvature (field
strength). Because varying the aspect ratio varies the strain on
the system, this response can be related to a non-dissipative
viscosity coefficient \cite{asz}, which we call {\em Hall
viscosity} (earlier it was termed ``odd'', or ``antisymmetric''
viscosity \cite{asz}, or ``Lorentz shear modulus'' \cite{tv}). Hall
viscosity is so-named because it is the natural analog in viscosity
of the Hall conductivity.  We show that this viscosity arises in quantum
Hall fluids and in paired superfluids, and its magnitude is always
proportional to a spin density. This result should be of wide interest. For trial wavefunctions given by conformal blocks, the spin per particle is
given by the conformal weight.

In the remainder of this Introduction, we explain the central
issues in more depth, introduce some basic arguments that will be
crucial later, and also review some earlier work relevant to the
problems of adiabatic statistics and Hall viscosity.


\subsection{Adiabatic transport and quasiparticle statistics}
\label{adtrans}

Here for the time being we are concerned primarily with quantum
Hall states. The basic examples are Laughlin's trial wavefunctions
for a ground state with $n$ quasiholes (each of the lowest
possible non-zero fractional charge), which in their original form are \cite{laugh}%
\be%
\prod_{k,l}(z_k-w_l)\prod_{i,j}(z_i-z_j)^Q
e^{-\frac{1}{4}\sum_i|z_i|^2}.\label{laughqhole}\ee%
Here the exponent $Q$ determines the filling factor of the state
in the thermodynamic limit as the particle number $N$ goes to
infinity, $z_i$ ($i=1$, \ldots, $N$) are the complex coordinates
of particles, and $w_l$ are the complex coordinates of the
quasiholes. When $Q$ is even, the particles are bosons, and when
$Q$ is odd they are fermions, such as electrons. In either case,
the functions are for particles in the lowest Landau level (LLL).

Now we consider the adiabatic calculation of statistics. The
adiabatic calculation is meaningful if the excitation spectrum of
the system above the possibly degenerate quasiparticle state
subspace is gapped (with the gap going to a constant as the system
size goes to infinity). It can also be valid if any gapless
excitations are sufficiently weakly coupled to the quasiparticles.
For example, in a quantum Hall system with an edge, there are
gapless edge excitations. When quasiparticles are present within
the region occupied by the particles, we will assume throughout
this paper that the system (i.e.\ $N$) is large enough so that
during the exchange the distance of the quasiparticles from the
edge is large compared with their separation, which is itself
large compared with the microscopic correlation length scale
(which frequently is of order the particle spacing). In this case
the edge excitations effectively decouple from the calculation.

The pioneering calculation of adiabatic statistics in a QH system
was performed by Arovas, Schrieffer and Wilczek
\cite{asw,stonebook}. (An interesting alternative approach has
been formulated recently \cite{sl}, and applied
to non-Abelian states \cite{seidel} as we were finishing this paper.) We follow the framework of their method,
but generalize to allow for non-Abelian statistics. Thus suppose
that we have a space of degenerate states with basis $|\Psi_a(w)\rangle$,
where $a$ runs over some finite set, and we
will use $w$ to stand for the set of $n$ quasihole coordinates
$w_l$, $l=1$, \ldots, $n$. We assume that the basis is
orthonormal, $\langle\Psi_a(w)|\Psi_b(w)\rangle=\delta_{ab}$ for
each $w$. As $w$ varies, these vectors sweep out a subspace of a
common multiparticle Hilbert space in which they all live
\cite{berry,bs}. The states for any fixed $w$ are supposed to be
degenerate in energy, and we will compute the adiabatic statistics
within this subspace. We neglect the ``dynamical phase'' which
depends on the energy and the total time taken for an exchange.
The desired adiabatic phase or matrix is
the holonomy, which can be written \cite{berry,bs,asw,wz}%
\be%
{\cal P}\exp i\oint_C (A_w\cdot dw +A_{\overline{w}}\cdot d\overline{w})
\ee%
where $C$ is a closed directed path that begins and ends at a base
point $w_{(0)}$, in the configuration space of the quasiholes.
(This configuration space can be thought of as ${\bf C}^n$ where
$\bf C$ is the complex plane or Riemann sphere, minus the
``diagonal'' on which $w_k=w_l$ for some $k$, $l$, $k\neq l$,
modulo the action of the permutation group $S_n$ for the identical
quasiholes; if the quasiparticles are not all of the same type
there are obvious modifications of the permutation group.) ${\cal
P}$ is the path ordering operator, in which the matrix for a later
point on the path is to the left of earlier ones, $A_w\cdot
dw=\sum_l A_{w,l}dw_l$, and $A_{w,l}$ is a matrix. It is possible
to change to a different basis set
$|\Psi'_b(w)\rangle=|\Psi_a(w)\rangle M_{ab}(w)$ as a function of
position on the path $C$, where $M(w)$ is a unitary matrix for
each point $w$ on the path $C$, in order to preserve
orthonormality. This is referred to as a change of gauge. The
expression is correct as written when the vector potential $A$
refers to a single gauge choice for the whole path $C$ [as in the
case of the states in eq.\ (\ref{laughqhole}) above]. In general
such a choice may not be convenient, and then one must use patches
with a gauge choice on each patch. The patches overlap, and in the
overlap region there is a transition function (gauge
transformation), which is a unitary matrix $M$. The holonomy is
then the matrix composite of the path-ordered exponentials for the
part of the path within each patch, with gauge transformations by
the transition matrices inserted in between. It is frequently the
case that a single transition function $M$ is sufficient, and the
transition can be located at the basepoint $w_{(0)}$. Then the
adiabatic transport maps the basis states
$|\Psi_b(w_{(0)})\rangle$ to $\sum_a|\Psi_a(w_{(0)})\rangle
B_{ab}$, where the holonomy is
given by%
\be%
B=M{\cal P}\exp i\oint_C (A_w\cdot dw +A_{\overline{w}}\cdot
d\overline{w}).
\ee%
Defined this way, the holonomy is gauge invariant in all cases, up
to conjugation by a unitary matrix that corresponds to a change of
orthonormal basis at $w_{(0)}$. We note also that the use of a
different base-point on the same path $C$ conjugates the holonomy
by some unitary matrix, and has no effect on the structure of the
braid-group representation.

The Berry connection is given in components by%
\bea%
A_{w,l,ab}(w)&=&i\left\langle\Psi_a(w)\left|\frac{\partial\Psi_b(w)}
{\partial w_l}\right.\right\rangle,\\
A_{\overline{w},l,ab}(w)&=&i\left\langle\Psi_a(w)\left|\frac
{\partial\Psi_b(w)}{\partial\overline{w_l}}\right.\right\rangle.\eea%
Then $A_{\overline{w},l}(w)=A_{w,l}(w)^\dagger$, which ensures
that the holonomy is unitary ($A_w$ is not holomorphic in $w$ in
general). For functions in the LLL, the inner product is the usual
$\langle\Psi_a|\Psi_b\rangle=\int\prod
d^2z_i\,\Psi_a^\ast\Psi^{\vphantom{\ast}}_b$. In the calculation
in Ref.\ \cite{asw}, the trial quasihole states used were the
wavefunctions (\ref{laughqhole}) which are not orthonormal with
respect to this inner product (for Laughlin quasiholes, there is
only a single state for $n$ quasiholes, so the label $a$ is
dropped). Arovas {\it et al.}\ appeared to neglect this point
\cite{stonebook}, but arrived at the correct answer nonetheless.
The result is that if two quasiholes are exchanged around a
counterclockwise path that does not enclose any other quasiholes
(the others stay at fixed positions throughout), the adiabatic
phase change is $e^{i\pi/Q}$.

The central point of the calculation is that the necessary inner
products can be evaluated by using the plasma mapping of Laughlin
\cite{laugh}, plus the fact that screening occurs in the Coulomb
plasma provided $Q$ is not too large. We reformulate this as
follows (following a line of argument begun by Halperin
\cite{halp}). Consider instead the following wavefunctions, which
represent the same quantum states as (\ref{laughqhole}) because
they differ only by functions of the parameters
$w$:%
\bea%
\lefteqn{\Psi(w_1,\ldots,w_n;z_1,\ldots,z_N)=}\non\\
&&\prod_{k<l}(w_k-w_l)^{1/Q}\cdot
\prod_{i,k}(z_i-w_k)\cdot\prod_{i<j}(z_i-z_j)^Q\cdot\nonumber\\
&&{}\times
e^{-\frac{1}{4Q}\sum_k|w_k|^2-\frac{1}{4}\sum_i|z_i|^2}
\label{halpqhole}\eea%
(here and below we take the notational liberty of using some
indices more than once in distinct factors, which have been
separated here by the dots $\cdot$). The modulus-squared is the
Boltzmann weight for a classical plasma of charges $Q$ at all
$z_i$, $1$ at all $w_k$, with a neutralizing background of density
$1/2\pi$, impurities at fixed positions $w_k$, with the
two-dimensional Coulomb interaction potential between unit charges
taken to be minus the natural logarithm of the distance-squared,
and with temperature $Q$. Thus the overlap integral is the
corresponding partition function:
\bea%
\lefteqn{{\cal Z}(w_1,\ldots,w_n)=\left\|
|\Psi(w_1,\ldots,w_n)\rangle
\right\|^2=}\non\\
&&\int\prod_id^2z_i\,\exp\frac{1}{Q}\left[\sum_{k<l}\ln|w_k-w_l|^2+
Q\sum_{i,k}\ln|z_i-w_k|^2\right.\non\\&&{}\left.+Q^2\sum_{i<j}
\ln|z_i-z_j|^2
-\frac{1}{2}\sum_k|w_k|^2-\frac{Q}{2}\sum_i|z_i|^2\right].\eea%
Given the partition function, we can define the free energy $F$ as
$F=-\ln {\cal Z}$ (or as this times the temperature
$Q$, but this makes no difference for us). Then by screening, for
sufficiently large separation of the quasiholes, this free energy
goes to a constant at sufficiently large separation of the
quasiholes. This constant has the form%
\be%
F=Af_0(Q)+nf_{\rm qh}(Q),\ee%
consisting of an extensive background term, which is the area $A$
occupied by the particles plus quasiholes times a constant
$f_0(Q)$, plus a ``defect free energy'' $f_{\rm qh}(Q)$ for each
quasihole. (In a more general situation with different types of
quasiparticles, these latter terms would be different for each
type of quasiparticle, as well as depending on the underlying
ground state.) Both constants $f_0(Q)$ and $f_{\rm qh}(Q)$ are
well-defined but $Q$-dependent, given the definition above, but
have no universal significance.

Now we can reformulate the statistics calculation. Suppose that,
in addition to being (ortho-)normalized, the wavefunctions of the
states $|\Psi_a(w)\rangle$ are holomorphic in $w$, as the Halperin
functions above are (except for the Gaussian factors, which we
ignore here, but comment on afterwards, and except on the
diagonal $w_k=w_l$, which is to be avoided). Then it follows that%
\be%
A_{w,l,ab}(w)=i\frac{\partial}{\partial
w_l}\left\langle\Psi_a(w)\left|\Psi_b(w)\right.\right\rangle=0.\ee%
Then the holonomy is given entirely by the transition matrix or
matrices: $B=M$. For the wavefunctions (\ref{halpqhole}), the
transition matrix is just a number $M$ of modulus $1$ that is
required to transform back to the original sheet (or gauge) after
making an exchange, due to the wavefunctions not being single
valued in the $w_k$s. For the exchange of two quasiholes along a
path not enclosing any others (which stay at fixed positions
throughout), this gives the phase $M=e^{i\pi/Q}$. It is important
to emphasize that when applying this argument to wavefunctions,
they should be functions in a common Hilbert space for all $w$, and with
the integration measure independent of the parameter $w$ being
varied.

The fact that the normalizing Gaussian factors are not holomorphic
in $w$ means that the calculation also produces a phase factor
$e^{iA(C)/Q}$, where $A(C)$ is the area enclosed by the loop $C$
\cite{asw}. This can be interpreted as the fractionally-charged
quasiparticles detecting the background magnetic field \cite{asw}
through an Aharonov-Bohm phase, though in fact it is $Q$ times the
particle density that they detect, which happens to give the same
result because the particle density is uniform, with filling
factor $1/Q$, outside the quasiholes. (If the particle density
were not uniform, the normalizing Gaussian factors in $w$s would
be modified.) This effect is ubiquitous for QH states. We will comment
on it further in connection with systems that are not QH systems.
Other than this effect, the holonomy depends only on the
homotopy class of the path $C$ in the configuration space, not on
the precise path; that is, it is invariant under small
deformations of the path such that $C$ does not pass through the
diagonals $w_k=w_l$ during the deformation. Then the holonomy
gives a unitary representation of the braid group, acting in the
space of degenerate states at $w_{(0)}$, times the path-dependent
factor $e^{iA(C)/Q}$ which we will frequently just ignore.

The transition matrices $M$ result from the behavior of the
wavefunctions under analytic continuation of the quasiholes, with
the particle coordinates held fixed (or even if they are not
fixed). In the context of solutions of differential equations,
this is called monodromy, and we follow this terminology here.
Halperin \cite{halp} noted that the monodromy of his functions
(\ref{halpqhole}) suggested fractional statistics, but did not
perform the adiabatic calculation. The use of the present approach
in the adiabatic calculation gives the statistics without the
further calculations performed by Arovas {\it et al.}\ \cite{asw},
in particular, confirming the sign of the phase (alternatively,
their calculation \cite{asw} can also be used to produce the
normalization of the Laughlin quasihole functions
\cite{stonebook}).

The statistics calculation for the Laughlin quasiholes generalizes
easily to wavefunctions that correspond to multicomponent Coulomb
plasmas \cite{mr}. Here we wish to generalize it to non-Abelian
cases, specifically those in which the trial wavefunctions are
conformal blocks from CFT. These are holomorphic in the quasihole
coordinates, and have non-trivial, sometimes non-Abelian,
monodromy; the Halperin form (\ref{halpqhole}) of Abelian
quasihole wavefunctions can also be viewed as conformal blocks
\cite{mr}. The goal will be to see whether the effect noted above
for the latter states, that {\em holonomy equals monodromy}, holds
for these trial functions, as conjectured by MR. The result will
hinge on whether the screening property in the plasma mapping for
the Laughlin states, which makes the Halperin functions (conformal
blocks) orthonormal, also generalizes when other conformal blocks
are used as trial wavefunctions. The answer will be yes in some
cases.

In addition, we apply the same formalism to consider adiabatic
variation of the geometry of the system. We will show later that
the curvature of the Berry connection is non-zero, and this determines the Hall viscosity. We also consider the holonomy around a loop corresponding to a modular transformation. In some cases we can apply a similar argument,
that the orthonormalized wavefunctions are almost holomorphic in
the relevant complex parameter $\tau$, except for a very simple
non-holomorphic part.


\subsection{Conformal blocks as trial wavefunctions}
\label{cft}

Next we discuss how conformal blocks from CFT can be used as trial
wavefunctions, following MR. For background on CFT, see Ref.\
\cite{dfms}.

The trial wavefunctions for QH systems that we will study in this
paper take the form%
\bea%
\lefteqn{\Psi_a(w_1,\ldots,w_n;z_1,\ldots,z_N)=}\\
&&\Psi_{\rm
charge}\cdot\langle\psi(z_1)\cdots\psi(z_N)\tau(w_1)\cdots\tau(w_n) \rangle_{a,\,\,\rm CFT}
.\non\eea%
Here again coordinates $z_i$ are those of particles (either bosons
or fermions), so $\Psi_a$ is single valued and either symmetric or
antisymmetric in these variables, while coordinates $w_l$ are
those of quasiholes. The label $a$ again runs over a basis for a
space of functions. On the right hand side, $\Psi_{\rm charge}$ is
independent of $a$, and is a function of the same coordinates of
similar form to eq.\ (\ref{halpqhole}), though the exponents (or
charges of the particles and impurities in the corresponding
plasma) may take other rational-fraction values. We note that the
exponent in the particle-particle factors is the inverse of the
filling factor, $\nu=P/Q$, and the Gaussian in the particle
coordinates is always as in (\ref{halpqhole}). The values of the
exponents in the charge part are determined by the requirement
that the whole function $\Psi_a$ be single valued and (anti-)
symmetric in the particle coordinates; this may always be done
consistently with a plasma form, thanks to consistency properties
of conformal blocks. The exponents are determined only up to
addition of integers, and we usually consider the smallest
possible positive values, which gives the highest value of the
filling factor, and the lowest value of the quasihole charge;
further the field $\tau$ is usually also chosen, given the CFT, to
obtain the lowest possible quasihole charge for a given possible
filling factor.

The expectation value ${\cal F}_a(w_1,\ldots;z_1,\ldots)= \langle\cdots\rangle_{a,\,\,\rm CFT}$
stands for a conformal block in some CFT, in which $\psi$ and
$\tau$ stand for fields. The notation is somewhat formal, because
the function is generally not single-valued, and a sheet should be
specified. The function is holomorphic in the $z$s and $w$s off
the diagonal on which some $z$s and/or $w$s coincide. The field
$\psi$ must have the property that its monodromy is Abelian,
which means that there is a single primary field (of some chiral
algebra) in its operator product expansion (ope) with itself (we
omit non-zero ope coefficients, which play no role here):%
\be%
\psi(z)\psi(0)\sim z^{-2h_\psi+h_{\psi^*}}\psi^\ast(0)+\ldots,\ee%
as $z\to 0$, where $\psi^*$ is another field, and the $\ldots$
stands as usual in CFT for terms smaller by positive integer
powers of $z$ as $z\to0$, which are descendants of $\psi^*$ under
the chiral algebra. Moreover, $\psi$ must generate in this way
only fields that are also Abelian. Further, there must be exactly
one term in the ope of $\psi$ with $\tau$:
\be%
\psi(z)\tau(0)\sim
z^{-h_\psi-h_\tau+h_{\tau^*}}\tau^\ast(0)+\ldots,\ee%
where again $\tau^*$ is another field, $h_\psi$ is the conformal
weight of $\psi$ (and similarly for other fields), and the same is
true for $\tau^*$ and so on. These requirements mean that $\psi$
is what is called a ``simple current'': the operation of taking
the ope of $\psi$ with the primary fields in the theory (including
$\psi$ itself) just permutes the primary fields. This has the
effect of guaranteeing that the full function $\Psi_a$ has the
stated properties. The ope of $\tau$ with itself may be
nontrivial, containing terms that do not differ in conformal
weight only by integers:%
\be%
\tau(z)\tau(0)\sim z^{-2h_\tau+h_{\tau^{(1)}}}\tau^{(1)}
+\ldots+z^{-2h_\tau+h_{\tau^{(2)}}}\tau^{(2)}+\ldots+\ldots,\ee%
where, depending on which CFT is used, any number of distinct
primaries could appear on the right. Those appearing with nonzero
coefficients may be summarized by the {\em fusion rules}\/
(analogous to the Clebsch-Gordan formulas for SU(2) tensor
products), which
generally have the form%
\be%
\phi_\alpha\times\phi_\beta=\sum_\gamma {\cal N}
_{\alpha\beta}^\gamma
\phi_\gamma,\ee%
in which the fields $\phi_\alpha$ run over the full set of
primary fields in the CFT used ($\psi$, $\psi^*$, \ldots,
as well as $\tau$, $\tau^*$, $\tau^{(1)}$, \ldots, will be among
these), and the ${\cal N}_{\alpha\beta}^\gamma={\cal
N}_{\beta\alpha}^\gamma$ are non-negative integers (which may be
larger than $1$ in some cases). The product here is formal and
simply refers to terms in an actual ope. We define $\phi_0=1$, the
identity operator. For an Abelian field (one obeying Abelian
fusion rules), ${\cal N}_{\alpha\alpha}^\beta$ is equal to $1$ for
one value of $\beta$ (and zero otherwise), and similarly for
iterated products; otherwise the field is non-Abelian. (As
explained in Ref.\ \cite{mr}, $\Psi_{\rm charge}$ can also be
viewed as a kind of conformal block in the CFT of a single scalar
field $\varphi$, with the role of the fields $\psi$, $\tau$ played
by charged fields (exponentials of $\varphi$), but which also
includes a chiral version of the neutralizing background charge
density \cite{laugh}.) A {\em rational} CFT is one in which there is a {\em chiral algebra}, which is either the Virasoro algebra or an algebra extension of it, obtained from ope's of a finite set of Abelian conformal fields, and a {\em finite} set of primary fields, defined generally as conformal fields that generate irreducible highest-weight representations of the chiral algebra (by operator products) \cite{ms,dfms}. Examples of non-trivial chiral algebras include the affine Lie algebras. At many places we will need to assume that the CFT is rational, as is the combined one that includes the charge part. Irrational examples will be discussed at the end of the paper.

The conformal blocks are multisheeted functions when the exponents
such as $2h_\psi-h_{\psi^*}$ are not integers, and because of
non-Abelian fields $\tau$. The multisheetedness due to fractional
exponents from $\psi$ is no worse than in the case of the Halperin
functions (\ref{halpqhole}), and produces only a phase factor when
these are exchanged or encircle a $\tau$, so not a
linearly-independent function of the particle coordinates (as the
latter vary over some open set that lies on a single sheet).
Finally, $a$ labels a basis for the space of linearly-independent
conformal blocks associated with a given correlation function; the
range of $a$ is the dimension of the space of blocks, which can be
calculated for the case of CFT on the sphere by repeated use of
the fusion rules. By definition, this is the number of
linearly-independent functions of all the variables $z$ and $w$ as
they vary over some open set. When viewed as many-particle
wavefunctions, we would instead count the number of
linearly-independent functions of the particle coordinates $z$ for
fixed $w$. In general, the latter number might be less than the
number of blocks (for example, when $N=0$). However, in examples
these numbers do coincide when the particle number $N$ is large
enough, and we will assume this from here on, as large $N$ is the
case of interest anyway. More generally, of course, one could have
more than two types of primary field in the correlator, which
could represent the particles and more than one type of quasihole;
such could result from use of the ope starting from one type of
``basic'' quasihole.

In CFT, correlation functions are constructed from combinations of
blocks and their conjugates, for example%
\bea%
&&\langle\psi(z_1,\bar{z}_1)\cdots
\psi(z_N,\bar{z}_N)\tau(w_1,\bar{w}_1)\cdots\tau(w_1,\bar{w}_1)
\rangle_{\rm
CFT}\non\\
&&\qquad{}={}\sum_a|{\cal F}_a(w_1,\ldots;z_1,\ldots)|^2,\eea%
where the sum is over the basis for the space of blocks. Such an
expression is supposed to represent a single-valued correlation
function of local operators in a CFT; the local operators
$\psi(z,\bar{z})$ and so on that appear here differ from the
chiral versions $\psi(z)$ that appeared before, and are related
roughly by $\psi(z,\bar{z})=\psi(z)\overline{\psi}(\bar{z})$, but
also require the sum over the blocks. The ``diagonal'' form given
is single-valued if the monodromy of the space is given by unitary
matrices $B_{ab}$ in the basis used. Such a diagonal form is
always available in a unitary theory, and even in non-unitary
rational theories, such as the Virasoro minimal models of Belavin,
Polyakov, and Zamolodchikov (BPZ) \cite{bpz,dfms}. More generally,
when the CFT is formulated on a surface of higher genus, the
number of blocks is in most cases larger than one even when no
fields are inserted in the correlator. For $N$, $n \geq0$, the
single label $a$ labels the full space of blocks. A diagonal
theory, in which all correlators are given by diagonal expressions
like that above exists in most cases, and is valid for surfaces of
any genus; it is referred to as the diagonal modular invariant
theory, where modular invariance refers to the case of the torus
(genus 1), with no fields inserted.  At this point we should
emphasize that what we call a correlation function (or correlator)
here is really a partition function for the CFT on the given
surface, with fields $\psi$ and $\tau$ inserted at specified
points. This differs from the more usual use of normalized
correlation functions, in which one divides by the partition
function for the case of no field insertions. For example, when
$\psi$ is a basic field of the theory, described by an action
$S[\psi]$, the theory can be
described by a functional integral,%
\be%
Z=\int {\cal D}[\psi]\,e^{-S[\psi]},\ee%
and the unnormalized correlation function is %
\bea%
&&\langle\psi(z_1,\bar{z}_1)\cdots
\psi(z_N,\bar{z}_N)\tau(w_1,\bar{w}_1)\cdots\tau(w_1,\bar{w}_1)
\rangle_{\rm
CFT}\non\\
&&\quad= Z(w_1,\bar{w}_1,\ldots;z_1,\bar{z}_1,\ldots)\non\\
&&\quad=\int {\cal
D}[\psi]\,\psi(z_1,\bar{z}_1)\cdots\tau(w_1,\bar{w}_1)\cdots
e^{-S[\psi]},\eea%
whereas usually one would define the correlation function to be
$Z(w_1,\bar{w}_1,\ldots;z_1,\bar{z}_1,\ldots)/Z$. Even leaving
aside a factor of the form exponential of minus a free energy
proportional to the area of the surface, this makes a difference
for genus $>1$, because the denominator is in most CFTs a sum of
more than one mod-square conformal blocks. The exceptions to the
latter are ``holomorphic'' CFTs, in which the only primary field
of the chiral algebra is the identity operator $1$; an example is
the current algebra or Wess-Zumino-Witten theory for the Lie group
E$_8$ at level $1$.

As an example of the conformal blocks and their use as trial
wavefunctions, we give the MR state, for the case of no
quasiholes, in which $\psi$ is a Majorana fermion field, with
conformal weight $h_\psi=1/2$. The ground state trial wavefunction
for the sphere or infinite plane contains the conformal block \cite{mr}%
\be%
\langle\psi(z_1)\psi(z_2)\cdots\rangle_{\rm Ising}={\rm
Pf}\,\frac{1}{z_i-z_j},\ee%
where the Pfaffian is defined for any even-by-even antisymmetric
matrix with matrix elements $M_{ij}$ by%
\be%
{\rm Pf}\,M_{ij}={\cal A}(M_{12}M_{34}\cdots M_{N-1,N}),\ee%
where the antisymmetrizer $\cal A$ sums over all permutations that
produce distinct pairings $(i,j)$, times the sign of the
permutation. For quasiholes, one uses \cite{mr} $\tau=\sigma$, the
spin field of the critical Ising model; $1$, $\psi$ and $\sigma$
are the only primary fields in the Ising (or Majorana fermion)
CFT. The scaling dimensions are $h_\psi=1/2$, and $h_\sigma=1/16$.
Explicit conformal blocks for this case will be quoted later in
the paper. The fusion rules for this CFT (other than for products
with the identity $1$) are%
\bea%
\psi\times\psi&=&1,\\
\psi\times\sigma&=&\sigma,\\
\sigma\times\sigma&=&1+\psi.
\eea%
The fusion rules imply that the number of conformal blocks for $n$
quasiholes on the sphere is $2^{n/2-1}$, and $n$ must be even.
Other examples include the RR states \cite{rr2}, in which the
field $\psi=\psi_1$, one of the parafermion currents in the ${\bf
Z}_k$ parafermion CFT; these fields are good examples of ``simple
currents''. One can also consider QH systems in which the
particles carry SU(2) spin greater than $0$, and then inner
products involve the inner product of spin states as well as
spatial integrals.

We may now define the overlap integrals, or inner products of the
trial states with wavefunctions $\Psi_a$, as %
\bea%
\lefteqn{\langle\Psi_a(w_1,\ldots,w_n)|\Psi_b(w_1,\ldots,w_n)\rangle}
\non\\
&=&{\cal Z}_{ab}(w_1,\ldots,w_n)
\non\\
&=&\int\prod_{i=1}^N d^2z_i
\overline{\Psi_a(w_1,\ldots;z_1,\ldots)}\non\\&&\qquad{}\times
\Psi_b(w_1,\ldots;z_1,\ldots).\eea %
(The coordinates, and the integration measure are written for the
plane, though the definition applies to other geometries with
modifications that are hopefully obvious. We do not imply that the
overlap matrices ${\cal Z}_{ab}(w_1,\ldots,w_n)$ are holomorphic in the
$w$s.) The difference between ${\cal Z}_{ab}$ and $Z$ above should
be noted: both are sesquilinear in conformal blocks, but
$Z(w_1,\ldots,z_1,\ldots)$ depends on $z$'s as well as $w$'s, and
is a diagonal sum of mod-square conformal blocks, while ${\cal
Z}_{ab}$ is integrated over $z$'s, but not summed over the indices
$a$, $b$.

{}From the discussion we now see that when conformal blocks are used
as (or in) trial wavefunctions as described above, then the
holonomy under adiabatic transport will equal the monodromy
provided that the overlap matrix ${\cal Z}_{ab}(w_1,\ldots,w_n)$
is proportional to $\delta_{ab}$ with proportionality constant
independent of the positions $w$, asymptotically for large
separations of the $w$'s, in a basis for the conformal blocks in
which the braiding matrices $B$ are unitary. This orthonormality
of the conformal blocks is then the desired statement generalizing
screening in the Coulomb plasma. (MR noted in Abelian examples
such as the Laughlin states that when conformal blocks are used as
trial wavefunctions, the gauge is such that the Berry connection
vanishes, and the holonomy is given entirely by the monodromy.)

Let us also point out here that the so-called shift for the ground state
on the sphere or disk geometries can be obtained from the CFT as well.
The shift $\cal S$ in the number of flux $N_\phi$ piercing the sphere
is defined by%
\be%
N_\phi=\nu^{-1}N-{\cal S}.\ee%
The flux $N_\phi$ can be obtained from the degree of the wavefunction
in each coordinate $z_i$, which itself can be obtained by letting
$z_i\to\infty$ and extracting the leading power of $z_i$ (neglecting
 the Gaussian factor). The CFT
contributes $-2h_\psi$ to this, as particle $i$ and the $N-1$
remaining particles must fuse to give the identity \cite{rr2}.
The charge sector contributes $\nu^{-1}(N-1)$. Then the shift is%
\be%
{\cal S}=2(\nu^{-1}/2+h_\psi).\ee%
We write it in this form because when the charge sector is interpreted as a CFT
also \cite{mr}, the conformal weight of the field contributing to
the particle is $\nu^{-1}/2$, so that ${\cal S}/2$ is simply the
total conformal weight of the field representing the particle, including
the charge sector.


\subsection{Bose-Einstein condensates and paired states}
\label{bec}

Some similar trial wavefunctions also have applications outside of
the QH effect. Let us start again with the simplest case, that of
a Bose-Einstein condensate (BEC), with $n\geq 0$ vortices included
in infinite space with no background potential. A trial function
for this is similar to the Laughlin function divided by its
modulus (but with the particle-particle factors completely removed)%
\be%
\prod_{i,l}\frac{(z_i-w_l)}{|z_i-w_l|}\ee%
Unlike the Laughlin states, in such a condensate, while the
average charge (or number) density is uniform, there are large
(Poissonian) fluctuations in the density, or in the number in a
subregion. This is connected with the infinite compressibility of
this BEC in non-interacting particles. One may wonder if the
vortices possess fractional statistics. It is simple to perform
the adiabatic calculation, using expressions similar to those in
Arovas {\it et al.}\ \cite{asw}. There is a part of the holonomy
phase factor related to the expectation of the charge density
times the area of the loop enclosed by the path, which is related
to the Magnus force on the vortex. When vortices are exchanged,
there is a correction due to the charge deficiency at the vortex.
However, for this trial function, the latter is clearly zero, as
the vortices disappear from the density calculation when the
mod-square is taken. Thus, the vortices are bosons. The screening
effect of the Coulomb plasma that arose on taking the mod-square,
which produced a net deficit of charge around each quasihole,
which was so important in the discussion of the QH functions, is
simply absent here.

However, for a BEC, the trial wavefunction above is not very
physical. For one, it has an unpleasant singularity at the
locations of the vortices. Even for non-interacting particles, it
does not solve the Schr\"odinger equation (a function that does is
$\prod_{i,l}(z_i-w_l)$, which is not normalizable; the situation
is better for trapped atoms in e.g.\ a harmonic potential, when
the LLL can again be used). Further, in an interacting Bose
superfluid, the circulation of the fluid around the vortex
produces a ``centrifugal'' force effect, and due to the finite
compressibility of the fluid there is a long-range tail in the
deficiency of density compared with the background, going as $\sim
\xi^2/r^2$, where $r$ is the distance from a vortex, and $\xi$ is
the healing length (this result may also be obtained from a
Gross-Pitaevskii equation analysis---see Ref.\ \cite{fw}---and
shows that the simple form above is not even valid asymptically
outside the vortices). As pointed out by Haldane and Wu
\cite{haldwu}, this leads to a charge deficit in a circle of
radius $R$ centered at the vortex that increases logarithmically
with $R$. Thus, the fractional statistics phase is not path
independent; it depends on the separation of the vortices.
(Compared with the QH case, it also depends differently on the
signs of the vorticities, since the deficiency of particle number
at a vortex is independent of the sign of its vorticity; the above
wavefunction represents vortices all with positive vorticity.)
Clearly, this is connected with being in a phase of matter that is
not ``topological'', due to the existence of gapless Goldstone
(density) modes of the superfluid, as required by the broken
symmetry. Ref.\ \cite{haldwu} also pointed out that if the
interaction between the particles falls off slower than $1/r^2$
(in which case a neutralizing background potential will be
required), then the fluid exhibits screening, and as for the trial
function above, there is no net charge accumulated at the vortex,
and the net phase for exchange of two vortices is zero (in the
borderline case of $1/r^2$ interaction, a non-zero result is
possible).

In this paper we will consider BCS paired states of fermions in
addition to QH states. In these, the vortices carry vorticity in
multiples of half the usual unit, due to the pairing. If the
fermions are not coupled to any gauge field (either the
electromagnetic field, or the Chern-Simons field that arises in
composite particle theory \cite{comppart}), then the fermion
wavefunctions must be
single-valued even in the presence of vortices, while the local
gap function (or condensate wavefunction, or pairing function)
must wind in phase by $2\pi$ on making circuit around a
minimum-vorticity vortex. The charge sector contribution to the
adiabatic statistics is expected to come from viewing the system
as made of composite pairs of particles, and the pairs behave as
bosons, similar to the BEC. The point we wish to emphasize
\cite{rg} is that, like the BEC wavefunction above, the ``nice''
trial wavefunctions that will be considered here do not include
the result of the self-consistent calculation of the gap. That is,
the gap function for the pairing should be calculated by solving
self-consistency conditions from BCS mean-field theory that
incorporates the presence of the vortices. This is difficult, and
in general these details should not be relevant to the topological
properties of a topological phase.  The result of such a
calculation should be similar to the hydrodynamic or
Gross-Pitaevskii calculation for the Bose superfluid, and (for
short range interaction of the fermions) the density deficiency at
the vortices, and consequently the charge sector contribution to
the adiabatic statistics, will not be well defined. Again, the
neutral paired fermion superfluid is a gapless phase with a
Goldstone mode, if there is no long-range interaction. Thus, while
there may be a non-Abelian contribution to statistics from the
paired wavefunctions, as discussed in Ref.\ \cite{rg} and the
present paper, the Abelian contribution is not well-defined; in
this sense, these systems are not in a topological phase. The same
applies to other gapless degrees of freedom, for example when
spin-rotation symmetry is broken. This should be kept in mind when
considering the use of non-QH paired systems for topological
quantum computation, such as the half-flux vortices in He$^3$
\cite{ivanov}. These will most likely only be successful if the
Abelian phases drop out of computations. (Even in QH systems, the
Aharonov-Bohm phase is an inconvenience.)

For interactions falling slower than $1/r^2$, with a neutralizing
background added, the contribution to adiabatic statistics from
the charge sector is well-defined, but is zero. Thus for
interacting electrons with $1/r$ Coulomb interactions, only the
effects other than the charge sector will be left. These can be
calculated using the trial wavefunctions, if there are no other
Goldstone modes in the system apart from the charge mode.

Returning to trial paired states in the presence of vortices but
without the self-consistent calculation of the gap function, one
can make a singular gauge transformation as a function of the
fermion coordinates that turns the trial state into one with a gap
function that does not wind in phase, with the many-particle
wavefunction for the fermions changing in sign on making a circuit
around the vortex. If all the vortices have positive vorticity,
the transformation is multiplication by the inverse square root of
the above BEC trial function. Then for this second gauge choice,
it was shown in Ref.\ \cite{rg} that the long-distance behavior of
a $p+ip$ paired state of spinless or spin-polarized fermions is
the same as that of the MR trial state, as given by the Ising
conformal block (or Pfaffian) above. More generally, we might view
conformal blocks, now without the charge sector factor $\Psi_{\rm
charge}$, as trial wavefunctions for Abelian anyons in zero
magnetic field. These represent superfluid states of anyons, and
we will include this possibility in the following discussion. We
should point out that this relation of QH wavefunctions to those
for particles (possibly of different, though still Abelian,
statistics) in zero magnetic field is precisely the idea of
``composite particles'' \cite{comppart}, here specialized to
composite particles in zero net magnetic field. Later we will
argue when performing the calculations of overlaps for the QH
trial functions that the charge part may as well be removed,
implying that the results also have application to particle
systems in zero magnetic field, consistent with the composite
particle point of view. However, as we have just seen for the
boson case (which is related to composite bosons
\cite{comppart}), in practice there can be important differences
in the behavior in the charge sector that distinguishes these two
types of physical systems.

For such trial functions (conformal blocks) that are not
single-valued in the particle coordinates, some technical issues
must be dealt with in order to discuss the adiabatic statistics of
the vortices. The monodromy of the functions is well-defined if
one keeps track of all the $z$s and $w$s. But for our
purposes we wish to move $w$ only, with $z$'s fixed, and the
result depends on the precise path taken relative to the $z$'s,
due to the square roots in the trial function (this generalizes
directly to particles that are Abelian anyons). As we wish to
compare with holonomy calculated by integrating out the particles
in each infinitesimal time step, this dependence on particle
positions is unacceptable, even though it leads only to an
ambiguity in sign in the present case (more generally, to some
root of unity). This effect was not present for the QH trial
wavefunctions above, which included the charge sector and were
single valued in the particle coordinates (and the particles were
bosons or fermions). [We note that the Berry connection is
well-defined even for the non-single valued wavefunctions, because
the dependence on the particle coordinates is just a square (or
other) root, and the phase change cancels in the overlap (thus
relies on the particles obeying Abelian statistics).] One solution
when the charge sector is removed from the QH wavefunctions to
obtain the blocks considered here is to retain the
gauge-transformation phase factor so that the functions are single
valued in the particle coordinates, which amounts to using the
original gauge. Then we expect still to obtain the charge sector
contribution that is the area enclosed by the loop (because the
particle density is uniform), in addition to the contribution of
the conformal blocks. Another simple solution, which we adopt
here, is that when the exchange of vortices is made along a path
$C$ that is contractible to a limit point on the intersection of
some diagonals, then we may define it as made along a different
path, homotopic to the original in the vortex (or quasihole)
configuration space, that does not enclose any particles. This is
acceptable for the monodromy as it does not require well-separated
vortices (or quasiholes). This approach is not available in the
case of exchange by non-contractible paths, for example on a
surface of non-trivial topology, but we will not enter into this
in this paper.

The preceding applies to fermions that either are not coupled to a
gauge field, or are but the penetration depth for the gauge field
is large. If the penetration depth is instead small, and we
consider exchange of vortices at separations larger than the
penetration depth, then there is a circulating pure-gauge vector
potential outside a penetration depth from the vortices, and the
gap function is covariantly constant. This corresponds to the use
of the second gauge choice above. The calculation of adiabatic
statistics may be made well defined by the gauge transformation
technique as described above.


\subsection{Earlier work}

There have been various earlier steps towards demonstrating that
non-Abelian adiabatic statistics occurs in trial QH wavefunctions
based on conformal blocks, and in certain BCS paired states. The
idea that in the MR wavefunction, the holonomy equals the
monodromy was re-emphasized (though not using this terminology) by
Nayak and Wilczek (NW) \cite{nw}, who also emphasized that this
generalizes screening in the plasma mapping for the Laughlin
states. They also found explicitly the two conformal blocks
corresponding to any even number $N$ of particles and $n=4$
quasiholes. Even though they did not find these in the form of
Majorana fermion zero mode states on the quasiholes (which was
found at around the same time in Ref.\ \cite{rr1}), they guessed
that this interpretation was correct for any number of quasiholes.
This led them to conjecture the form of the braid group
representation in the monodromy, which apart from Abelian factors
(i.e.\ tensor product with an Abelian representation of the braid
group) can be viewed as an image of the braid group in the spinor
representation of the rotation group in $n$ dimensions \cite{nw}.
This representation of the braid group was known \cite{jones}, and
was also known to occur in the Ising CFT (its structure is
described in more detail in Ref.\ \cite{proj}). A similar argument
was spelled out in greater detail in Ref.\ \cite{ivanov}, after
the work of Ref.\ \cite{rg} on the Majorana zero modes in $p+ip$
paired states. However, the argument gives only the monodromy of
the states (modulo Abelian factors), and it is not clear if
adiabatic transport is actually considered in Ref.\ \cite{ivanov}
(no expression for the Berry connection appears there). Another
argument of NW is somewhat similar to the one we will give in
Sec.\ \ref{gencon} below.

Adiabatic transport of quasiholes or vortices in the paired state
was considered further more recently in Ref.\ \cite{stern}, and
especially clearly in Ref.\ \cite{stone} (Appendix A), where it is
shown that the Berry connection is proportional to the identity
matrix, thus proving that the holonomy is given by the monodromy
found by NW, up to some Abelian factor.

Other approaches to the problem for the MR QH state should also be
mentioned. Gurarie and Nayak \cite{gn} used another Coulomb gas
method from CFT to represent the overlap integrals. For the case
of only two quasiholes, they succeeded in obtaining the vanishing
of the Berry connection, and hence that the holonomy equals the
monodromy in this case. For four quasiholes their result depended
on some assumptions, the validity of which does not appear to be
obvious. Other groups \cite{fntw,fns} formulated field-theoretic
arguments, but seem to assume that the edge theory is the expected
CFT. Tserkovnyak and Simon \cite{ts} evaluated the holonomy
numerically for two and four quasiholes by Monte Carlo methods,
finding agreement with the expected result, at some degree of
accuracy.

For most other states, such as those of Read-Rezayi \cite{rr2},
much less has been shown. But there is a series of spin-singlet
trial states due to Blok and Wen \cite{bw} for particles of SU(2)
spin $k/2$ ($k=1$, $2$, \ldots ; $k=1$ is Halperin's Abelian
spin-singlet state, see e.g.\ Ref.\ \cite{mr}), in which the CFT
is SU(2) level $k$, which have many nice properties. These authors
were able to show that the Berry connection vanishes, and so
holonomy equals monodromy, for these states by using the
Knizhnik-Zamolodchikov equation from CFT \cite{dfms}, and making
an assumption that screening holds in an SU(2) generalization of
the Coulomb plasma, in a manner closely parallel to the work of
Arovas {\it et al.}\ \cite{asw}. The screening assumption implies
that the trial ground state has short-range spin correlations. We
will comment on this further in Sec.\ \ref{gencon} below. For recent further
results on monodromy of blocks in the RR series, see Refs.\ \cite{cs,as}.

We conclude from this survey that with few exceptions existing
results in the literature are either only partial ones for the MR
state (not calculating the Abelian factors), or else depend on
unproven assumptions, or apply only to particular states. By
contrast, the results presented below for the MR state are
complete in that they yield the full holonomy for up to four
quasiholes, or for the ground states on a surface of any genus. An
argument that we invoke frequently, which is supported by
renormalization group arguments herein, is that the charge sector
factor of the QH wavefunctions can be dropped without jeopardizing
the results for holonomy in the CFT sector; this is related to
conventional lore about composite particle methods, and has of
course also been used by others, for examples, Refs.\
\cite{stern,stone}. The results for quasiholes rest on an
assumption, that screening occurs in a certain very conventional
two-component plasma, which will be accepted by most physicists.
The arguments given in Sec.\ \ref{gencon} apply to any trial state
given in terms of conformal blocks as explained here, and show
that holonomy equals monodromy under some general conditions that
can be checked for each particular trial state as a well-posed
physical question in two-dimensional field theory. Further, there
is a simple easily-checked criterion (relevance or irrelevance of
a perturbation) that may provide important clues as to whether or not
the general conditions hold.

Now we turn to earlier work on adiabatic variation of the aspect
ratio of a QH system on a torus. In the complex plane, the torus
is defined by identifying points under $z\to z+L$ and $z\to z+
L\tau$, where $\tau$ is in the upper half plane ${\rm Im}\,
\tau>0$; thus the upper half plane is the parameter space on which
we may study adiabatic transport. This was considered for the
filled LLL in an elegant paper by Avron {\it et al.}\ \cite{asz}.
They showed that there is a contribution that is not holonomy, but
curvature (anholonomy or field strength) of the Berry connection
(vector potential) on the upper half plane. (This is somewhat
analogous to the Aharonov-Bohm phase that is proportional to the
area enclosed by a loop, in the Arovas {\it et al.}\ \cite{asw}
calculation for moving one quasihole.) For the case of the filled
LLL, the result is proportional to the total flux through the
torus \cite{asz}. Physically, for a homogeneous many-particle
fluid state, this adiabatic curvature divided by the system area
represents the {\em Hall viscosity}, here denoted $\eta^{(A)}$, a
non-dissipative transport coefficient \cite{avron,odd} that is
known in plasma physics but often overlooked elsewhere in fluid
dynamics; $\eta^{(A)}$ is the only coefficient of viscosity that
can be non-zero in a two-dimensional isotropic incompressible
fluid (but must vanish if time-reversal symmetry is present)
\cite{asz}. The result for the filled LLL can also be extracted
from the detailed single-particle results of L\'{e}vay
\cite{levay}; this paper clarifies many aspects of this problem.
More generally, for the state in which the lowest $\nu$ Landau
levels are filled ($\nu$ integer), the result $\eta^{(A)}=\hbar \nu
\overline{n}/4$ is quoted for integer $\nu$, $|\nu|\geq 1$ in
Ref.\ \cite{avron}, again using results from Ref.\ \cite{levay}
($\overline{n}$ is the particle density $\overline{n}=|\nu|/(2\pi\ell_B^2)$; $\hbar$ and the magnetic length $\ell_B$ are set to 1 elsewhere in this paper).
[Note that following comments from the authors of Ref.\ \cite{tv}, we have corrected the coefficient to $1/4$ to account for apparent typos in Ref.\ \cite{avron}.]

A recent paper \cite{tv} has argued that the result of Ref.\
\cite{asz} generalizes to an arbitrary QH state in the LLL to give
$\eta^{(A)}=\hbar \nu/(8\pi\ell_B^2)$ for arbitrary (possibly
negative) values of $\nu$, independent of the state (this paper
\cite{tv}, and references therein, use the Hall viscosity in an
interesting hydrodynamic approach to collective modes in QH fluids).
Unfortunately, the end of the calculation uses the incorrect argument that all
particles are close to the $x$-axis. The single-particle
results of L\'{e}vay \cite{levay} can be applied to obtain a seemingly similar result. He finds, for the Berry calculation in the $N_\phi$-fold
degenerate space of single-particle states in a single Landau
level ($N_\phi$ is the number of flux piercing the torus), that
the Berry connection and its curvature are proportional to the
identity matrix in this space, with a coefficient related to the
kinetic energy of the Landau level. This
implies (e.g.\ by using the Slater determinant basis) that for the
space of all many-particle states with all the particles in a
single Landau level, the Berry connection and curvature are again
proportional to the identity {\em matrix}, and so in this sense
are independent of the state. (This generalizes further if we
consider a space consisting of all the many-particle states with a given
number of particles in each Landau level.) However, this adiabatic
transport is not what should be considered for a quantum fluid state,
in which because of the presence of interactions we take a family
consisting of a single ground state (as considered in Ref.\ \cite{tv})
for each value of $\tau$, or more generally a space of ``degenerate'' states for each $\tau$, but generally not the full space of many-particle single Landau level states. This is an example of the general set-up for adiabatic
transport, in which a family of subspaces within a common Hilbert
space is considered, as discussed above in Sec.\ \ref{adtrans}. (The
$\tau$-dependent Landau level states themselves arise in this way,
as all are subspaces of the space of square-integrable functions
on the torus with given boundary conditions.) The curvature of the
Berry connection definitely does in general depend on the choice
of a subspace for each value of $\tau$ (though not on the gauge,
that is the basis for the subspace), and cannot be reduced to a
calculation that ignores restriction to the subspace. Hence the
result that is independent of the state chosen, as claimed in Ref.\
\cite{tv}, cannot be as general as was stated. Later, we will show
that the correct result depends on the form of the ground state,
not only on the density, though it is universal within a
topological phase. (We will also find related results for paired
superfluids. We are not aware of any earlier results for Hall
viscosity of paired states.)

{}From L\'{e}vay \cite{levay}, the result of Ref.\ \cite{tv} can be viewed as the correct one for non-interacting particles in the LLL. Also, using L\'{e}vay's results for non-interacting particles but at non-zero temperature $T$, and going to high temperature using classical equipartition of energy, we find that the result agrees with that in Ref.\ \cite{odd}, which is $\eta^{(A)}=k_B T \bar{n}/(2\omega_c)$, where $\omega_c$ is the cyclotron frequency.


\subsection{Structure of paper}

In Sec.\ \ref{genpair}, we first review the essentials of paired states at the BCS mean-field level, then specialize to ground states on the torus. We calculate the normalization factors, then examine the monodromy under modular transformations, and then the curvature and holonomy of the Berry connection for changes in the aspect ratio $\tau$; this determines the Hall viscosity of BCS paired systems, for which we also give a simple direct calculation. We also discuss higher genus surfaces, and the strong-pairing phases. In Sec.\ \ref{halllaugh}, we examine similar questions for the Laughlin QH states, relate the Hall viscosity for trial states given by conformal blocks to the conformal weight of the field for the particle, and conclude the general discussion for Hall viscosity. In Sec.\ \ref{statcalc}, we present direct arguments for the non-Abelian adiabatic statistics of two and four quasiholes in the MR state on the sphere or plane in the thermodynamic limit. The calculations work by ``doubling'' (taking two copies) of the system with the charge part removed, and using an argument that a plasma is in a screening phase. In Sec.\ \ref{gencon}, we present general arguments that amount to necessary and sufficient conditions for trial wavefunctions given by conformal blocks to describe a topological phase with adiabatic statistics given by the monodromy of the blocks. The condition is that related correlation functions in {\em two} dimensions should go to a constant, as in a ordered phase. We discuss numerous examples in this light. Cases not obeying the condition are argued to be gapless phases or critical points. We argue that use of non-unitary rational CFTs in our way cannot produce such a topological phase if there are any negative quantum dimensions in the theory (which follows if there are any negative conformal weights). The argument assumes that the {\em twist} in the theory, defined adiabatically, is also the same as in the CFT, which has not been shown. In an Appendix we discuss this, which is the last step in deriving a modular tensor category from the construction, and show that the consistent twists are very limited, so that the argument does go through in at least one family of examples.


\section{Paired states in a closed finite system}

\label{genpair}

In this section, we consider the $p+ip$ paired states on compact
two-dimensional surfaces (no boundary). The set-up was already
described in Ref.\ \cite{rg} (see also Ref.\ \cite{stone}), but we
review some essential steps, and add a few points. The basic case,
other than the sphere, is the torus, or equivalently periodic
boundary conditions on a parallelogram in the plane. The other
cases, surfaces of genus higher than one, require more work to set
up, and we will be more brief. For all these problems, we can show
starting from the general pairing problem, that at long
wavelengths the orthonormalized wavefunctions (within the BCS
mean-field formulation) are the conformal blocks of the Majorana
fermion (Ising) CFT. This allows us to calculate explicitly the
adiabatic transport of the states as the aspect ratio of the torus
is varied; this leads to the Hall viscosity and the modular
transformation group. We also consider the generalization to other
paired phases of fermions, including the strong-pairing phases.


\subsection{General equations for pairing}

We begin with BCS mean field theory \cite{schrieffer}. The
effective grand-canonical Hamiltonian for the fermions is, in the
most general case (we assume for the present that the fermions are
in a finite set of orbitals labeled by Greek indices like
$\alpha$, taking values $\alpha=1$, \ldots, $M$)%
\be%
K_{\rm
eff}=\sum_{\alpha,\beta}\left[h_{\alpha\beta}c_\alpha^\dagger
c_\beta+\frac{1}{2}\left(\Delta_{\alpha\beta}c_\alpha^\dagger
c_\beta^\dagger+\overline{\Delta}_{\alpha\beta}c_\beta
c_\alpha\right)\right].\ee%
Here the creation and annihilation operators, $c_\alpha^\dagger$
and $c_\alpha$, which are adjoints of one another, obey the
canonical anticommutation relations
$\{c_\alpha,c_\beta^\dagger\}=\delta_{\alpha\beta}$,
$\{c_\alpha,c_\beta\}=0$. The $M\times M$ matrix $h$ must be
Hermitian, $\overline h_{\alpha\beta}=h_{\beta\alpha}$ (the
$\bar{}$ is complex conjugation), and represents the usual kinetic
energy and any one-body potential terms, while $\Delta$ is an
antisymmetric $M\times M$ matrix,
$\Delta_{\alpha\beta}=-\Delta_{\beta\alpha}$ and corresponds to
the gap function. The former condition ensures that $K_{\rm eff}$
is Hermitian, while the latter means that there are no redundant
components of $\Delta$.

It will be convenient to write everything in terms of a
$2M$-component column vector of field operators $C$, in which the
first $M$ components are $c_\alpha$, the remaining $M$ are
$c_\alpha^\dagger$. Then, up to an additive constant, %
\be%
K_{\rm eff}=\frac{1}{2}C^\dagger
\left(\begin{array}{cc}h&\Delta\\
                   -\overline\Delta&-\overline{h}
                   \end{array}\right)C
 =\frac{1}{2}C^\dagger{\cal H}C.          \label{calH}     \ee%
The $2M$-dimensional vector space can be viewed as a tensor
product of $M$-dimensional space with a two-dimensional
``particle-hole'' space. Then we may use a tensor product notation
for matrices, with Pauli matrices $\sigma_\mu$, $\mu=x$, $y$, $z$,
acting in the two-dimensional space. Then the
$2M\times 2M$ matrix $\cal H$ obeys %
\be%
{\cal H}^\dagger={\cal H}=-\Sigma_x {\cal H}^T\Sigma_x,\ee%
where $^T$ is transpose, $\Sigma_x=\sigma_x\otimes {\bf 1}_M$, and
${\bf 1}_M$ is the $M\times M$ identity matrix.

We will now diagonalize $K_{\rm eff}$ and find the ground state,
by performing a Bogoliubov transformation. First we note the
anticommutation relations for $C$, in terms of
$\alpha$, $\beta=1$, \ldots, $2M$, %
\bea%
\{C_\alpha,C_\beta\}&=&\Sigma_{x,\alpha\beta},\non\\
\{C_\alpha,C_\beta^\dagger\}&=&\delta_{\alpha\beta}.\eea%
To diagonalize the Hamiltonian, we require creation and
annihilation operators for quasiparticle modes, $\chi_r$,
$\chi_r^\dagger$ ($r=1$, \ldots, $M$), which we combine into a
$2M$-component column vector $\widetilde{\chi}$, similar to $C$.
These operators $\widetilde{\chi}$ must obey the canonical
anticommutation relations of the same form as those for
$C$ above. If we define explicitly %
\be%
C=\left(\begin{array}{c}c\\
c^\dagger\end{array}\right)=U\widetilde{\chi}=
\left(\begin{array}{cc}\overline u&v\\
\overline v&u\end{array}\right)
\left(\begin{array}{c}\chi\\
\chi^\dagger\end{array}\right),
\ee%
where $u$, $v$, are complex $M\times M$ matrices, then requiring
$C$ and $\widetilde{\chi}$ to satisfy the canonical
anticommutation relations, we must have%
\bea%
\overline u v^T+v u^\dagger&=&0,\non\\
\overline u u^T+vv^\dagger&=&{\bf 1}_M.\label{Urels2}\eea%
Given the above form for $U$, these equations are equivalent to
$UU^\dagger={\bf 1}_{2M}$, and it follows that
$U^{-1}=U^\dagger=\Sigma_x U^T \Sigma_x$. These imply that $U$ is
an element of the Lie group O$(2M)$. We note that the condition
$U^\dagger U={\bf 1}_{2M}$ leads to additional relations%
\bea%
u^Tv+v^T u&=&0,\non\\
u^\dagger u+v^\dagger v&=&{\bf 1}_M,\label{Urels}\eea%
which may be useful, but are equivalent to those above as we are
dealing here with finite matrices for which the right and left
inverses of a matrix are equal.

To make $K_{\rm eff}$ diagonal, we require that%
\be%
K_{\rm eff}-K_0=\frac{1}{2}\widetilde{\chi}^\dagger
\left(\begin{array}{cc}E&0\\
0&-E\end{array}\right)\widetilde{\chi}
=\frac{1}{2}\widetilde{\chi}^\dagger\widetilde{E}\widetilde{\chi},
\label{diagK}
\ee%
where $E$ is an $M\times M$ diagonal real matrix, $\widetilde{E}$
is a $2M \times 2M$ matrix, and $K_0$ is the ground state energy,
where by ``ground state'' we mean the state annihilated by all
$\chi_r$. The conditions for $E$ to be diagonal, which can be
found by commuting $\chi_r$ with $K_{\rm eff}$, are the matrix
equations (a complete eigenvalue problem) ${\cal
H}U=U\widetilde{E}$, or more explicitly%
\bea%
\overline h u+\overline\Delta v&=&uE,\non\\
\Delta u+h v&=&-vE,\label{bdgi}\eea%
which are the BdG equations for this problem. (The BdG equations
may look more familiar if one takes the $r$th column of each of
these, which gives the eigenvalue problem for the $r$th eigenvalue
$E_r$, for $r=1$, \ldots, $M$.) We notice that the equations for
$\overline u$, $\overline v$, which are the complex conjugates of
these, are in fact the same, by replacing $\overline u\to v$,
$\overline v\to u$, $E\to-E$. This can also be applied to an
individual column of $u$, $v$, say the $r$th, to effectively
exchange the corresponding $E_r$ with $-E_r$.

The BdG equations are not easy to solve in closed form in this
general case. Fortunately we do not require these general
solutions. For the moment, all we need is the form of the ground
state wavefunction. We will concentrate on the case in which the
ground state contains only states of even particle number. The
ground state can be written in the unnormalized BCS form %
\be%
|\Omega\rangle = \exp\left(\frac{1}{2}\sum_{\alpha,\beta}
g_{\alpha\beta} c_\alpha^\dagger c_\beta^\dagger\right)|0\rangle,\label{Omega}\ee%
where $|0\rangle$ is the vacuum, which is annihilated by all
$c_\alpha$s, and $g$ is an antisymmetric $M\times M$ matrix. The
condition that all $\chi_r$ annihilate $|\Omega\rangle$ yields the
relation%
\be%
u^Tg=-v^T,\ee%
as a matrix equation, and so $g=vu^{-1}$ by taking the transpose,
or by using the first of eqs.\ (\ref{Urels}). The wavefunction of
the ground state is then a Pfaffian, up to an overall sign:%
\be%
\Psi(\alpha_1,\ldots,\alpha_N)=\langle0|c_{\alpha_1}\ldots
c_{\alpha_N}|\Omega\rangle=\pm{\rm
Pf}\,g_{\alpha_i\alpha_j}.\ee%
Note that for any state $|\Omega\rangle$, with wavefunctions
$\Psi$ defined in this manner for each $N$-particle subspace,
the norm square is %
\be%
\langle\Omega|\Omega\rangle=\sum_N\sum_{\alpha_1,\ldots,\alpha_N}
|\Psi(\alpha_1,\ldots,\alpha_N)|^2/N!.\ee%

We now obtain the normalization of the ground state wavefunction.
To calculate $\langle\Omega|\Omega\rangle$, we expand the
exponentials and use Wick's theorem. With the help of some simple
combinatorics, we obtain%
\be%
\langle\Omega|\Omega\rangle=\exp\frac{1}{2}{\rm Tr}\,\ln({\bf
1}_M+g^\dagger g)=\sqrt{\det({\bf 1}_M+g^\dagger g)}.\ee%
Using the second of eqs.\ (\ref{Urels}), we obtain %
\be%
uu^\dagger = ({\bf 1}_M+g^\dagger g)^{-1}.\ee%
Hence $(\det uu^\dagger)^{1/4}|\Omega\rangle$ is normalized, a
generalization of the usual normalization for BCS theory in
momentum space (see e.g.\ Refs.\ \cite{schrieffer,rg}). Also the
expectation value $\overline{N}$ of the particle number
$N=\sum_\alpha c_\alpha^\dagger
c_\alpha$ is%
\be%
\overline{N}={\rm Tr}\, v^\dagger v={\rm Tr}\,g^\dagger g({\bf
1}_M+g^\dagger g)^{-1}.\ee

By writing the BdG equations in terms of $\widehat{E}=uEu^{-1}$ in
place of $E$, one finds that%
\be%
g=\overline\Delta^{-1}(\widehat E-\overline h),\ee%
and then by eliminating $\widehat{E}$, $g$ obeys
a matrix quadratic equation,%
\be%
\Delta+g\overline h+hg+g\overline \Delta g=0,\ee%
so that $g$ can be found without first solving for $E$ or
$\widehat{E}$. (This is similar to a diagonal equation in Ref.\
\cite{vollwolf}, p.\ 68.) This equation may also be obtained more
quickly by applying $K_{\rm eff}$ to $|\Omega\rangle$, whence one
also finds that the ground state energy is $K_0=-{\rm
Tr}\,\overline{\Delta}g/2$.

In some situations of interest, $h$ is a constant multiple of the
identity matrix, say $h=-\mu$. In this case the BdG equations
(\ref{bdgi}) can be analyzed further. By multiplying the first of
equations (\ref{bdgi}) on the right by $(E-\mu{\bf 1}_M)$, and the
second on the left by $\overline\Delta$, then
eliminating $v$, we obtain%
\be%
uE^2=(\mu^2{\bf 1}_M+\Delta^\dagger\Delta)u.\ee%
Here we used also $-\overline\Delta=\Delta^\dagger$, and we note
that $\Delta^\dagger\Delta$ is Hermitian and positive. Thus we see
that the columns of $u$ are the eigenvectors of
$\Delta^\dagger\Delta$. Similarly, the columns of $v$ are the
eigenvectors of $\Delta\Delta^\dagger$. It follows that the
diagonal entries of $E^2$ are greater than or equal to $\mu^2$. In
fact, a standard result from linear algebra allows a complex
antisymmetric matrix $\Delta$ to be written in the form%
\be%
\Delta=w^T D w,\ee%
where $w$ is unitary, and $D$ is an antisymmetric $M\times M$
block-diagonal matrix of $[M/2]$ ($[x]$ means the largest integer
$\leq x$) $2\times 2$ blocks of the form $\lambda_\alpha
\sigma_y$, where $\lambda_\alpha$ ($\alpha=1$, \ldots, $[M/2]$)
obey $\lambda_\alpha\geq0$, and for $M$ odd also a single $1\times
1$ block containing $0$. The transformation can be considered as a
basis change among the orbitals $C$ only, or as a Bogoliubov
transformation with $u=w$ and $v=0$, that brings $\cal H$ to a
form with $\Delta$ replaced by $D$, so that the gap function links
orbitals in pairs only (similar to the case of BCS theory for a
translationally-invariant system in $\bf k$ space, where the gap
functions links $\bf k$ and $-\bf k$ only). It follows that the
eigenvalues in $E^2$ are $\mu^2+\lambda_\alpha^2$ with
multiplicity $2$ for each $\alpha$, and an additional $\mu^2$ if
$M$ is odd.

Taking the positive solution for $E$, we have
\be%
E
=w\sqrt{\mu^2{\bf 1}_M+\Delta^\dagger\Delta}w^{-1},\ee%
and the positive square root of a positive Hermitian matrix can be
defined through spectral theory. At the same time, we also have
\be%
E=u^{-1}\widehat{E}u=u^{-1}\sqrt{\mu^2{\bf 1}_M+\Delta^\dagger\Delta}
u,\ee%
and consequently $g$ is%
\be%
g=\overline\Delta^{-1}\left(\mu{\bf 1}_M + \sqrt{\mu^2{\bf 1}_M
+\Delta^\dagger\Delta}\right),
\ee%
which solves the quadratic equation for the present case, which is%
\be%
\Delta-2\mu g+g\overline\Delta g=0.\ee%
This solution leads to %
\bea%
g^\dagger g&=&{}\left(\mu+\sqrt{\mu^2+\Delta^\dagger\Delta}\right)
(\Delta^\dagger\Delta)^{-1}\left(\mu+\sqrt{\mu^2
+\Delta^\dagger\Delta}
\right)\non\\
&=& {}\left(\mu+\sqrt{\mu^2+\Delta^\dagger\Delta}\right)^2
(\Delta^\dagger\Delta)^{-1},\eea%
The two expressions for $E$ are consistent, because we can write
the polar decomposition of $u$, $u=|u|w'^{-1}$, where $|u|$ is
defined as $|u|=\sqrt{uu^\dagger}=({\bf 1}_M+g^\dagger g)^{-1/2}$
and $w'$ is unitary. Then we see that $|u|$ commutes with
$\sqrt{\mu^2{\bf 1}_M+\Delta^\dagger\Delta}$, and we can take
$w'=w$. As $E$ is positive, it follows that the solution for $g$
describes the lowest-energy ground state.

Next we will pass to a continuum limit, in which the finite
matrices are now operators as in an infinite-dimensional single
particle Hilbert space. This applies to systems in continuous
space, and also for the large distance limit of a lattice. In all
cases, our interest will be in scales larger than the coherence
length, which can be thought of as the characteristic size of a
Cooper pair (and which we will assume does not depend
significantly on position). Then the terms in $\cal H$ will be
assumed, first for the translationally-invariant case, to be of
the form $h=k^2/(2m^*)-\mu$ and $\Delta=\hat{\Delta}(k_x-ik_y)$
when written in momentum space; at present $m^*$ (the effective
mass of the fermions), $\mu$ (the chemical potential), and
$\hat{\Delta}$ (the magnitude of the gap function) are constant.
We have assumed for $\Delta$ the characteristic form for
$p_x-ip_y$ pairing, as in Ref.\ \cite{rg} (the coefficients of
$p_x$ and $p_y$ are the same, but a different non-zero ratio could
be removed by a rescaling once and for all). Then in position
space these become differential operators; to represent these as
functions of two positions ${\bf r}_1$, ${\bf r}_2$ as in the
above formalism, with $\alpha$ replaced by ${\bf r}=(x,y)$
everywhere, one would consider them as integral kernels, by
including suitable
$\delta$-functions. They are %
\bea%
h&=&-\nabla^2/(2m^*)-\mu\non\\
\Delta&=&-2i\hat{\Delta}\partial\eea%
where $\partial\equiv\partial/\partial z\equiv
\frac{1}{2}(\partial_x-i\partial_y)$. In fact, the $k^2$ term in
$h$ can also be dropped for our purposes, even at $\mu=0$ which is
the transition point between weak and strong pairing \cite{rg}.
The behavior of both operators is implicitly assumed to change at
large wavevectors, so as to include suitable physical cutoffs,
including the coherence length. For long-wavelength purposes, the
above results for $h=-\mu$ a constant are now applicable.


\subsection{Periodic boundary condition ground states}

The case of periodic boundary conditions was already considered in
Ref.\ \cite{rg}, so we will be brief, and just make some
additional points. For $h=-\mu$, $\hat{\Delta}$ both constant, we
can apply the above results in position space. {}From here on, we
use units in which $|\hat{\Delta}|=1$ in this case. Then as $g$
acts from the left on $c^\dagger$, use of the standard Fourier
transform definitions, and by comparison with
\be%
 K_{\rm
eff}=\sum_{\bk}\left[-\mu c_\bk^\dagger c_\bk +\frac{1}{2}\left(
\Delta_\bk^\ast c_{-\bk}c_\bk+\Delta_\bk
c_\bk^\dagger c_{-\bk}^\dagger\right)\right], \ee%
we obtain $\Delta_\bk=k_x-ik_y$, and%
\be%
g_\bk=\left(\mu+\sqrt{\mu^2+|\bk|^2}\right)/(k_x+ik_y), \ee%
which agrees with Ref.\ \cite{rg} up to an unimportant minus sign
\cite{rg}. In position space, for $\mu>0$, we find
$g(\br_1,\br_2)\equiv g(\br_1-\br_2)\sim 2\mu/(z_1-z_2)$ for
$|\br_1-\br_2|\gg 1/|\mu|$, where $z_1=x_1+iy_1$, etc. For
distances less than of order $1/|\mu|$, the behavior changes
\cite{rg}.

This behavior of $g(\br)$, which is what appears in the
many-particle wavefunction (the Pfaffian), might surprise us as we
usually think of the fermions as forming bound Cooper pairs,
whereas the integral of our $|g(\br)|^2$ over infinite space
diverges. However, the usual physical way to think of these pairs
is to examine the order-parameter function $\langle
c(\br_1)c(\br_2)\rangle$, which is a function of $\br_1-\br_2$
only. This is the Fourier transform of $\langle c_\bk
c_{-\bk}\rangle=u_\bk^\ast v_\bk=-\Delta_\bk/E_\bk$ (where
$E_\bk=\sqrt{\mu^2+|\bk|^2}$). Using the same approximations, this
is analytic near $\bk=0$, and so has no long-range tail at
distances $>O(1/|\mu|)$, though it does at shorter distances. If
$\mu=0$, one finds the same behavior $\sim 1/(z|z|)$ as
$|z|\to\infty$ (as for $g(\br)$ in this case). This function is
normalizable, and so the size of the Cooper pair is still the
coherence length $\xi_0\ll 1/|\mu|$ that must be defined using
shorter-distance or larger wavelength behavior of the pairing
(note, however, that these arguments show that if one attempts to
use the expectation of $|\br|^2$ in this function to define the
size of a pair, one encounters a logarithmic divergence at
$\mu=0$). Then in the topological (weak-pairing, $\mu>0$) phase of
interest, the interesting effects will be transmitted by the
long-range behavior of $g(\br)$, yet this does not show up if one
looks at the natural order parameter as above. This is to be
expected for a topological property of a phase.

We now consider the ground states on the torus, as this is a
warm-up problem for the issues of most interest in this paper.
Here we can take the torus to be the $\br$ plane, modulo
translations by two linearly-independent vectors
{\boldmath$\omega$}$_1$, {\boldmath$\omega$}$_2$, with
{\boldmath$\omega$}$_1$ along the positive $x$-axis and of length
$L_x$. If we identify these vectors with complex numbers, then the
first becomes $L_x$, and we define the second to be $L_x\tau$. We
assume that ${\rm Im}\, \tau>0$, and define $L_y=L_x{\rm
Im}\,\tau$. When we vary the aspect ratio $\tau$ we will assume
that $L_xL_y$ is fixed.  First we assume that the fermions obey a
periodic boundary condition
$c(\br+\hbox{{\boldmath$\omega$}$_1$})=
c(\br+\hbox{{\boldmath$\omega$}$_2$})=c(\br)$, and that there is
no vector potential. Then from the underlying physics that
produces $p+ip$ pairing, we can assume that there is a ground
state with $\hat{\Delta}$ constant. In the weak-pairing phase
$\mu>0$, this has odd particle numbers $N$ in all components of
the state \cite{rg}. By boosting all the fermions by a momentum of
order $2\pi/L_x$, we can imagine that there is also a low energy
state with a net circulation around the $x$ cycle parallel to
{\boldmath$\omega$}$_1$, or similarly for the $y$ cycle parallel
to {\boldmath$\omega$}$_2$, or both. In terms of the small
wavevector behavior of the gap function $\Delta$, these involve
$\hat{\Delta}(\br)$ that winds by $2\pi$ along the $x$ cycle, or
the $y$ cycle, or both, respectively. Because the change in the
state is small, locally, these should exist as low energy states.

In the case when $\hat{\Delta}(\br)$ varies with position, some
generalization of the above expressions is necessary. The most
important possible variation is in the phase of
$\hat{\Delta}(\br)$, though the magnitude also varies in some
important situations. Then in either case it becomes%
\be%
\Delta=-i\{\hat{\Delta}({\br}),\partial\}\ee%
which involves the anticommutator of one-body operators. Thus the
effective Hamiltonian becomes%
\be%
K_{\rm eff}= \int
d^2r\left[c^\dagger(\br)hc(\br)+\frac{1}{2}\left(
c^\dagger(\br)\Delta
c^\dagger(\br)+c(\br)\overline\Delta c(\br)\right)\right]\ee%
In the present section we can see that $\Delta$ is antisymmetric
by integrating by parts. We can
also see that $K_{\rm eff}$ is gauge invariant under %
\bea%
c(\br)&\to& e^{i\Lambda(\br)}c(\br),\non\\
\hat{\Delta}(\br)&\to& e^{2i\Lambda(\br)}\hat{\Delta}(\br),\eea%
with no need to include a vector potential. The physical
motivation for the form of $\Delta$, at least for $\hat{\Delta}$
constant in magnitude, is to ensure this gauge invariance and
antisymmetry. Thus these forms are valid at long wavelengths
whether the superfluid is neutral (no U$(1)$ vector potential), or
charged (coupled to the scalar and vector potentials of
electromagnetism, or a Chern-Simons (CS) gauge potential in the CS
formalism for the FQH effect). In the charged or CS cases, the
vector potential would enter in the higher-derivative terms that
we neglect here. (We correct here some slight mis-statements about
the role of the vector potential in Ref.\ \cite{rg}.)

{}From the form of $\Delta=-i\{\hat{\Delta}({\br}),\partial\}$, we
can see that in the states in which $\hat{\Delta}$ winds in phase
around the torus, instead of $\bk$ pairing with $-\bk$, $\bk$ is
paired with $-\bk$ shifted by a small amount. These states
correspond to the minimum quantized amounts of vorticity or
circulation of the superflow around the two cycles (we do not call
these flux as there is no vector potential present at the moment).
They can be generalized further to states with $m_1$, $m_2$ units
of quantized circulation around the two cycles. These states do
not all have the same energies in the present situation, because
there will be contributions to the total energy which can be
expressed as a Ginzburg-Landau (GL) type functional of $\Delta$,
and this energy density will presumably include a term like
$|\nabla\hat{\Delta}|^2$ (we assume that the gap function, which
after all is actually a function in $\bk$ space, can be adequately
characterized by a single complex number for this purpose, as in
the usual GL theory for $s$-wave paired states). This term arises
from the extra kinetic energy of the fermions in these states, for
example. For example, for the case $m_2=0$, the total energy cost
will go as $m_1^2L_y/L_x$, multiplied by constants independent of
system size. When $m_1$, $m_2$ are both even, these states occur
at odd particle numbers in the weak-pairing phase, while in the
other cases (and in all cases in the strong-pairing phase $\mu<0$)
they occur at even particle numbers.

An alternative point of view is also useful. As mentioned in Sec.\
\ref{bec}, for each of the winding states, we may perform a gauge
transformation to bring $\hat{\Delta}$ back to a constant, which
will presumably be close in magnitude to that in the $(0,0)$
ground state. Acting on the fermions, this is a phase change by
$-\pi m_{1,2}$ along the two cycles, and produces a corresponding
change in the boundary condition. Thus the boundary conditions on
the fermions can be described by whether or not there is a sign
change around each of the two cycles. We will write the boundary
condition as $+$ if the corresponding $m$ is even, $-$ if it is
odd. Thus there are four possible boundary conditions
corresponding to the four values of $(m_1,m_2)$ modulo $2$. For
each of the four choices, there is a tower of states in which the
winding numbers relative to the lowest energy state in the tower
are $0$ (mod 2). We note that the gauge transformations used here
are  ``large'' transformations (though not singular anywhere), as
for $m_1$, $m_2$ not both zero they are not continuously connected
to the trivial transformation by a path in the space of gauge
transformations.

Returning to the original gauge (boundary condition), for
superfluids in which a U$(1)$ vector potential is present, the
same winding states still exist, but their energies are now
shifted because of the possible holonomy of the vector potential
along the two cycles (again, this effect is due to the entrance of
the vector potential into the larger wavevector aspects of the
physics). This holonomy is the usual gauge-invariant notion of the
flux of the vector potential threading the two cycles on the
torus. Although the holonomies for the two cycles can take any
real values, when they are equal to $\pi (m_1,m_2)$ for some
integers $m_1$, $m_2$ (in units of $hc/e$, in ordinary units) then
the corresponding winding state has the lowest energy, and this
energy is independent of $(m_1, m_2)$ by a (large) gauge
transformation argument. This is referred to as flux quantization
\cite{schrieffer}. When the gauge field is viewed as an additional
quantum degree of freedom, all of these states are distinct in the
combined Hilbert space of particle and electromagnetic field
states. Note that physical states can be viewed as invariant under
gauge transformations that are continuously connected to the
identity, but not under ``large'' gauge transformations, because
the holonomies represent real physical flux.

In the QH applications, the physics is slightly different. In this
case the vector potential is not a truly independent degree of
freedom, but instead is tied to the total momentum of the
fermions. Further this momentum space is compactified; it is
topologically a torus \cite{hald85}. While the different winding
states corresponding to the four choices $\pm$ for each cycle are
still distinct valid ground states \cite{rg} (when the signs are
not both $+$ they lie on the boundary of the Brillouin zone in
pseudomomentum \cite{hald85}), it appears that the towers of
winding states over each of these are not distinct states, but can
all be identified with the lowest one.

In the gauge-transformed view of the $(m_1,m_2)$ winding states,
whether a vector potential is present or not, the long-distance
part of the pairing function $g$ is independent of the vector
potential (if present), as we have shown, and so the long-distance
wavefunction only depends on $m_1$, $m_2$ (mod 2), that is on the
signs $\pm$. The pairing functions $g$ in the four cases have been
determined previously, and in the weak-pairing phase at scales
larger than $1/\mu$ are essentially Jacobi elliptic functions
(except for some slight complications in the ++ case) \cite{rg}.
At large lengths (i.e. separation $|{\bf r}_1-{\bf r}_2|$ of the
particles), $g_a$ in boundary condition $a$ is $2\mu$ times the
inverse of $\overline{\Delta}$, and thus of $\overline\partial$,
up to a constant factor independent of the system size, aspect
ratio, and boundary condition. These functions may be determined
using complex analysis arguments, and for $m_1$, $ m_2$ not both
zero (mod 2) are (up to such a constant)%
\bea%
\lefteqn{g_{m_1,m_2}(\br|\tau)\propto}\non\\
&&\frac{\vartheta_{(m_1+1)/2,(m_2+1)/2}(z/L_x|\tau)
\partial\vartheta_{1/2,1/2}(0|\tau)}
{L_x\vartheta_{1/2,1/2}(z/L_x|\tau)\vartheta_{(m_1+1)/2,
(m_2+1)/2}(0|\tau)},
\label{ellip}\eea%
where $m_1$, $m_2=0$, $1$, and $\vartheta_{a,b}(z|\tau)$ are
elliptic theta functions with characteristics, defined by%
\be%
\vartheta_{a,b}(z|\tau)=\sum_n e^{i\pi\tau(n+a)^2+2\pi
i(n+a)(z+b)}.\ee%
We note that $\vartheta_{1/2,1/2}(z|\tau)$ is the odd theta
function (odd under $z\to-z$), and so has a zero at $z=0$ and
points related by translation by $1$ or $\tau$, while the others
we use are even about $z=0$ (see e.g.\ \cite{mumtheta}). We define
$\partial\vartheta_{1/2,1/2}(0|\tau)=\partial\vartheta_{1/2,
1/2}(z|\tau)|_{z=0}$. Thus the elliptic functions on the
right-hand side of eq.\ (\ref{ellip}) have a simple pole with
residue $1$ at $z=0$, and the required (anti-)periodicities. For
$m_1$, $m_2$ both zero, we have instead%
\be%
g_{0,0}(\br|\tau)\propto
\frac{\partial\vartheta_{1/2,1/2}(z/L_x|\tau)}
{\vartheta_{1/2,1/2}
(z/L_x|\tau)}
+\frac{2\pi i {\rm Im}\, z/L_x}{{\rm Im}\,\tau},\label{ellip00}\ee %
(with the same proportionality constant as before) which is
periodic, but not analytic. The non-analytic term cancels in the
ground states with odd particle number \cite{rg}, in which the $\bk=0$
state is occupied by an unpaired fermion. That is, the
wavefunction for $N$ fermions is the product of $g$s for the pairs,
times $1/\sqrt A$ for the unpaired fermion ($A=L_x^2{\rm
Im}\, \tau$ is the area), antisymmetrized over all fermions. For
each pair $i$, $j$, the non-analytic term is proportional to
${\rm Im}\,z_i-{\rm Im}\,z_j$. It breaks into terms ${\rm Im}\,z_i$
(independent of $z_j$) and the same with $i$, $j$ interchanged. The
former of these two puts $j$ in the constant mode, which is already
occupied by at least one other particle, so this contribution vanishes
on antisymmetrization. (For a long proof, see Ref.\ \cite{stone}.)
Hence all such terms give zero, and the non-analytic term can be
dropped for all pairs; the wavefunction of course is still periodic
\cite{rr1}.

One further step will be useful for later purposes. The factor $1/L_x$
in each pairing function can be rewritten as $({\rm Im}\,\tau)^{1/2}/
A^{1/2}$, where $A=L_x^2{\rm Im}\,\tau$ is the area.

Now we turn to the norms of these paired states, in the gauge
transformed point of view. {}From the general theory above, the
factor that is needed in the state to normalize it
is %
\be%
\left[\det({\bf 1}+g^\dagger g)\right]^{-1/4}
\ee%
(up to an overall phase), which equals%
\be%
\frac{\left[\det\Delta^\dagger\Delta\right]^{1/4}}%
{\left[\det\left(\Delta^\dagger\Delta+(\mu+\sqrt{\mu^2
+\Delta^\dagger\Delta})^2
\right)\right]^{1/4}}.\label{normfac}\ee%
In our case, $\Delta=-2iD$, and
$\Delta^\dagger=-\overline\Delta=-2i\overline{D}$ [here $D$ is
$\partial$, with the understanding that it acts on functions of
the chosen (anti-)periodicity]. On passing to momentum space, we
find that these expressions are identical to those in Ref.\
\cite{rg}. For long wavelengths, the denominator is non-singular
and we can ignore it; it depends exponentially on the area of the
system, but not on its aspect ratio $\tau$ because effects of
finite size are exponentially decaying corrections (formally this
is shown by applying Euler-Maclaurin methods to the logarithm of
the product of factors, and noting the analytic behavior in $\bk$
near $\bk=0$). But the numerator is $\det\overline D$ times its
conjugate (times a constant), and is nonanalytic near $\bk=0$.
With a momentum cutoff of order $\mu$, this product can be
evaluated (for $L_x\gg 1/\mu$). The discreteness of $\bk$ is
important here, and hence the boundary conditions enter. For the
three cases other than $++$, $\bk=0$ is absent from the product,
and we consider these first. The logarithm of the product is%
\be%
\frac{1}{4}\sum_\bk\ln |\bk|^2=L_xL_y\int
\frac{d^2\bk}{(2\pi)^2}\ln|\bk|^{1/2}+\ldots,\ee%
where the subleading terms represent the corrections due to the
discreteness of the sum. The first (integral) term is again
extensive (proportional to the area $L_xL_y$) and independent of
$\tau$, and will be discarded; this corresponds to renormalizing
the determinant.

At this stage it is helpful to realize that for each of the three
boundary conditions stated, the product of interest,
$\det(D\overline D )$, can be interpreted as the partition
function of a massless Dirac fermion on a Euclidean torus. (The
operators $\overline D$ and $D$ are parts of the massless Dirac
operator in two Euclidean dimensions, see Ref.\ \cite{rg}.) The
square root is the partition function of a massless Majorana
fermion, which has half as many degrees of freedom. This theory
arises in connection with the two-dimensional Ising model on a
torus, at criticality. The finite-size corrections have been
computed in that context (Ref.\ \cite{ff}, or see e.g.\ Refs.\
\cite{dfms,id}). To describe these, we introduce some functions
$F_{m_1,m_2}$ of $q=e^{2\pi i\tau}$ that correspond to the
boundary conditions, $m_{1,2}=0$, $1$ (mod $2$) %
\bea%
F_{11}&=&q^{-1/48}\prod_{n=0}^\infty(1+q^{n+1/2}),\non\\
F_{10}&=&q^{-1/48}\prod_{n=0}^\infty(1-q^{n+1/2}),\non\\
F_{01}&=&\sqrt{2}q^{1/16-1/48}\prod_{n=1}^\infty(1+q^n).\eea%
These products are convergent for ${\rm Im}\,\tau$ sufficiently
large, and it is convenient to take the limit of large $L_x\mu$ at
fixed $\tau$.

For the $++$ boundary condition, the ground state of the paired
superfluid is at odd fermion number, and there is an unpaired
fermion occupying the $\bk=0$ state \cite{rg}. The normalization
of the ground state is given by a similar product, but with
$\bk=0$ omitted. The interesting part is then related to the
function \cite{id}%
\be%
F_{00}=\sqrt{2}({\rm Im}\,\tau)^{1/4} q^{1/16-1/48}\prod_{n=1}^\infty(1-q^n).\ee%
(The $\sqrt{2}$ here could be dropped, though not in $F_{01}$.)

In each of the four cases, the partition function of the Majorana
fermion field theory with fixed boundary condition is
$|F_{m_1,m_2}|^2$. The normalizing factor for our paired
wavefunctions is $|F_{m_1,m_2}|$, up to an overall phase and
factors independent of $\tau$ and the boundary conditions. We now
notice that the phase can be chosen so that the normalizing
factors are simply the respective $F_{m_1,m_2}$ themselves, which
are holomorphic in $\tau$ [except for the factor $({\rm
Im}\,\tau)^{1/4}$ in $F_{00}$]. We view these as the chiral
partition functions, proportional to $\sqrt{\det\overline D}$
(with the zero mode deleted for $++$). As $\overline D$ is
antisymmetric, this can be viewed as a Pfaffian, ${\rm
Pf}\,\overline D$. Note that we are saying that
$\sqrt{\det\overline D}$ is holomorphic in $\tau$ [again, except
for the factor $({\rm Im}\,\tau)^{1/4}$ in the $++$ case].

These partition functions can also be conveniently calculated from
a Hamiltonian point of view of the (non-chiral) massless Majorana
field theory in imaginary time \cite{dfms}. Then
$|F_{m_1,m_2}|^2$ are given by, respectively,%
\bea%
{\rm Tr}\,q^{L_0-c/24}\bar{q}^{\overline{L}_0-c/24}&&\quad(\mp-), \non\\
{\rm Tr}\,(-1)^{\hat{F}} q^{L_0-c/24}\bar{q}^{\overline{L}_0-c/24}
&&\quad(-+), \eea%
where $L_0\pm\overline{L}_0$ are proportional to the Hamiltonian
(with the energy of the ground state in finite size for $-$
spatial boundary condition subtracted off) and momentum of the
system, $\hat{F}$ is the fermion number, $c=1/2$ is the central
charge, and in the first two cases the trace is over states in the
sector with either the $-$ or $+$ boundary condition in the space
direction, respectively. The factor of $\sqrt{2}$ in $F_{01}$
arises because for $+$ spatial boundary condition there is a zero
energy mode shared between right and left-moving fermions that can
be occupied at most once, giving two states. These functions arose
in the same way as partition functions of edge Majorana fermion
modes of the MR state on the cylinder \cite{milr}, which
emphasizes again the relation between bulk and edge of the QH
states. For the final function $|F_{00}|^2$ for $++$, one must
consider the trace weighted with $(-1)^{\hat{F}}$ as for
$|F_{10}|^2$, but with fermion operators $\bar\psi\psi$ inserted
also, so that the zero mode does not give a zero answer. This
treatment of the zero mode produces the factor $({\rm
Im}\,\tau)^{1/4}$.

The wavefunction of the $N$-particle component of the {\em
normalized}\/ BCS state is then, in the three even cases, $m_1$,
$m_2$ not both zero,%
\be%
F_{m_1,m_2}(\tau){\rm Pf\,}g_{m_1,m_2}(\br_i-\br_j|\tau).\ee%
We emphasize that the normalization is correct (up to a
$\tau$-independent constant at least) even on passing to the
fixed-particle number components, provided the particle number is
chosen close to (say, within $\sqrt{N}$ of) the expected value in
the BCS ground state (which we assume is large), by standard
arguments about the equivalence of canonical and grand canonical
approaches. For these even cases, these functions are precisely
the three conformal blocks for $N$ Majorana fermions on the torus,
for all even $N$ including $N=0$, whereas the function without the
factor $F_{m_1,m_2}$ is the ``normalized'' chiral correlator, in
which the block is here divided by the corresponding one for $N=0$
(see the discussion in Sec.\ \ref{cft}). For $m_1=m_2=0$, there
are similar statements for odd $N$, but the ``normalized'' chiral
correlator involves dividing by the $N=1$ function $F_{00}$, which
seems less natural. Note that the function $F_{00}$ itself is the
conformal block for $N=1$, a single Majorana fermion on the torus,
which is independent of the coordinate $z_1$.

Now we introduce the familiar fact that the parametrization of
torii of given area by $\tau$ is redundant: infinitely many
different $\tau$s describe the same torus, up to an isometry. The
transformations from one to another of these are of the form
$\tau\to(a\tau+b)/(c\tau+d)$, where
$\left(\begin{array}{cc}a&b\\c&d\end{array}\right)$ is a matrix of
integers of determinant $1$. These matrices form a group,
SL$(2,{\bf Z})$, called the modular group. (The transformations
arise by changing to a different basis for the lattice generated
by {\boldmath$\omega$}$_1$, {\boldmath$\omega$}$_2$.) SL$(2,{\bf
Z})$ is generated by the elements
$T=\left(\begin{array}{cc}1&1\\0&1\end{array}\right)$, and
$S=\left(\begin{array}{cc}0&-1\\1&0\end{array}\right)$, which
correspond to the transformations $T:\tau\to\tau+1$ and
$S:\tau\to-1/\tau$; it can be defined abstractly as the group
generated by elements $S$, $T$ with relations $(ST)^3=S^2$,
$S^4=1$ (where $1$ is the identity element). Note that the action
on $\tau$ of any matrix in SL$(2,{\bf Z})$ is unchanged by
multiplying it by $-{\bf 1}_2$, so that strictly speaking the
group of modular transformations on $\tau$ is the quotient
PSL$(2,{\bf Z})\equiv$ SL$(2,{\bf Z})/{\bf Z}_2$, where ${\bf
Z}_2=\{\pm{\bf 1}_2\}$, defined by generators $S$, $T$ with
relations $(ST)^3=S^2=1$ [we will not distinguish between $S$ and
$T$ in SL$(2,{\bf Z})$ and their images in PSL$(2,{\bf Z})$].

Given our states $|m_1,m_2,\tau\rangle$ on the torus for each
$\tau$ and for each boundary condition $(m_1,m_2)$ (the system
size will not play a role here, provided $\mu L_x$, $\mu L_y$ are
large, and will be dropped from the notation), we can consider the
effect on them of modular transformations. This is a warm-up
exercise for the statistics calculation later. First we consider
the monodromy of the states viewed as functions of $\tau$.
Starting from any state $|m_1,m_2,\tau\rangle$, we can vary $\tau$
along a path to reach either $\tau+1$ or $-1/\tau$, keeping
$(m_1,m_2)$ fixed. Note that $(m_1,m_2)$ are defined using the
parametrization of the torus determined by $\tau$. The final
result is a torus equivalent to the original one, but the boundary
conditions in terms of the original $\tau$ are now different. The
effect is $T:(m_1,m_2)\to(m_1,m_2-m_1)$, and
$S:(m_1,m_2)\to(m_2,-m_1)$. When reduced modulo $2$, these also
describe the effect on the boundary conditions around the two
cycles that are important in the long-distance part of the state.
The three even cases are permuted by these transformations, while
the odd ($++$) case maps to itself (it was clear this must be so,
because fermion number is conserved by these operations).

If we consider the winding numbers only modulo $2$, then the three
even states form some representation of the modular group, while
the odd state is a one-dimensional representation. First we consider
the normalizing factors $F_{m_1,m_2}$.  The
transformations of these under monodromy, simply comparing
$F_{m_1,m_2}(\tau+1)$ or $F_{m_1,m_2}(-1/\tau)$ with
$F_{m_1',m_2'}(\tau)$, are known \cite{dfms}; the action of $T$ is
easily found to be%
\bea%
F_{11}(\tau+1)&=&e^{-2\pi i/48}F_{10}(\tau),\non\\
F_{10}(\tau+1)&=&e^{-2\pi i/48}F_{11}(\tau),\non\\
F_{01}(\tau+1)&=&e^{2\pi i/24}F_{01}(\tau),\eea%
while the action of $S$ is%
\bea%
F_{11}(-1/\tau)&=&F_{11}(\tau),\non\\
F_{10}(-1/\tau)&=&F_{01}(\tau), \non\\
F_{01}(-1/\tau)&=&F_{10}(\tau).\eea%
Thus the coefficients in these transformations form unitary
matrices, which generate a unitary representation of the modular
group. The sum $\sum_{m_1,m_2}|F_{m_1,m_2}|^2$ (over $m_{1,2}=0$, $1$,
not both zero) is a modular invariant, and is the partition
function of the critical Ising model.

The function $F_{00}$ is $\sqrt{2}({\rm Im}\,\tau)^{1/4}$
times the Dedekind eta function,%
\be%
\eta(\tau)=q^{1/24}\prod_{n=1}^\infty(1-q^n),\ee %
which is a modular form of weight $1/2$ (with multiplier
system---see e.g.\ Ref.\ \cite{iwaniec}) for the modular group:
$\eta(\tau+1)=e^{2\pi i/24}\eta(\tau)$,
$\eta(-1/\tau)=\sqrt{-i\tau}\eta(\tau)$. As ${\rm
Im}\,(-1/\tau)={\rm Im}\,\tau/|\tau|^2$, $({\rm
Im}\,\tau)^{1/4}\eta(\tau)$ transforms purely by phases under the
modular group, like the other $F_{m_1,m_2}$ above, though in this
case the phase depends on $\tau$. The significance of these facts
will be explained below.

Next we turn to the transformation of the pairing functions $g$ under
modular transformations implemented simply by monodromy. The
transformation of the various $\vartheta$ functions can be easily
obtained \cite{mumtheta}, and for $z=0$ is similar to that of
$\eta(\tau)$ (i.e.\ the weight is $1/2$), but also involves the
appropriate permutation of boundary conditions. For general $z$,
the $\vartheta$s are said to be automorphic forms under the larger
group of translations of $z$ and modular transformations that act
on both $z$ and $\tau$ \cite{mumtheta}; for example, one has%
\bea%
\vartheta_{1/2,1/2}(z|\tau+1)&=&e^{i\pi\tau/4}\vartheta_{1/2,1/2}
(z|\tau),\non\\
\vartheta_{1/2,1/2}(z/\tau|-1/\tau)&=&-(-i\tau)^{1/2}e^{i\pi z^2/\tau}
\non\\
&&{}\times\vartheta_{1/2,1/2}(z|\tau).
\eea%
In the ratio of $\vartheta$s giving the pairing function, most such
factors cancel, the exception being due to the derivative of $\vartheta_{1/2,1/2}$ in the numerator. Consequently, one finds for
all $m_1$, $m_2$ that the pairing functions transform under the
generators $S$, $T$, simply by the permutation of boundary conditions,
already described above, and by a factor of $\tau/|\tau|$ in the
case of $S$. The factor of $\tau/|\tau|$ arises from the interplay
of the elliptic function and the factor $({\rm Im}\,\tau)^{1/2}$,
similarly to the case of $\eta(\tau)$ discussed above. Let us
point out that the net number of factors $({\rm Im}\,\tau)^{1/4}$
in the $N$-particle component of the wavefunction is just $N$,
for all cases, including the $++$ boundary condition in which
$N$ is odd. These factors are essential in order that the
monodromy matrices be unitary, as they clearly must be when
orthonormalized states are analytically continued.

The interpretation for the phases in the transformation under
$S$ is then that, apart from the $N$- and $\tau$- independent
factors arising from the $F_{m_1,m_2}$s, the phase
$(\tau/|\tau|)^{N/2}$ is connected with the rotation of the
geometry that brings the torus back to an equivalent one under
$S$; the rotation angle is $-\arg \tau$, and the factor for
each particle reflects the conformal weight $1/2$ for each
Majorana fermion in the conformal block. Put another way,
each Cooper pair carries angular momentum $-1$, and thus
each particle effectively carries (orbital) angular momentum
$-1/2$. This is true even for the unpaired particle in the
case of $N$ odd, though there the factor arises in the BCS
point of view through the normalization factor. No such
phase arises from $T$ as no rotation of the system is involved.

Thus we have determined the monodromy representation of the
modular group on our ground states. Readers are cautioned that
because the full states depend on
$(m_1,m_2)$ as integers, and not just modulo $2$, we cannot say
that the representation on the states is three-dimensional in the
case of paired superfluids. In the QH case, this would be true,
but there one must also consider the charge sector or Laughlin
factors in the wavefunctions, which we will do in the following
section.


\subsection{Hall viscosity and adiabatic modular transformations}
\label{oddvis}

Finally we are ready to consider the more physical operation of
adiabatically dragging the states to perform a modular transformation
of $\tau$. Because of the intrinsic interest of the result, we will
present two arguments that both give the same result. In the first,
we will take the long-distance forms derived above, and assume
that they apply at all length scales (down to a short distance
cutoff, so that the state can be normalized). The functions are
then conformal blocks, and an elegant calculation is possible.
In the second argument, we instead consider a general $p-ip$
paired state, and show that the result can be obtained quite
directly in the language of the pairing function. Thus the
second argument is very general, and applies even for the
strong-pairing phase, and for other paired states, and even
in higher dimensional space.

Now we begin the first argument. We have already constructed
for each $\tau$ a set of four ground states that are orthonormal,
and depend holomorphically on $\tau$, except for a factor $({\rm Im}\,\tau)^{N/4}$. The transition functions are determined by
the $S$ and $T$ transformations given above. The adiabatic
transport can be considered separately for the one odd and
three even cases, as these two types clearly cannot mix with
one another; here we will denote the four as $\Psi_a(\tau)$ for
brevity.

We can now apply the reasoning of Sec.\ \ref{adtrans}, varying $\tau$
instead of quasihole coordinates. If the wavefunctions were completely holomorphic in $\tau$, then the Berry connection would vanish.
As they are not holomorphic, this is not the case, but as in the
quasiparticle transport calculation, the non-holomorphic dependence
takes a simple form, and one can easily take it into account, while
still exploiting the holomorphy of the remaining more complicated
factors in the wavefunction. (We should also point out here that
the integration measure should be independent of $\tau$. If we
write $z=L_x(\rho_1+\tau\rho_2)$, where $\rho_1$ and $\rho_2$ are
in the unit interval $[0,1]$, then the integration measure for
each $\br_i$ becomes $d^2\br_i=Ad\rho_1 d\rho_2$, and the
variation of $\tau$ is always performed with $A$ fixed. The
crucial point is that in this parametrization, the wavefunctions
are still holomorphic in $\tau$ up to the same factor as before.)
Then for the Berry connection we have%
\bea%
A_{\tau,ab}&=&i\langle\Psi_a|\partial_\tau\Psi_b\rangle\non\\
&=&i\partial_\tau\langle\Psi_a|\Psi_b\rangle-\frac{iN\partial_\tau{\rm Im}\,\tau}{4{\rm Im}\,\tau}\langle\Psi_a|\Psi_b\rangle,\eea%
where we used the holomorphy of $\Psi$ up to a known factor. Because
the states are orthonormal, this reduces to %
\be%
A_{\tau,ab}=A_{\overline{\tau},ab}=-\frac{N\delta_{ab}}{8{\rm Im}\,\tau}.
\ee%
(In the paired states with indefinite particle number, $N$ can clearly be replaced by its quantum average $\overline{N}$.) The result happens to be
the same as in Ref.\ \cite{asz}, in which the filled LLL state has the same form, $({\rm Im}\,\tau)^{N/4}$ times a function holomorphic in $\tau$
(however, we will see that it is something of a coincidence that the
power of ${\rm Im}\,\tau$ is exactly the same). Following the reasoning
of that reference, the curvature (or field strength) of this connection,
which is $-N/(2{\rm Im}\,\tau)^2$, corresponds to a Hall viscosity of %
\be%
\eta^{(A)}=\frac{\hbar\overline{n}}{4}.\ee%
Here we restored $\hbar$ to obtain physical units for viscosity, and $\overline{n}$ is the average particle density. We emphasize that the
viscosity came out independent of the aspect ratio $\tau$ and system
size $L_x$, as is appropriate for an intrinsic local property of a fluid.

We may now also consider modular transformations, applied adiabatically.
Because of the curvature of the Berry connection that we just calculated,
the adiabatic effect (holonomy) depends on the path taken in the upper-half plane). To remove this effect, we can consider special paths that enclose vanishing area. That is, the ``moduli space'' of distinct torii is the
upper half plane, ${\rm Im}\,\tau>0$, modulo the action of PSL$(2,{\bf Z})$, which is generated by $S$ and $T$. As $S$ sends $\tau$ to $-1/\tau$, we
see that $\tau=i$ is a fixed point of $S$. We may consider adiabatic
transport along a path lying in the fundamental domain $|{\rm Im}\,\tau|
\leq 1/2$, $|\tau|\geq 1$, that connects two points on the unit circle that
are mapped to one another by $S$. If the path is a semicircle centered at
$i$ with radius shrinking to zero, then it implements the $S$ transformation
but the line integral of the Berry connection along the path vanishes (we
note that the Berry connection is smooth at $i$). Consequently, for this limiting path, the holonomy is once again given solely by the monodromy of
the conformal blocks. We may make a similar argument for $T$, using a
straight path from $-1/2+i{\rm Im}\,\tau$ to $1/2+i{\rm Im}\,\tau$, and
let ${\rm Im}\,\tau\to\infty$. Then once again, the line integral of the
Berry connection gives zero, and in the limit the action of $T$ by holonomy
is the same as the monodromy.

These results then fully characterize arbitrary modular transformations implemented adiabatically. To describe the results succinctly, we can check the defining relations of the modular group in our representation. This requires that we compose $S$ and $T$ operations. To do so, it is convenient to view all group elements as implemented by paths in the upper half plane that start and end at the same point, which we will take to be $\tau=e^{i\pi/3}$. To reach the points $i$ and $1/2+i\infty$ at which we defined $S$ and $T$ above, we follow a path along the unit circle (resp., ${\rm Re}\,\tau=1/2$), apply $S$ (resp., $T$), and retrace the path afterwards. Then the Berry phases related to the Hall viscosity cancel. [In similar calculations for QH states, a gauge transformation is sometimes required as part of the adiabatic transport (holonomy), because of a net rotation of space, to return to the original basis set. But for the operations $(ST)^3$ and $S^2$ we wish to check, the rotation is by $\pi$, under which commonly-used gauge choices are invariant, and this simplifies the calculation.] We find then that as holonomy equals monodromy, we can read off the transformations, and verify that $(ST)^3=S^2={\bf 1}$ in the present case. Thus we do have a representation of the modular group.

In the argument so far, we treated the long-distance wavefunction as if
it were valid for all particles, and so wavefunctions are just conformal
blocks. One may wonder if the result is different when the form of the
pairing functions is different at short distances, or just how general
(or universal) the result is.
For this reason we now include a second argument that shows that the
result for the Hall viscosity is, in fact, completely general. The
adiabatic modular transformations are then also invariant, within a given topological phase.

We will calculate the Berry connection in the upper-half plane directly
from the normalized BCS ground states. As the different boundary-condition states are orthonormal and the final result is diagonal, we will simply
drop the index $a$ for simplicity. Using the expressions for the
normalized BCS ground state $|\widetilde{\Omega}\rangle=[\det
({\bf 1}+g^\dagger g)]^{-1/4}|\Omega\rangle$ (note that this may differ
from the $\tau$-dependent family of states used above by a $\tau$-dependent phase), we obtain%
\be%
A_\tau=\frac{i}{4}{\rm Tr}\,[g^\dagger\partial_\tau g-(\partial_\tau g^\dagger)g]({\bf 1}+g^\dagger g)^{-1}.\ee
We evaluate this in $\bk$ space, in which case $g$ is a diagonal function
of $\bk$, and the values of $\bk$ are determined by $k_x+ik_y=2\pi i(n_2-n_1\tau)/(L_x{\rm Im}\,\tau)$. The values of $n_{1,2}$ are $n_1=m_1/2$ (modulo integers), and similarly for $n_2$ (for $00$, i.e.\ $++$ boundary conditions, $n_1$, $n_2$ both zero is omitted from the sum). Now for a $p-ip$ state, the gap function $\Delta_\bk$ is of the form $k_x-ik_y$ times a function of $|\bk|^2$, while the kinetic energy is a function of $|\bk|^2$ only. Hence the pairing function is of the form $(k_x+ik_y)^{-1}$ times a function of $|\bk|^2$. Further, the only dependence on $\tau$ is through the discrete
values of $\bk$. In $A_\tau$, we see that the $\partial_\tau$ acting on the function of $|\bk|^2$ cancels between $g$ and $g^\dagger$ contributions, and
the only remaining term is from the prefactor $(k_x+ik_y)^{-1}$ in $g$. The result then reduces to%
\be%
A_\tau= \frac{i}{4} \sum_{n_1,n_2}\frac{n_1}{n_2-n_1\tau}\frac{g^\dagger g}{1+g^\dagger g}.\ee
As $L_x\to\infty$ with $\tau$ fixed, the leading part of this is %
\bea%
A_\tau&=&-\frac{A}{8{\rm Im}\,\tau}\int \frac{d^2k}{(2\pi)^2}\,\frac{g^\dagger g}{1+g^\dagger g}\non\\
&=&-\frac{\overline{N}}{8{\rm Im}\,\tau},\eea
which agrees with the previous result. We should mention that the result here for $++$ boundary condition has $N-1$ in place of $N$, as the unpaired fermion is not counted by ${\rm Tr}\,g^\dagger g({\bf 1}+g^\dagger g)^{-1}=N-1$ for $++$, and this is a discrepancy with the calculation in terms of conformal blocks. This is because we were somewhat sloppy in the last step, replacing the sum by the integral, and a more careful treatment of the region near $\bk=0$ would lead to similar calculations as the preceding argument. But in any case, in the thermodynamic limit this discrepancy is negligible.

We can see from this calculation that the Hall viscosity is unchanged
throughout the weak-pairing phase, and is the same in the strong-pairing
phase. For different angular momentum values in the pairing, and even for mixtures of different values, it generalizes easily to give $\hbar/2$ times the
average angular momentum per pair, times half the particle density. A
similar form also holds for pairing in higher space dimensions, in which
the direction of the orbital angular momentum of the Cooper pairs is a
vector. Thus the Hall viscosity is a consequence of the net orbital angular momentum of each pair about its center of mass. The effect appears to
be overlooked in standard references \cite{vollwolf} on He$^3$.


\subsection{Higher genus surfaces and mapping class group}
\label{mapcl}

The discussion for the torus generalizes to oriented surfaces
(without boundary) of higher genus $\cal G$ (genus $1$ being the
torus). The set-up was discussed in Ref.\ \cite{rg}. The fermions
should experience a net magnetic field of ${\cal G}-1$ flux, in
order that the ground state contain no vortices. That is,
$\hat\Delta$, which can be defined within coordinate patches, is
non-vanishing everywhere. The gap function $\Delta$ is then a part
of the Dirac operator on the surface. It depends on a choice of a
boundary condition, which can be either $+$ or $-$ for each of the
$2{\cal G}$ cycles on the surface; we will label these by $a$, and
write $\Delta_a$. Of these $2^{2\cal G}$ boundary conditions or
``spin structures'', $2^{{\cal G}-1}(2^{\cal G}+1)$ are ``even'',
and the remaining $2^{{\cal G}-1}(2^{\cal G}-1)$ are ``odd''. In
the odd cases, the Dirac operator $\overline{\Delta}_a$ has one
zero mode (a holomorphic function, really a section of the
bundle), and in the even case it has none. Consequently, there is
a paired ground state with even particle number in the even cases,
and with odd particle number in the odd cases \cite{rg}. The pairing
function $g_a$ is proportional to $1/\overline\Delta_a$, as for
the plane, sphere, and torus. The general theory for $h=-\mu$
pairing still applies, and so the important long-distance part of
the normalizing factor in the ground state wavefunctions can be
taken to be $\sqrt{\det{\overline{\Delta}_a}}$ (with the zero mode
deleted in the odd cases). This again is the (chiral) partition of
a massless Majorana fermion field, and the normalized $N$-particle
wavefunctions are chiral partition functions with $N$ insertions
of the Majorana field $\psi$, which are essentially the conformal
blocks of the Majorana CFT.

The last statement contains the word ``essentially'' because one
further point was neglected. We saw for the case of the torus
(${\cal G}=1$) that the chiral partition function (without
insertions), $\sqrt{\det{\overline{\Delta}_a}}$, tends to a limit
which is a conformal block, apart from a factor of the form of the
exponential of the area times a constant independent of the
parameter $\tau$; this factor is all that remains of the cutoff.
For ${\cal G}\neq 1$, the corresponding statement does not quite
hold. There is an additional residual dependence on the scale size
$A^{1/2}$ of the system (relative to the cutoff scale). It is
given for a general conformal field theory by a factor
$A^{c\chi/24}$, where $c$ is the central charge ($c=1/2$ for the
Majorana fermion CFT), and $\chi$ is the Euler characteristic of
the surface. This originates from the trace anomaly, which relates
the expectation value of the trace of the stress tensor,
$T_{z\overline{z}}$, of the CFT, which generates a change of
scale, to $c$ times the local curvature of the manifold evaluated
\cite{dfms}. When integrated, the latter gives the Euler
characteristic $\chi=2-2{\cal G}$ of the surface (see e.g.\ Ref.\
\cite{carpes}, which includes also the more general case of a
surface with a boundary). Notice that this result applies to the
sphere, with ${\cal G}=0$. More generally, there is a dependence
on the local cutoff scale given by the exponential of $c$ times
the Liouville action \cite{poly,alv}; this is independent of the
boundary condition choice labeled by $a$. Such factors are not
expected to contribute to the holonomy, when we vary the geometry
holding the area $A$ constant, so we ignore them.

A genus ${\cal G}>1$ surface endowed with a complex structure is
described by $3{\cal G}-3$ complex parameters called moduli, in
place of the single $\tau$ for ${\cal G}=1$ \cite{alv}. The moduli
space ${\cal M}_{\cal G}$ of inequivalent surfaces of genus
$\cal G$ is described by a domain denoted ${\cal T}_{\cal G}$
(Teichmuller space) in ${\bf C}^{3{\cal G}-3}$ modulo the action
of a group $\Gamma_{\cal G}$ of equivalences analogous to the
modular group, called the mapping class group (or Teichmuller
modular group): ${\cal M}_{\cal G}\cong{\cal T}_{\cal
G}/\Gamma_{\cal G}$ \cite{mumford,bk}. The structure of
$\Gamma_{\cal G}$ is difficult to describe for ${\cal G}>1$,
especially for ${\cal G}>2$. The conformal blocks are holomorphic
in the moduli, up to a factor analogous to $({\rm Im}\,\tau)^{1/4}$
which is raised to the power $N$ (when considered at fixed $A$). While
the explicit functions are difficult to obtain, we expect that the
curvature of the Berry connection is again given by the Hall viscosity,
which is independent of the genus of the surface, and that the
representation (which is actually projective,
and unitary) of the mapping class group defined by the holonomy of
the adiabatic transport is the same as the monodromy
representation, up to path-dependent phases determined by
the Berry curvature.


\subsection{Generalizations and strong-pairing phase}

In this subsection we briefly consider other paired states of
fermions from a similar point of view, temporarily leaving behind
trial wavefunctions that are conformal blocks. The paired states
we consider are states that (except for $\mu=0$) are fully-gapped,
and result from a gap function that is a rotational eigenstate in
momentum space, like the $p+ip$ state above. These include pairing
of any odd angular momentum $\ell$ for spinless or spin-polarized
fermions. In the weak-pairing phase, the ground state on the torus
can be normalized by a factor $\sqrt{\det{\overline{\Delta}_a}}$,
where again $a$ labels the possible boundary conditions. By the
same argument as above, this equals $({\rm Pf\,}\overline
D)^\ell$. This is clearly equivalent to $\ell$ Majorana fermions.
For $\ell$ odd, as required here, these phases are non-Abelian, as
pointed out in Ref.\ \cite{rg}. However, since all the Majorana
fermions see the same boundary condition, the theories are not
simply direct products of $\ell$ copies of the Majorana fermion
theory above; rather, they possess a hidden SO$(\ell)$ symmetry.
Similarly, fermions with spin, or other multi-component fermions
have other possibilities, such as the 331 p-wave state related to
the Dirac fermion, which is Abelian. Another case is spin-singlet
pairing for spin-1/2 fermions, where the angular momentum $\ell$
of the pairing must be even. In the weak-pairing phase, the
long-wavelength normalization factor for the ground states on the
torus is $(\det\overline D)^\ell$. This may be viewed as $2\ell$
Majorana fermions, and there is a hidden SO$(2\ell)$ symmetry
[thus SO(4) for $\ell=2$, the $d+id$-wave case; the argument given
here is more direct than that given in Ref.\ \cite{rg}]. These
cases are all Abelian. The hidden symmetries under SO$(M)$ for
some $M$ are present in the topological properties of the bulk of
the paired states (e.g. in counting numbers of conformal blocks),
but may be fully or partially broken by the dynamics at the edge.

By contrast, in the corresponding strong-pairing phases, the
pairing function $g$ is non-singular at ${\bf k}\to0$ (and short
ranged in coordinate space), and the normalization factor
(\ref{normfac}) has no singular behavior at ${\bf k}=0$. Moreover,
all four ground states occur at even particle number. Thus the
ground state wavefunctions are concentrated on configurations in
which all fermions have positions equal in pairs (within a
coherence length), and the long-wavelength part of the ground
state wavefunctions is trivial. These correspond to conformal
blocks in a CFT that is trivial (except for the charge sector).
The orthonormal ground states are not holomorphic in $\tau$, but
have $\tau$-dependence related to the Hall viscosity already
discussed above. They also have non-trivial monodromy (and
hence also holonomy) in so far as they are simply permuted
by modular transformations, without the phase factors that
originated from the non-trivial normalization factors $F_a$
(under $T$). This is not the same as
possessing no holonomy whatsoever. We also point out here that the
normalization factor is non-singular at ${\bf k}=0$ at the
transition point $\mu=0$, even though $g$ has weakly singular
behavior. (The leading singular behavior in $g$ at ${\bf k}\to0$
cancels in $g^\dagger g$; possibly there is some subleading
effect.) Thus this point appears to have the same holonomy as the
corresponding strong-pairing phase. Of course, adiabatic transport
is presumably irrelevant here, because the excitation spectrum of
the fermions is gapless at the transition point.

The general framework for describing topological phases is that of
modular tensor categories (MTCs) \cite{turaev,bk,wang} (strictly, for QH systems, this is for bosons; some modifications are required for QH systems of fermions, which are not fully modular whenever the chiral algebra of the CFT contains the fermion, becoming a
chiral superalgebra \cite{mr}). For paired $s$-wave superfluids
(when viewed as a fully-gapped topological phase, such as when a
Coulomb interaction is present), Kitaev has argued \cite{kitferm} that the
correct MTC is the ``toric code'' or ${\bf Z}_2$ gauge theory,
with Abelian fusion rules corresponding to the group ${\bf
Z}_2\times{\bf Z}_2$, and we find that this applies here for
strong-pairing phases (without the charge sector) also; it
accounts for the four ground states. Other than the identity, or
trivial quasiparticle, this theory contains two quasiparticle
types that are bosons, call them ``electric'' and ``magnetic''
charges (these charges do not refer to particle number), and one
that is a fermion, under exchange of a type with itself.
Adiabatically making a circuit of a magnetic around an electric
charge gives a phase of $-1$, and hence the composite of an
electric with a magnetic charge makes the fermion, which picks up
a $-1$ factor under a circuit around either an electric or a
magnetic charge. (This behavior was calculated from trial
wavefunctions in a lattice model that resembles the strong-pairing
phase \cite{rc}.)

When the charge (particle number) sector is included to make a QH
system, from say the $p+ip$ state, the filling factor is
$\nu=1/Q$, where $Q$ is a positive integer, and is even when the
particles are fermions, and odd for bosons. It was argued in Ref.\
\cite{rg} that the strong-pairing phase is equivalent to a
Laughlin phase of charge-$2$ bosons at filling factor $1/(4Q)$.
This relation predicts the statistics of the family of
fractionally-charged excitations generated by the quasihole of
charge $1/(2Q)$ under fusion; those of particle number $m/(2Q)$
have fractional statistics phase $m^2\pi/(4Q)$ under an exchange.
The fusion rules for this theory are ${\bf Z}_{4Q}$. However, at
first sight it does not appear to account for the excitation that
consists of an unpaired composite (neutral) fermion, which exist
in this phase as well as in the weak-pairing phase. Under
adiabatic transport in a circuit around the quasihole of charge
$1/(2Q)$ (corresponding to the vortex in the paired state), the
neutral fermion should pick up a phase $-1$ (in the weak-pairing
$p+ip$ phase, this follows from the calculation in Section
\ref{statcalc};
we will assume this here in general, and also in the
strong-pairing phase), and this phase clearly does not come from
the charge sector.

These statements can, however, be reconciled. The neutral fermion
times the quasihole of charge $1$ gives us back the underlying
particle (a boson if $Q$ is odd, and a fermion if $Q$ is even).
Further, using the assertions just made, this particle is local
with respect to all the quasiholes (that is, a phase $+1$ is
produced by a circuit of one around the other). Therefore,
quasiparticles that are related by fusion with the particle should
be identified. This leaves a total of $4Q$ distinct types, with
fusion rules, charges and statistics as in the charge sector part
alone (however, the formal charge which was defined modulo $2$
equals the actual particle number charge only modulo $1$). Usually
we would have said that the particle is in the chiral algebra,
which leads to such identifications. However, in the CFT of the
edge excitations, the bulk particle does not appear (and is not
part of the chiral algebra), and the theory is simply that for the
charge sector as in Ref.\ \cite{rr1,rg}. We believe this is the
correct description of the strong-pairing phases that involve
pairing of composite fermions. (For spin-singlet phases, the
fermion carries spin-1/2, but this does not affect the MTC
properties.)


\section{Hall viscosity of the Laughlin states}
\label{halllaugh}

In this section we calculate the Hall viscosity of the Laughlin
states. Using the ideas already presented above, the main point is
to evaluate the normalization of the Laughlin states on the torus,
up to constants independent of the size $L_x$ and aspect ratio $\tau$.
We will find that the result is again given by a spin density, where
in this case the spin of each particle is $Q/2$, the conformal weight
of the field representing the particles in the CFT point of view
\cite{mr} (these states include the filled LLL, with $Q=1$, as a
special case \cite{asz}, providing a check on the calculation). We
discuss the general situation for QH states, and adiabatic modular transformations.

The key to obtaining the normalization of the Laughlin states in
this sense is the plasma mapping from Laughlin's original paper
\cite{laugh}, which here will be generalized to the torus. Thus we
will rely on the physical argument that screening holds in the plasma
when $Q$ is not too large. Most of the calculation is aimed at
establishing that the wavefunctions on the torus \cite{hr1} have
modulus-squared that is the Boltzmann weight in a plasma with a
neutralizing background, with control over $L_x$- and $\tau$-dependent
factors that are commonly omitted. Then the normalization integral
equals $e^{-Af_0(Q)}$, independent of $\tau$. A particular issue to be
dealt with is the neutralizing background charge (indeed, if the plasma
were neutral when only the mobile charges were considered, the results
would be almost immediate, as we will describe later). We will first
obtain the Boltzmann weight for $N$ point charges $Q$ and
$\alpha^{-1}N=M^2$ ($M$ integer) smaller charges of $-\alpha Q$ on
the torus. The latter will be fixed on the points of a grid. As
$\alpha\to0$, these become the uniform background. By taking this
limit carefully, we can determine the necessary normalizing factors.
Even though the potential due to the neutralizing background appears
to be a constant because of translational symmetry, its dependence
on $\tau$ has to be evaluated, with the area $A$ fixed at the correct
value for $N$ particles.

The Boltzmann weight for a Coulomb plasma in two dimensions can be
obtained from a 2D conformal field theory, the (non-chiral) free
massless scalar field $\varphi$ \cite{id,dfms}. Its partition
function is given by the functional integral%
\be%
Z=\int {\cal D}[\varphi]\, e^{-S[\varphi]}\ee
where the action is %
\be %
S=\frac{1}{8\pi}\int d^2r (\nabla \varphi)^2.\ee%
The normalization is such that the two-point function in the plane is%
\be%
\langle\varphi(\br)\varphi(0)\rangle=-\ln|\br|^2.\ee%
Some of the important operators in this theory are the electric charges,
given by ${\cal O}_e(\br)=e^{ie\varphi(\br)/R}$ for some charge $e$
($R$ is a parameter that will be used later). These have conformal weights $h=\overline{h}=e^2/(2R^2)$.

We now consider this theory on the torus of side $L_x$ and aspect ratio
$\tau$ as before. The field $\varphi$ will be compactified, which means
that the configurations $\varphi$ to be integrated over must be allowed
to obey  generalized periodic boundary conditions (for convenience, we
write $z$ for $\br$, even though the fields are not holomorphic)%
\bea%
\varphi(z+L_x)&=&\varphi(z)+2\pi R n,\\
\varphi(z+L_x\tau)&=&\varphi(z)+2\pi Rn'.\eea
Here $R>0$ is a parameter, the compactification radius. The meaning of
this is that, if $\nabla\phi$ is an ``electric'' field in two dimensions,
these boundary conditions allow for electric fields that circulate the
system, without any source. This is important in obtaining periodic
correlation functions when charges are present. A general field
configuration for these boundary conditions can be written as %
\be%
\varphi=\varphi^{\rm cl}_{nn'}+\widetilde{\varphi},\ee%
where $\varphi^{\rm cl}_{nn'}$ is a fixed solution to the boundary
conditions, say \be%
\varphi^{\rm cl}_{nn'}=2\pi R\,{\rm Im}\,\left(\frac{z(n'-n\overline{\tau}}{L_x{\rm Im}\,\tau}\right),\ee%
while $\widetilde\varphi$ is a periodic function. For any $n$, $n'$,
we have%
\be%
G(\br)=\langle\widetilde{\varphi}(\br)\widetilde{\varphi}(0)\rangle=
-\ln\left|\frac{{\cal E}(z/L_x|\tau)}{L_x}\right|^2.\label{gfntor}\ee%
The function ${\cal E}(z/L_x|\tau),$  %
\be%
{\cal E}(z|\tau)=\frac{\vartheta_1(z|\tau)}{\partial_z\vartheta_1(0|\tau)}
e^{-\pi({\rm Im}\, z)^2/{\rm Im}\, \tau},\ee%
where we have used the more familiar notation $\vartheta_1(z|\tau)=-\vartheta_{1/2,1/2}(z|\tau)$, is periodic up to
phase factors, and is called the prime form for the torus (or else this
name is used for the ratio of $\vartheta$s without the Gaussian). The
function $G(\br)$ is the Green's function for the Laplacian on the torus, defined as the inverse of $-\nabla^2$ in the space orthogonal to its zero
mode,
\be%
-\nabla^2G(\br)=4\pi\left[\delta(\br)-\frac{1}{A}\right].\ee%
At short distances, it is asymptotically the same as in the plane.

Now we evaluate the (unnormalized) correlation function of $N$ charges
$e=Q$  at positions $z_i$, and $N/\alpha$ charges $-\alpha Q$ at
positions $w_k$, where it is convenient to choose $N/\alpha=M^2$,
with $M>0$ an integer. We take $R^2=Q=\nu^{-1}$. (The terminology for
the charges thus agrees with that used earlier, with $R^2$ as the
temperature.) We will assume that $Q$ is even, and we will recover
the Laughlin states for bosons. The states for fermions require a
slightly different sum over boundary conditions, and one eventually
obtains a unitary representation of a subgroup of the modular group
\cite{milr}, whereas in the bosonic case one obtains a representation
of the full modular group (hence the correlation functions we are
calculating are invariant under these respective groups). The
unnormalized correlation function can be evaluated by standard
techniques  \cite{id,dfms}, including a Poisson summation which
replaces the $n'$ sum by a sum over integers $m$. Omitting some
algebra, we obtain%
\be%
\langle\prod_i {\cal O}_Q(z_i)\prod_k {\cal O}_{\alpha Q}(w_k)
\rangle_{\varphi}
=\sum_e|\Psi_e(z_1,\ldots,w_1,\ldots)|^2,\ee%
where $e$ runs over the $Q$ integer values $-Q/2+1$, $-Q/2+2$,
\ldots, $Q/2$,
and %
\bea%
\lefteqn{\Psi_e(z_i,w_k|\tau)=}&&\non\\
&&\frac{\vartheta_{e/Q,0}(QZ/L_x|Q\tau)}{\eta(\tau)}e^{-\pi Q\,({\rm Im}\,Z)^2/A}
\prod_{i<j}{\cal E}(z_{ij}/L_x|\tau)^Q\non\\
&&{}\times \prod_{ik}{\cal E}([z_i-w_k]/L_x|\tau)^{\alpha Q}\prod_{k<l}{\cal E}(w_{kl}/L_x|\tau)^{\alpha^2Q}\non\\
&&{}\times L_x^{-NQ/2-\alpha NQ/2},
\label{laughtor1}\eea%
where $Z=\sum_iz_i-\alpha\sum_kw_k$, $z_{ij}=z_i-z_j$, and $w_{ij}=w_i-w_j$ (compare Ref.\ \cite{cmmn}).
The evaluation of the sums over $n$, $n'$ has led to a sum of modulus-squares
of functions $\Psi_e$ that we will see are, when $\alpha\to0$, essentially
the $Q$ Laughlin ground states on the torus. The sums over $n$, $n'$ render
the correlation function doubly periodic in the $z_i$ variables, though
not in the $w_k$ variables, because of the fractional values of the corresponding charges which enter through $Z$. Hence we should be careful
to specify precisely the positions $w_k$. We will assume that $\sum_k w_k=0$.

Now we relate the functions $\Psi_e$ to the standard forms for the
Laughlin states on the torus \cite{hr1,rr1}, and take the $\alpha\to0$
limit. First we deal with the Gaussian factors. As they stand, all of
them actually cancel. This was expected for a conformal block of a
collection of charges with zero total charge. But we wish to separate
out the $w_k$s, and take a limit. So let us assume that $\sum_k w_k=0$,
and omit the $ik$ and $kl$ Gaussian factors. The others then simplify to %
\be%
e^{-\pi NQ\sum_i({\rm Im}\,z_i)^2/A}=e^{-\frac{1}{2}\sum_i ({\rm Im}\,z_i)^2},\ee%
which is the usual Gaussian in the LLL in the Landau gauge (we used $N/A=\nu/(2\pi$; note we implicitly selected a gauge when extracting
$\Psi_e$ from $|\Psi_e|^2$)). In addition, the factors $\vartheta_1(z_{ij}/L_x|\tau)^Q$ and the center of mass factor $\vartheta_{e/Q}(QZ/L_x|Q\tau)$ also occur in the Laughlin states \cite{hr1,rr1}. The periodic Laughlin states can be defined for
arbitrary phases in the boundary condition in each direction \cite{hr1},
and we have obtained particular values of these phases, which depend on
$\sum_k w_k$, so that other values are also available if desired. Thus
the $z_i$ dependence of these functions is exactly that of the Laughlin
states.

We can also see that the powers of $L_x$ become $A^{-NQ/4-\alpha NQ/4}
({\rm Im}\,\tau)^{NQ/4+\alpha NQ/4}$. As $\alpha\to0$ (with $N$ fixed),
this yields a power of $A$ times $({\rm Im}\,\tau)^{NQ/4}$, in which the exponent is related to the conformal weight $Q/2$ of each ${\cal O}_Q$
that represents the particles in the same way as in the paired states.
The correlation function is modular invariant, and this entails that the
vector of functions $\Psi_e$ transforms by a unitary matrix. As we saw
in connection with the paired states, the factors ${\rm Im}\,\tau$ do
not transform by phases under $S$, and there must be modular forms of
non-zero net weight to compensate them, in order that the transformations involve $\tau/|\tau|$ (to the power of conformal weights). Noting that
the prime forms contain $\partial_z\vartheta_1(0|\tau)=2\pi \eta(\tau)^3$ \cite{mumtheta}, we can extract $\eta(\tau)^2$ from each $\cal E$, that
is a net factor $\eta(\tau)^{NQ+\alpha NQ}$, and this combines with the
power of ${\rm Im}\,\tau$ to give the expected transformation behavior.
(Another way to see this is to realize that each factor $|{\cal E}(z/L_x|\tau)/L_x|^2$ must be modular invariant when the area is held
fixed, so that ${\cal E}(z/L_x|\tau)/L_x$ must be invariant up to a
phase, and the exponent of $\tau/|\tau|$ is related to conformal
weight.) The remaining factors, including the $e$-dependent one in
the center of mass, are all ratios of the form $\vartheta/\eta(\tau)$,
which transform with weight zero. Now we want to show that, of these
remaining factors, the product of those that depend on $w_k$ tends to
$1$  as $\alpha\to0$ (up to $\tau$- and $L_x$-independent factors).

As we already stated, the uniform neutralizing background will be
simulated by placing the charges on a grid, for which we choose the
lattice sites $L_x(n_1+n_2\tau)/M$, for integers $n_1$ and $n_2$. The
ratio $\vartheta_1/\eta$ of interest can be written as%
\bea%
\widehat{\cal E}(z|\tau)&=&
\frac{i\vartheta_1(z|\tau)}{\eta(\tau)}e^{-\pi({\rm Im}\,z)^2/
{\rm Im}\,\tau}\\
&=&e^{-\pi({\rm Im}\,z)^2/{\rm Im}\,\tau}e^{2\pi i\tau/12}(e^{i\pi z}
-e^{-i\pi z})\non\\
&&{}\times\prod_{n=1}^\infty(1-e^{2\pi in\tau+2\pi i z})(1-e^{2\pi
in\tau-2\pi i z}),\non\eea
using the product formula for $\vartheta_1$ \cite{mumtheta}. Then we
can show that%
\bea%
\lefteqn{\prod_{n_1=0}^{M-1}\prod_{n_2=0}^{M-1}\widehat{\cal E}\left.\left(z+\frac{n_1+n_2\tau}{M}\right|\tau\right)=}&&\non\\
&&e^{-i\pi {\rm Re}\,\tau(M-1)(2M-1)/6-i\pi M(M-1){\rm Re}\,z-i\pi(M-1)(3M-2)/2}\non\\
&&{}\times \widehat{\cal E}(Mz|\tau).\label{thetaid}\eea
Similar results, but differing by a phase, hold if any of the set of
$M^2$ points on the grid in the complex plane are shifted by integers
or by integer multiples of $\tau$; this is actually needed if we use the
above choice of a set of $w_k$ such that $\sum_k w_k=0$. However, these
phases are not important. We now apply the above formula, with $z$ replaced
by $z_i/L_x$ for each $i$, in the above expression for $\Psi_e$. In that
case, it must be raised to the power $\alpha Q=NQ/M^2$. Ignoring the
overall phase, we see that $\widehat{\cal E}(Mz_i/L_x|\tau)^{1/M^2}$
vanishes when $z_i$ is at any of the points on the grid, but (roughly
speaking) tends to $1$ as $M\to\infty$ away from these points. This is
a little crude, as the points of the grid themselves depend on $M$. If
we rescale $z_i$ so that the grid points in the $z_i$ plane are
independent of $M$, then the statement is clear. Moreover, for large
$M$, the {\em fraction} of the plane (or of a unit cell) of the grid
at which the function is smaller than $1-\epsilon$ [for any $\epsilon$
in $(0,1)$] is of order $(1-\epsilon)^{2M^2}$ for large $M$, which is exponentially small in $M$. Without the rescaling, it follows that $\widehat{\cal E}(Mz_i/L_x|\tau)^{1/M^2}$ tends to $1$ at almost
every $z_i$, that is except on a set of measure zero; the latter set is essentially the set of points $L_x(a_1+a_2\tau)$ where $a_1$ and $a_2$
are rational numbers.

The factors containing $w_{kl}$ can be handled similarly, with both
$w_k$ and $w_l$ on the grid. There is the slight complication that
$k=l$ must be excluded. This can be handled by taking the derivative
of the identity (\ref{thetaid}) at $z=0$, and extracting a factor $\partial_z\vartheta_1(0|\tau)$ from both sides.

Putting all these results together, and dropping phase factors and $\tau$-independent factors (including $A$), we find that the wavefunctions
are
\bea%
\lefteqn{\Psi_e(z_i,w_k|\tau)=}&&\non\\
&&\frac{\vartheta_{e/Q,0}(QZ/L_x|Q\tau)}{\eta(\tau)}
\prod_{i<j}\left(\frac{\vartheta_1(z_{ij}/L_x|\tau)}{\eta(\tau)}
\right)^Q\non\\
&&{}\times[({\rm Im}\,\tau)^{1/4}\eta(\tau)]^{NQ}e^{-\frac{1}{2}
\sum_i({\rm Im}\,z_i)^2}\label{laughtor2}
\eea%
which essentially agrees with Refs.\ \cite{hr1,rr1} up to $z$-independent factors, and we emphasize again that different boundary condition phases
on the particles can be incorporated by changing the value of $-\lim_{M\to\infty}\alpha\sum_k w_k$
in $Z$. This was derived for bosons ($Q$ even), but the result is similar
for fermions ($Q$ odd).

Let us now summarize, and complete the calculation of the Hall viscosity.
The form (\ref{laughtor1}) for the wavefunctions (before the $M\to\infty$
limit) shows that for particles at separations much less than $L_x$, the interactions  in the plasma are independent of the system geometry, as
they should be. Hence the sum of the norm-squares reduces to $e^{-Af_0(Q)}$, independent of the geometry. The other form (\ref{laughtor2}) exhibits the effect of a modular transformation (in particular $S$) most clearly. Also,
it shows that the dependence on $\tau$ is holomorphic, apart from the power
of ${\rm Im}\,\tau$ and the Gaussian. Using $z_i=L_x(\rho_{1i}+\tau
\rho_{2i})$, we see that%
\be%
e^{-\frac{1}{2}({\rm Im}\,z_i)^2}=e^{-\pi N_\phi {\rm Im}\,\tau \rho_{2i}^2}\ee%
which differs only by a phase from $e^{i\pi N_\phi\tau \rho_{2i}^2}$.
Hence by using an appropriate gauge choice, the wavefunctions are
holomorphic in $\tau$ except for the explicit factor $({\rm Im}\,
\tau)^{NQ/4}$. We have actually calculated only the sum of the
modulus-squares of the wavefunctions, while we want to show that
the set is orthonormal. Because the set transforms by unitary matrices
under modular transformations (by monodromy), and these form an irreducible representation, we can borrow an argument from Sec.\ \ref{gencon}
below, and conclude that they are in fact orthonormal (up to a constant). Alternatively, we can obtain orthonormality from the known fact
that the functions are mutually orthogonal, which follows as they
have distinct quantum numbers under certain translations under
which they are eigenfunctions  \cite{hald85}. Then it follows from
use of center-of-mass translations, which map them to each other,
that they have the same normalization constant, and so are orthonormal
up to a common constant.

By a similar calculation as the first one in Sec.\ \ref{oddvis}, it
now follows that the Berry connection on the space of $Q$ states as
a function of $\tau$, and its curvature, are proportional to the
matrix ${\bf 1}_Q$. The value of the Hall viscosity of the Laughlin
states is finally%
\be%
\eta^{(A)}=\frac{\hbar Q\overline{n}}{4},\ee%
where the density here is $\overline{n}=1/(2\pi Q \ell_B^2)$ if we
restore $\ell_B$ which is $1$ elsewhere. The result agrees with the
special case $Q=1$ (the filled LLL) which was treated in Ref.\
\cite{asz} by using a Slater determinant of single-particle
wavefunctions; there should be a version of the Vandermonde
determinant identity for the torus, which will show that the
wavefunctions themselves are the same, up to an $L_x$- and
$\tau$-independent constant. For $Q\neq 1$, the result
disagrees with a more general result that was claimed
recently in Ref.\ \cite{tv}. Further, adiabatic modular
transformations can be performed using the ideas of Sec.\
\ref{oddvis}, and are given by the monodromy, up to the
path-dependent phases.

Finally, we address the more general QH trial states that
include a conformal block from a CFT. On the torus, the
charge part is similar to that in the case of the Laughlin
states above, the differences being in the center of mass
factor, which plays only a minor role here anyway, and in t
he exponent $Q$ which more generally is replaced by $\nu^{-1}$
(while the denominator $Q$ of the filling factor still appears
in the center of mass factor). Under modular transformations,
the conformal blocks behave as automorphic forms, and at fixed
area a factor ${\rm Im}\,\tau$ must be included to an appropriate
power, so that the $S$ transformation acts by unitary matrices
that include factors $\tau/|\tau|$ to the total conformal
weight of the fields. This is rather more clear for these blocks
than for the Laughlin states, where we had to deal with the
neutralizing background. For general blocks, the factor
$L_x^{-h_\psi}$ must appear for each particle, by scaling
considerations. This produces the desired power of ${\rm Im}\,
\tau$ when written in terms of $A$, which is fixed. (The paired
states in Sec.\ \ref{oddvis} are an example, in which $h_\psi=1/2$
for Majorana fermions.) Hence it becomes clear that the full functions
always include the factor %
\be%
({\rm Im}\,\tau)^{N(\nu^{-1}/2+h_\psi)/2}\ee%
times a function holomorphic in $\tau$. More generally, the trial
functions with quasiholes included contain similar factors depending
on their total conformal weight (including the charge sector). Under
monodromy, $S$ is given by $\tau$-dependent phases because of these
factors, and $\tau$-independent coefficients from the center of mass
factors, which are also $N$-independent and occur even for $N=0$, in
which case the wavefunctions are the conformal blocks for the CFT
(including charge sector) on the torus with no particle insertions.
We will argue in Sec.\ \ref{gencon} that these trial wavefunctions
are orthonormal at large $N$, at least in some cases. The arguments
presented here then show that the Hall viscosity is generally given
by%
\be%
\eta^{(A)}=\frac{\hbar (\nu^{-1}/2+h_\psi)\overline{n}}{2}.\ee%
This will be a universal result within a topological phase, when the
trial functions represent such a phase.  We emphasize that the result
here, and that for the paired states, is always a certain constant
times the average ``conformal spin density''. For the trial QH states
given by conformal blocks, the conformal spin of each particle is just
its total conformal weight, $\nu^{-1}/2+h_\psi$. This explains why the
$p+ip$ paired state, and the filled LLL, both gave the same Hall viscosity
(when written in terms of density): both involve the same conformal weight, $1/2$ (the value for Dirac or Majorana fermions; for the filled LLL, the associated CFT is the Dirac fermion theory, which appears e.g.\ as the
theory on the edge).

We recall that the conformal spin also enters in the shift ${\cal S}$. More generally, in non-chiral CFTs, the conformal spin of a field is the holomorphic minus the anti-holomorphic conformal weight, which equals the conformal weight in a chiral theory. For $\nu$ filled Landau levels, the result \cite{levay,avron} can also be viewed as the average spin per particle,
where the spin arises from the cyclotron motion [in the $\cal N$th
Landau level, ${\cal N}=0$, $1$, \ldots, it is $({\cal N}+1/2)\hbar$],
and this appears also to agree with the classical non-zero temperature
result \cite{odd}. For other topological QH phases of matter, such as the hierarchy or composite fermion phases, the conformal spin density in the ground
state can be readily calculated by utilizing the preceding results, and the average spin per particle always equals half the shift. For the gapless Fermi-liquid--like phases \cite{hlr}, for which, in the simplest cases ($\nu=1/Q$ with $Q$ even), trial wavefunctions are related to Laughlin wavefunctions, we expect that the Hall viscosity takes the same form as in the Laughlin states as calculated here; however, as these phases are gapless, other components of viscosity may be non-zero also. Note that ``spin'' here always refers to orbital effects, as we ignore the intrinsic spin of the electrons.
We expect similar effects involving the ``real'' or intrinsic spin of spinful particles.

It is worth emphasizing that the value $\hbar$ times density
is the natural quantum of viscosity, in any dimension, and so is
analogous to $e^2/h$ for electrical conductance, or $\pi k_B^2 T/\hbar$
for thermal conductance. Values of components of the viscosity tensor
in a quantum fluid (or respectively, of a conductivity tensor in two
dimensions) at some temperature might be either larger or smaller than
this value, but the non-dissipative part is quantized at zero temperature
(in the sense of taking a universal value) throughout a topological phase,
and helps characterize the phase. However, viscosity relates to momentum transport, and so the Hall viscosity probably does not mean much if translational symmetry is violated at short length scales, say by
impurities, unlike electrical conductivity, which relates to charge transport, and charge conservation is not violated by impurities (similarly for thermal conductivity).


\section{Adiabatic statistics calculation}
\label{statcalc}

In this section we present direct arguments for the adiabatic
statistics of two and four quasiholes in the MR state on the sphere or
plane in the thermodynamic limit. The calculations work by ``doubling''
(taking two copies) of the system, which turns the problem (the Ising, or Majorana fermion, CFT) into an Abelian one, the Dirac fermion CFT, which
can be bosonized, meaning turned into a neutral Coulomb plasma without a
uniform background charge density. (The Abelian problem can be viewed in
the QH context as the 331 state which applies for example to bilayer
systems, but with the charge part removed.) It is straightforward in the
Abelian problem to argue that various blocks are orthonormal (up to a $w$-independent constant) for well-separated quasiholes, as well as
holomorphic in $w$, and that will conclude the argument. Readers who
prefer a more conceptual approach that is also more general (and avoids doubling) may prefer to skip this section and go to the following
one.

We will consider here only the trial wavefunctions with the charge
sector removed. These presumably correspond to the long-distance behavior
of the $p+ip$ paired state in the presence of vortices, though this has
not been shown directly from the point of view of BCS theory. Arguments
that the charge part can be dropped from the QH trial wavefunctions for
the present purposes are considered in more detail in Sec.\ \ref{gencon}
below. The trial wavefunctions will be written in terms of conformal blocks
from conformal field theories. We will first recall the relation of two copies of the Majorana fermion field theory to a Dirac theory, beginning with zero quasiholes, then progressing to two, and finally to four quasiholes.

The relation of the MR Pfaffian wavefunction (with charge sector removed)
to the Ising (or Majorana) CFT was reviewed in the introduction. Here we
will also need the similar relation for the Dirac CFT. The Dirac fermion
CFT is defined using the fields $\psi$, $\psi^\dagger$ with the opes%
\bea%
\psi^\dagger(z)\psi(0)&\sim& \frac{1}{z}+\ldots,\\
\psi(z)\psi(0)&\sim& 0+\ldots,\eea%
(and similarly for $\psi^\dagger\psi^\dagger$), in which as usual omitted
terms tned to zero as $z\to0$. These expressions then also give the chiral correlators for the fermion fields, using Wick's theorem (and note that the fields anticommute). The theory can be related to two Majorana fermion field theories $\psi_1$, $\psi_2$ by writing %
\bea%
\psi^\dagger(z)&=&[\psi_1(z)+i\psi_2(z)]/\sqrt{2}\\
\psi(z)&=&[\psi_1(z)-i\psi_2(z)]/\sqrt{2}.\eea%
The relation also works in the presence of spin fields $\sigma_\pm$; by definition, the Dirac fields acquire a change of sign on continuation around
the location of a spin field. More precisely, these are defined by the opes%
\bea%
\psi^\dagger(z)\sigma_-(0)&\sim& \frac{1}{z^{1/2}}\sigma_+(0)+\ldots,\\
\psi(z)\sigma_-(0)&\sim& 0+\ldots,\eea
and similarly with $\psi^\dagger\leftrightarrow \psi$, $\sigma_+\leftrightarrow\sigma_-$. The spin field $\sigma_+$ ($\sigma_-$) carries charge $1/2$ ($-1/2$), in units where $\psi^\dagger$ has charge $1$,
and $\sigma_\pm$ both have conformal weight $1/8$. (The U($1$) charge is also referred to as pseudospin in the bilayer context.) In terms of the two Majorana theories, the Dirac spin fields become the product $\sigma_1\sigma_2$ of the Majorana (Ising) spin fields, which were defined earlier.

The basic property we wish to exploit is the relation of the Dirac fermion theory to a Coulomb plasma. In field theory language, this is bosonization,
in which by introducing a chiral scalar field $\varphi'(z)$, we represent $\psi^\dagger(z)=e^{i\varphi'(z)}$, $\psi(z)=e^{-i\varphi'(z)}$. We note that this is the chiral version of what we had for $Q=1$ (the filled LLL) in Sec.\ \ref{halllaugh} (but now with $\varphi'$ as this is not the charge sector).
Then the spin fields are represented by $\sigma_\pm(z)=e^{\pm i\varphi'(z)/2}$ (all these fields are present in the modular-invariant $Q=4$ {\em bosonic} theory in Sec.\ \ref{halllaugh}). These relations imply that conformal blocks
in the Dirac theory can be written as Coulomb gas functions, that is as products. This is exemplified by the Cauchy determinant identity. The conformal block for Dirac fermions on the plane with $N$ $\psi$'s, $N$ $\psi^\dagger$'s, at $z_i$, $i=1$, \ldots, $2N$, with $i$ even for $\psi$, $i$ odd for $\psi^\dagger$, is %
\bea%
\lefteqn{\langle \psi^\dagger(z_1)\psi(z_2)\cdots \psi(z_{2N})\rangle_{\rm Dirac}=\det \frac{1}{z_i-z_j}}&&\non\\
&=&(-1)^{N(N-1)/2}\frac{\prod_{i<i'}(z_i-z_{i'})\prod_{j<j'}(z_j-z_{j'})}
{\prod_{i,j}(z_i-z_j)}\eea
in which in the matrix with entries $1/(z_i-z_j)$, and in the products, all $i$'s run over odd values, all $j$'s over even values; we are not concerned about the overall sign. The last equality is the determinant identity, and exhibits the block as a Coulomb gas Boltzmann weight, similar to those
discussed earlier. In terms of the Majorana fermion theories, the determinant becomes a sum of products of Pfaffians,%
\be%
\det \frac{1}{z_i-z_j}=2^{-N}\sum_{S_1}{\rm Pf}_{S_1}\,{\frac{1}{z_i-z_j}}
{\rm Pf}_{S_2}\,{\frac{1}{z_i-z_j}}.\ee
Here the sum is over the distinct subsets $S_1$ of $\{1,2,\ldots,N\}$ with an even number of elements $|S_1|=N_1$, and $S_2$ is the complement, $S_1\cup S_2=\{1,2\ldots,N\}$; $S_{1,2}$ refer to the type $1$ and $2$ Majorana
fermions. ${\rm Pf}_{S_{1,2}}$ denotes the Pfaffian with the indices $i$, $j$, ranging over the set $S_{1,2}$. Evaluating a Pfaffian requires an ordering on the index set to fix the overall sign; $S_{1,2}$ are viewed as inheriting the lexicographic ordering from $\{1,2,\ldots,N\}$).

The sum over different numbers of Majorana insertions $\psi_1$ and $\psi_2$ in the last expression is clearly inconvenient. To remove the technical difficulty, we will treat all the states grand-canonically, which is natural anyway for the paired states, as in Sec.\ \ref{genpair}. To do this, let us introduce fermion creation operators $c_\uparrow^\dagger$, $c_\downarrow^\dagger$, for particles with pseudospin in two space dimensions, with canonical anticomuutation relations. These are related to two copies of the spinless fermion creation operators (as in Sec.\ \ref{genpair}) by %
\bea%
c_\uparrow^\dagger&=&[c_1^\dagger(\br)+ic_2^\dagger(\br)]/\sqrt{2}\\
c_\downarrow^\dagger&=&[c_1^\dagger(\br)-ic_2^\dagger(\br)]/\sqrt{2},\eea%
similar to those for the 2D fermion operators $\psi_{1,2}$ (and again, local in $\br$).
(If we think of the Dirac theory as part of the 331 QH states, then $c_\uparrow$ and $c_\downarrow$ represent fermions in the two layers with definite U($1$) ``pseudospin'' values, while $c_1$, $c_2$ are similar to the even and
odd combinations of the pseudospins that are also sometimes used \cite{rr1},
except for the factor of $i$ which will be convenient here.)
Now using the vacuum $|0\rangle$ annihilated by the $c_{\uparrow,\downarrow}$ operators, an unnormalized BCS state with pairing only between $\uparrow$ and $\downarrow$ has the form%
\be%
|\Omega_{\uparrow\downarrow}\rangle=e^{\int d^2\br\, d^2\br'\,M(\br,\br')c_\downarrow^\dagger(\br) c_\uparrow^\dagger(\br')}|0\rangle,\ee%
and the $2N$-particle wavefunction is %
\bea%
\Psi(\br_1,\ldots,\br_{2N})&=&\langle0|c_\uparrow(\br_1)\cdots c_\downarrow(\br_{2N})|\Omega_{\uparrow\downarrow}\rangle.\non\\
&=&{\rm \det}_N\, M(\br_i,\br_j),\eea%
where we have used the same convention that $i$ is odd and $j$ is even
as in the wavefunctions above. In this case, the norm square of any state
(not only a paired state) $|\Omega_{\uparrow\downarrow}\rangle$  with equal numbers of $\uparrow$ and $\downarrow$ particles is computed from the $2N$-particle functions $\Psi$ as %
\bea%
\langle\Omega_{\uparrow\downarrow}|\Omega_{\uparrow\downarrow}\rangle
&=&\sum_N\int\prod_i d^2\br_i\,
\frac{|\Psi(\br_1,\ldots,\br_{2N})|^2}{(N!)^2},\eea%
(which also generalizes to $N_\uparrow\neq N_\downarrow$, in which one divides by $N_\uparrow! N_\downarrow!$). For the paired states above, this norm square reduces to $\det({\bf 1}+MM^\dagger)$.

For these fermions, $M(\br,\br')$ need not have any particular symmetry
in general. If we choose $M(z,z')=\mu/(z-z')$ (switching to $z$'s
corresponding to $\br$'s), where $\mu$ is a fugacity similar to the chemical potential in Sec.\ \ref{genpair}, then we obtain the above wavefunctions.
If we now use the relation to fermions of types $1$, $2$, then the two types decouple in $|\Omega_{\uparrow\downarrow}\rangle$, which equals the (tensor) product of two BCS states for spinless fermions, as in eq.\ (\ref{Omega}),
with $g=M$:%
\be%
|\Omega_{\uparrow\downarrow}\rangle=|\Omega\rangle_1\otimes|\Omega\rangle_2.
\ee%
Clearly, this is consistent with the normalizing factors for these states,
also. In carrying out integrations over space in these states, we will
disregard boundary conditions and the integration domain, but one can think
of them as performed on the sphere, with suitable changes in the details of the functions to include rotational symmetry and so on. The vortices will always
be separated by much less than the system size, and we consider the thermodynamic limit at fixed density. In addition, short distance divergences as fermions come together must be cut off; this will not be shown explicitly. Because the leading term in the ope is independent of the positions of other fields (including the vortices), the cutoff effects should have no effect on the results.

With these few remarks, we have completed the explanation of the relation of
the Dirac and doubled Majorana theories and conformal blocks (wavefunctions) without spin fields (vortices). We should point out that the removal of the charge sector was technically necessary to obtain the tensor product form of paired states. If the charge sector were included in the wavefunction for $\uparrow$ and $\downarrow$ particles (as in the $331$ states), it would
couple the $1$ and $2$ particle types together, and the system would not
factor into decoupled MR states (however, one might argue that this does not matter, along similar lines to arguments for dropping the charge sector).

Now we turn to the introduction of spin fields. The simplest functions to
write down are those in the Coulomb gas form. For $N_\uparrow$ $\uparrow$ particles (with labels $i$ odd), $N_\downarrow$ $\downarrow$ particles (with labels $j$ even), with $n_+$ $+1/2$ charges (or $\sigma_+$'s) at positions $w_k$, $k$ odd, $n_-$ $-1/2$ charges (or $\sigma_-$'s) at positions $w_l$, $l$ even, and with $2N_\uparrow + n_+=2N_\downarrow+n_-$ for charge neutrality,
 the conformal
block (times $\mu$ to the power half the number of fermions) is%
\bea%
&&\mu^{(N_\uparrow+N_\downarrow)/2}\frac{\prod_{k<k'}(w_k-w_{k'})^{1/4}
\prod_{l<l'}(w_l-w_{l'})^{1/4}}{\prod_{k,l}
(w_k-w_l)^{1/4}}\non\\
&&{}\times\frac{\prod_{i,k}(z_i-w_k)^{1/2}\prod_{j,l}(z_j-w_l)^{1/2}}
{\prod_{i,l}(z_i-w_l)^{1/2}\prod_{j,k}(z_j-w_k)^{1/2}}\non\\
&&{}\times
\frac{\prod_{i<i'}(z_i-z_{i'})\prod_{j<j'}(z_j-z_{j'})}
{\prod_{i,j}(z_i-z_j)} \label{dirblock}\eea%
The $z$-dependent factors can also be obtained from a generalization of the Cauchy determinant identity. For $N_\uparrow=N_\downarrow$, the same block is equal to (up to a sign)%
\be%
\frac{\prod_{k<k'}(w_k-w_{k'})^{1/4}\prod_{l<l'}(w_l-w_{l'})^{1/4}}
{\prod_{k,l}(w_k-w_l)^{1/4}}\det M(z_i,z_j),\ee%
where $M(z_i,z_j)$ is given by%
\be%
M(z_i,z_j)=\mu\frac{\frac{\prod_{k}(z_i-w_k)^{1/2}\prod_{l}(z_j-w_l)^{1/2}}
{\prod_{l}(z_i-w_l)^{1/2}\prod_{k}(z_j-w_k)^{1/2}}}{z_i-z_j}.\ee%
This is easily proved as the $w$-dependent factors in $M$ can be taken outside the determinant. Hence, apart from a function of $w$'s only, the grand-canonical state takes the BCS form $|\Omega_{\uparrow\downarrow}(M)\rangle$ for this $M$. This special case of $N_\uparrow=N_\downarrow$ will be sufficient for our purposes, but can be generalized if required.

Now we can address the normalization (or overlap) integrals for the grand-canonical states with $N$-particle components (\ref{dirblock}). For the self-overlap, we can note that the sum (over $N$) of the integrals is simply
the partition function of a Coulomb plasma, with some ``impurities'' inserted
at the $w_k$'s (note the factors $1/(N_\uparrow!N_\downarrow!)$ are essential for this). For the values of the exponent used here, the plasma is in a screening phase. Then if the impurities at $w_l$ are far apart compared with
the screening length, the overlap integral takes the form %
\be%
e^{-Af_0-(n_++n_-)f_{\sigma}},\ee%
similar to that for the case with a neutralizing background. Here $f_0$ and $f_{\sigma}$ are constants, and $A$ is the area of the system. Note that this
is independent of the positions and of the charges of the spin fields.

Though the Dirac theory is Abelian, we can nonetheless obtain not just one but a set of functions from it, by choosing a set of positions $w_1$, \ldots, $w_{n++n_-}$, and varying the assignment of which are $+1/2$ and which are $-1/2$ charges. This gives us $n_++n_-\choose n_+$ distinct states.  We can form the overlap matrix for these states, and so far we discussed the diagonal elements. For
the off-diagonal overlaps, let us suppose we take the overlap of two states
in which the charge assignments are the same except for one $+1/2$ and one $-1/2$ which are switched in one state compared with the other. Then the phase of the integrand winds by $\pm 2\pi$ as one fermion is carried around just one of these two locations (by monodromy), instead of the integrand being purely real in all previous cases. We want to argue that when these locations are far apart compared with the screening length, the overlap goes to zero, because of the oscillations of phase.

One way to see this is to consider the sine-Gordon formulation of this Coulomb plasma, which is represented by a field theory with action%
\be%
S=\int d^2r[\frac{1}{8\pi}(\nabla \varphi')^2-2\mu\cos\varphi']\ee%
in which the screening is due to a term $-2\mu\cos\varphi'$ added to
the action of the (non-chiral) scalar field $\varphi'$. In this phase one can consider the latter term as simply producing a mass term (by expansion of the cosine around a maximum). In this language, the phase winding corresponds to the two $w$s in question being locations of ``magnetic'' charges that produce a ``vortex'' in $\varphi'$, instead of ``electric'' charges. That is, $\varphi'$ changes by $\pm 2\pi$ on going around a magnetic charge. This should be viewed as allowed in the functional integral over $\varphi'$, by identifying values $\varphi'+2\pi$ with $\varphi'$. But as the cosine term tries to pin $\varphi'$ to a multiple of $2\pi$, all of this winding has to be accomodated by a ``domain wall'' on a straight line connecting the two vortices (with exponentially decaying tail off the wall).
The system then incurs a positive domain-wall free energy proportional to the length of the wall (when these vortices are separated by more than the screening length, which is also the thickness of the wall), which means that the overlap we were computing decays exponentially with separation. This argument clearly generalizes to any of the off-diagonal overlaps. Hence the matrix of overlaps ${\cal Z}_{ab}$ (in the language of Sec.\ \ref{cft}) is proportional to the identity, and as the wavefunctions are holomorphic in $w$'s we can obtain the holonomy of these states under adiabatic transport of the $w$'s by reading it off from the monodromy of the wavefunctions, that is from the factors $(w_k-w_{k'})^{\pm1/4}$, where the exponent depends on the signs of the charges.

The main problem now is to relate the conformal blocks of the Dirac and doubled Majorana theories with spin field insertions. Once we do so, we can use the orthonormality of the blocks in the Dirac CFT to infer the same for the blocks in the Majorana CFT, which will complete the argument.

First we will consider the simple case of only two vortices (spin fields), at positions $w_1$ and $w_2$. Above, we took $+1/2$ at $w_1$, $-1/2$ at $w_2$. If we compare with the similar state with these charge assignments reversed, which means in the wavefunctions that we exchange $w_1$ and $w_2$, and take the sum of the two states (with a certain phase relation to be determined in a moment), then the normalization changes only by a factor of $\sqrt{2}$ when the spin fields are far apart, using the above argument. Now in the form of the functions as paired states, we can write for $M(z,z')$%
\be%
M(z,z')=\widetilde{M}(z,z')+L(z,z'),\ee%
where $\widetilde{M}$ is antisymmetric, $\widetilde{M}(z',z)=-\widetilde{M}(z,z')$, and $L$ is symmetric, $L(z',z)=L(z,z')$. Moreover, for the case here of $n_+=n_-=1$, $\widetilde{M}$ is symmetric under the exchange $w_1\leftrightarrow w_2$, while $L$ is antisymmetric. Explicitly,
\be%
\widetilde{M}(z,z')=\frac{1}{2}\mu\frac{
\frac{(z-w_1)^{1/2}(z'-w_2)^{1/2}}
{(z-w_2)^{1/2}(z'-w_1)^{1/2}}+
\frac{ (z-w_2)^{1/2}(z'-w_1)^{1/2}}
{(z-w_1)^{1/2}(z'-w_2)^{1/2}}}{z-z'},\ee%
and
\be%
L(z,z')=\mu\frac{w_1-w_2}{2\prod_{l=1,2}[(z-w_l)^{1/2}(z'-w_l)^{1/2}]}.\ee%
Now when we sum the states over the exchange of $w_{1,2}$, the phase is
chosen so that if the factor $(w_1-w_2)^{-1/4}$ is omitted, the functions are simply added. Then in the expansion in many-particle wavefunctions, all terms with an odd number of $L$'s drop out. $L(z,z')$ is non-singular as $z\to z'$.
Thus, in the language of Ref.\ \cite{rr1}, in the terms containing $L$'s some $\uparrow\downarrow$ pairs are broken, and the unpaired $\uparrow$ fermions occupy a mode with constant wavefunction (as do the $\downarrow$ fermions). But as only even numbers of $L$ factors can appear in our state, the mode must be occupied by more than one $\uparrow$ fermion, and so these terms vanish on antisymmetrization (i.e.\ by Fermi statistics).

The wavefunction of the state thus has the form of the above paired state, with the antisymmetric $\widetilde{M}$ in place of $M$. Hence, it is equal to the tensor product of two paired states of spinless fermions containing the same $g=\widetilde{M}$. The form of each of these paired states is therefore%
\be%
(w_1-w_2)^{-1/8}|\Omega(\widetilde{M})\rangle_{1,2}.\ee%
The form of $\widetilde{M}$ is simply $\mu$ times the propagator for the Majorana fermion in the presence of two spin fields. The factor $(w_1-w_2)^{-1/8}$ has exponent equal to minus twice the conformal weight $1/16$ of the spin field, and the $N$-particle wavefunction is exactly the conformal block for $N$ Majorana fermions and two spin fields (here $N$ is even) \cite{mr}. Because the doubled version of this state is normalized for well-separated spin fields, and holomorphic, we can conclude the same for this state. Hence we can read off the phase for adiabatic exchange of the two vortices, $e^{-i\pi/8}$. This is a simple phase factor, yet does not come from the charge sector of a QH state, which was removed here. This contribution was expected and emphasized early on \cite{mr}, and previously derived by a different method \cite{gn}. We note that the same result can be obtained for $N$ odd instead of even. This is obtained by taking the linear combination of the two charge assignments for $w_1$, $w_2$ in the Dirac theory that is orthogonal to that used above, in which case all the terms with exactly one $L$ survive, and give rise to blocks in the Majorana theory that all have odd particle number $N$ (these are non-vanishing in the presence of a positive number of spin fields). These correspond to occupying the zero mode with one fermion \cite{rr1}.

For the case of four vortices in the paired state, we use similar but more involved arguments. We begin by introducing notation for the six states with $n_+=n_-=2$. We will write the BCS states previously called  $|\Omega_{\uparrow\downarrow}(M)\rangle$ as $|\Omega_{13,24}\rangle$, and the remaining five are obtained by permuting the indices; thus the first two subscripts are the positions of the $+1/2$ charges, and the last two are the positions of the $-1/2$ charges. The grand-canonical states are in full%
\be%
|\Psi_{13,24}\rangle= \frac{\prod_{k<k'}w_{kk'}^{1/4}\prod_{l<l'}w_{ll'}^{1/4}}
{\prod_{k,l}w_{kl}^{1/4}}|\Omega_{13,24}\rangle,\ee%
in which again $k$, $k'=1$, $3$, and $l$, $l'=2$, $4$, and five other states obtained by permuting the indices; these form an orthonormal set when the spin fields are well separated. Once again, we can make use of the decomposition of $M(z,z')$ into antisymmetric and symmetric pieces under $z\leftrightarrow z'$, called $\widetilde{M}(z,z')$ and $L(z,z')$ respectively. These are respectively even and odd under exchange of all of the coordinates of $+1/2$ with coordinates of $-1/2$ charge spin fields, that is $w_1\leftrightarrow w_2$, $w_3\leftrightarrow w_4$. Explicitly,%
\bea%
\lefteqn{\widetilde{M}(z_i,z_j)_{13,24}=}&&\\
&&\frac{1}{2}\mu\frac{\left[\frac{(z-w_1)(z-w_3)(z'-w_2)(z'-w_4)}{(z-w_2)
(z-w_4)(z'-w_1)(z'-w_3)}\right]^{1/2}+(z\leftrightarrow z')}{z-z'}
\eea%
$L$ is non-singular as $z\to z'$ for symmetry reasons. When we write the states using the $1$, $2$ types of fermions instead of $\uparrow$, $\downarrow$, each $\widetilde{M}$ term involves types $11$ or $22$, while each $L$ term involves types $12$. Hence if we sum and form the combinations like%
\be%
|\Omega_{13,24}\rangle+|\Omega_{24,13}\rangle\ee
(and two similar ones), then these are sums of states that contain only {\em even} numbers $N_1$ and $N_2$ of the $1$, $2$ fermions. They are symmetric under exchanging $1$ with $2$ type fermions. There are just two Majorana conformal blocks for even fermion number (and also two for odd fermion number). We are able to show that these three states in the doubled theory are constructed from the symmetrized tensor products of the two states in each of the spinless theories that produce the Majorana blocks (similar to the construction of the triplet of spin $1$ states from the symmetrized tensor product of two spins $1/2$ in angular momentum theory). This is sufficient to deduce the orthonormality of the two Majorana blocks for $N$ even.

Explicitly, the two conformal blocks for four spin fields and $N$ fermions ($N$ even) in the Majorana theory were found by NW \cite{nw}, who used a related doubling technique. The expressions are %
\bea%
\mu^{N/2}{\cal F}_{0,1/2}&=&\frac{1}{\sqrt{1\pm \sqrt{1-x}}}\left(\frac{w_{13}w_{24}}{w_{12}w_{32}w_{34}w_{14}}\right)^{1/8}
\non\\
&&{}\times \left[{\rm Pf}\,\widetilde{M}(z_i,z_j)_{13,24}\right.\non\\
&&\quad\left.\pm\sqrt{1-x}\,{\rm Pf}\,\widetilde{M}(z_i,z_j)_{14,23}\right],\eea
where $x=w_{12}w_{34}/(w_{13}w_{24})$ is a cross-ratio, and the upper sign is for $0$, the lower for $1/2$.
For $N=0$, in which case the Pfaffians are replaced by $1$, these two blocks were found in Ref.\ \cite{iz} (see also Ref.\ \cite{id}). {}From these, the four-point spin correlation function of the critical Ising model is $\sum_{s=0,1/2}|{\cal F}_s|^2$. The same combination should give the correlation function in the presence of fermion (energy operator) insertions. Then if we are to obtain holonomy equal to monodromy of these blocks, these states should be an orthonormal pair \cite{nw}. We will denote the two grand-canonical states corresponding to these $N$-particle wavefunctions by $|\pm\rangle$.

The relations of the Majorana and Dirac blocks, when there are no fermions,
also dictate what the relations must be when fermions are present, most conveniently for the grand-canonical forms. After a considerable amount of algebra, we can verify that%
\bea%
\left(\frac{w_{13}w_{24}}{w_{12}w_{32}w_{34}w_{14}}\right)^{1/4}
\left(|\Omega_{13,24}\rangle+
|\Omega_{24,13}\rangle\right)&=&\non\\
\qquad|+\rangle_1\otimes|+\rangle_2+|-\rangle_1\otimes|-\rangle_2,&&\eea
\bea%
\left(\frac{-w_{14}w_{23}}{w_{12}w_{42}w_{43}w_{13}}\right)^{1/4}
\left(|\Omega_{14,23}\rangle+
|\Omega_{23,14}\rangle\right)&=&\non\\
\qquad|+\rangle_1\otimes|+\rangle_2-|-\rangle_1\otimes|-\rangle_2,&&\eea
\bea%
\left(\frac{-w_{12}w_{34}}{w_{13}w_{23}w_{24}w_{14}}\right)^{1/4}
\left(|\Omega_{12,34}\rangle+
|\Omega_{34,12}\rangle\right)&=&\non\\
\qquad|+\rangle_1\otimes|-\rangle_2+|-\rangle_1\otimes|+\rangle_2.&&\eea
The left-hand sides of these are three orthonormal states (in the usual sense), and it is then easy to see that $|+\rangle_1$ and $|-\rangle_1$ are an orthonormal pair (in the usual sense), and of course similarly for type $2$ fermions. As we have explained, this implies that the holonomy for an adiabatic exchange of well-separated vortices in the paired states of spinless fermions equals the (non-Abelian) monodromy.

Finally, we can consider more than four quasiholes, at the level of counting arguments. For $n=6$, similar to $n=4$, the number of symmetrized tensor products of Majorana conformal blocks with $N_{1,2}$ even equals the number of Dirac blocks with $n_+=n_-=3$, when they are symmetrized between $+$ and $-$ spin fields (i.e.\ both numbers equal $10$). In this case a similar argument may go through. But for $n\geq8$, the use of $n_+=n_-=n/2$ and $N_\uparrow=N_\downarrow$ and symmetrization produces a set of orthonormal states the number of which is less than the number of Majorana blocks (with $N_{1,2}$ both even) symmetrized between $1$ and $2$. We are then forced to consider all particle numbers together. In the Majorana theory, there are $2^{n/2-1}$ blocks for $N$ even, and to these we can add the number for $N$ odd. The square of this is $2^n$, the total number of blocks in the Dirac theory if we allow all possible assignments of the charges on the spin fields, with the numbers $N_\uparrow$ and $N_\downarrow$ chosen to ensure neutrality. The functions must agree because of the interpretation of the Dirac theory as the doubled Majorana theory, and hence we expect that a version of the argument goes through for all values of $n$.


\section{General conditions for holonomy to equal monodromy}
\label{gencon}

In this section we discuss general conditions for the holonomy of
trial wavefunctions given by conformal blocks, including those for
QH systems that include the charge sector, to equal their
monodromy. In view of the preliminary discussion in Sec.\
\ref{intro}, it is sufficient to find general conditions for the
conformal blocks to be orthonormal in the basis in which the
monodromy matrices are unitary. After giving the general
arguments, we turn to examine a number of examples, and discuss non-unitary and irrational CFTs.

\subsection{Relevant and irrelevant perturbations of the CFT}

We begin with the functions $\Psi_a$ from Sec.\ \ref{cft}, which
include the charge sector, or with the functions ${\cal
F}_a(w_1,\ldots;z_1,\ldots)$ which are the blocks of the CFT with
the charge part removed. In either case, we can view these as
related to a perturbation of an underlying CFT by some operators
(viewed within two-dimensional field theory); here by the
underlying CFT we mean including the charge sector contribution,
which as we mentioned is a massless scalar field $\varphi$
\cite{mr}. Then the particles are represented by $\psi_{\rm
e}(z)=e^{i\sqrt{\nu^{-1}}\,\varphi(z)}\psi(z)$, where $\nu=P/Q$ is
the filling factor (if we view the CFT with the charge sector
removed as a theory of Abelian anyons, then these particles are
represented by $\psi$ alone). Similarly, the quasiholes are
represented by some fields $\phi(z)=e^{iq_{\rm
qh}\varphi(z)}\tau(z)$ (or just by $\tau$, respectively; it should
be clear how to generalize to include several types of qhasihole).
(These ``fields'' are really ``chiral vertex operators''
\cite{dfms}.) Here $\varphi$ has two-point function
$\langle\varphi(z)\varphi(0)\rangle=-\ln z$, $q_{\rm qh}$ is
determined by single-valuedness of $\Psi_a$ in the particle
coordinates, and the charge (particle number relative to the
ground state) of the quasihole is $-q_{\rm qh}\sqrt{\nu}$. The
functions $\Psi_a$ are reproduced as a chiral correlator of these
fields together with the background charge density factor%
\be%
\exp[{-i\int d^2z'\,\sqrt{\nu}\,\varphi(z')/(2\pi)}];\ee%
the approach to handle this factor is discussed in MR \cite{mr}
(for the functions ${\cal F}_a$, the background charge density is
omitted).

The overlap integrals ${\cal Z}_{ab}$ represent the inner products
of these $\Psi_a$, viewed as quantum mechanical states of $N$
particles (there are similar ones without the charge sector). It
is convenient to introduce a grand-canonical point of view
\cite{read89}, as in the previous sections. Then the overlap
integral can be written formally as an expression%
\bea%
{\cal Z}_{ab}&=&\left\langle e^{\lambda\int d^2z\,\bar\psi_{\rm
e}(\bar{z})\psi_{\rm e}(z)}e^{-i\int
d^2z'\,\sqrt{\nu}(\varphi(z')+\bar\varphi(\bar{z'}))/(2\pi)}\right.
\non\\
&&\left.{}\vphantom{e^{\int}}\times\overline\phi(\bar{w}_1)
\phi(w_1)\cdots\right\rangle,\eea%
where the expectation is taken in the underlying CFT, and the
notation for a choice of blocks is suppressed. The fields with
bars $\bar{}$ over them are antiholomorphic partners of the chiral
fields. Expanding the exponential in $\bar{\psi}_{\rm e}\psi_{\rm
e}$ gives a series of integrals over $N$ particles, $N=0$, $1$,
\ldots. For the plane the background charge is correct as written,
but for the sphere one should define the series with the
appropriate background charge, spread uniformly on the sphere, for
each particle number, so as to ensure total charge neutrality
(related to the total number of magnetic flux quanta through the
sphere in the QH effect). For a given (real) value of $\lambda$,
the dominant terms in the sum cluster around some value of $N$,
and we choose $\lambda$ so that this is large, and the
fluctuations in $N$ will be of order $\sqrt{N}$. In some cases,
the correlator due to the non-charge part will be nonzero only for
$N$ congruent to some number, modulo another constant (for
example, $N$ even, for the case of no quasiholes in the MR state).

Leaving aside the background charge density for a moment, the
overlap integral has a form similar to a correlator in a {\em
perturbed}\/ CFT. In fact, the diagonal sum $\sum_a{\cal Z}_{aa}$
is the (un-normalized, in the sense of Sec.\ \ref{cft}) correlator
in the diagonal version of the underlying CFT, perturbed by adding
the term $\lambda\bar{\psi}_{\rm e}\psi_{\rm e}$ to the
Lagrangian. For example, omitting the charge sector, if $\psi$
were a Majorana fermion, this term would be the mass term that
drives the continuum theory of the Ising model off its critical
point. For the charge sector, we have in addition the background
charge density, which we will view as also part of the
perturbation. This notion of a perturbed CFT will be crucial to
our approach from here on.

In general, given a CFT and a perturbation, the basic question is
whether the perturbation is relevant, irrelevant, or marginal.
This analysis (and even referring to it as a perturbation)
presupposes that the perturbation is weak, i.e.\ that the
coefficient $\lambda$ is small. Let us once again ignore the
charge sector entirely for a moment, or assume that it has been
removed. Then the perturbation is by the operator
$\bar{\psi}\psi$. It is relevant if its scaling dimension
$2h_\psi$ is less than two, irrelevant if this is greater than
two, and marginal if it is two. This is because the scaling
dimension of $\lambda$ is the codimension
$d-2h_\psi=2-2h_\psi$, and the perturbation is relevant if the
codimension is positive, because then the coefficient grows under
renormalization group (RG) rescaling transformations to approach
larger length scales, with the opposite for the irrelevant case.
For marginal perturbations, one must go to higher (non-linear)
order in $\lambda$ to see whether it grows or shrinks under the RG
on going to larger length scales. If the perturbation is relevant,
then a crossover length scale
$\xi_\times\sim\lambda^{-1/(2-2h_\psi)}$ can be defined. We should
note that the conformal blocks are singular as two $\psi$s
approach one another, diverging as $z^{-2h_\psi+\psi^*}$ as
discussed in the introduction. This may be strong enough that a
cutoff must be introduced, by cutting tube-shaped regions out of
the integration along these diagonals, and also along those at
$z_i=w_l$, with radius say $\varepsilon$. Then the limit
$\varepsilon\to0$ requires renormalization of the operators. We
are suppressing dependence on $\varepsilon$ from these
expressions, by holding it fixed. The parameter $\lambda$ may be
viewed as defined at the cutoff length scale, and the crossover
length $\xi_\times$ is defined in units of $\varepsilon$. Then
$\xi_\times$ is much larger than $\varepsilon$ when $\lambda$ is
small (and relevant). For our purposes, $\xi_\times$ can be viewed
as the typical spacing of the insertions of $\bar\psi\psi$ in the
integral. Note that while the conformal blocks are viewed as
conformally covariant, conformal, and in particular scale,
invariance of the correlators is violated by the presence of the
perturbation.

It is now clear that in general we will need to examine the
relevance of the perturbation. When the charge sector is included,
the situation is different. The perturbation is certainly
relevant, but the scaling dimension of
$e^{i\sqrt{\nu^{-1}}(\varphi+\bar{\varphi})}$ in the underlying
CFT, which is $\nu^{-1}$, is apparently of little importance. {}From
the basic example of the Laughlin state \cite{laugh}, we know that
the mod-square of the Laughlin wavefunction is a one-component
plasma with a background charge density, which is in a screening
phase when the exponent $q$ is larger than about $70$. For larger
values of $\nu^{-1}$, a crystalline state is formed. A scaling
analysis does not seem to be of much use in finding this physics.
(We note that Ref.\ \cite{nw}, section 8, began to formulate
arguments about the relevance of the perturbation, but incorrectly
states that $\psi_{\rm e}$ has negative scaling dimension.) In
contrast, for the two-component plasma, which is charge-neutral
without a background, the physics is the Kosterlitz-Thouless
theory \cite{kt}, which at small $\lambda$ (``fugacity'') is
determined by the relevance of the perturbation; when it is
relevant (which would be $\nu^{-1}<2$, if we use the same
conventions but include charges of the opposite sign instead of
the neutralizing background), the plasma arrives at a screening
phase. This was emphasized in Ref.\ \cite{gmfrs} in a context
similar to the present one.

For the more general cases when the CFT includes more than the
charge sector, and the operator $\psi_{\rm e}$ contains
$\psi\neq1$, it appears that as long as $\nu$ is such that the
charge sector is in the screening phase when $\psi$ is ignored,
this will still be true in the presence of $\psi$. The correlators
in the CFT with the charge sector removed do not contain high
powers of $z$ as one $\psi(z)$ is moved away to large distance,
and so are unable to affect the screening properties very much
(except perhaps near the transition to the crystal). This is
generally taken for granted in treatments of trial wavefunctions
in the fractional QH effect.

It is now natural to adopt the following analysis. To consider the
effect of the perturbation of the underlying CFT, including the
charge sector, we begin by noting that if $\nu^{-1}$ is less than
about $70$, the RG flow will be to screening behavior of the
charge sector. This sets in by the time the screening length is
reached; this scale is of order the particle spacing for filling
factors not too far (say within a factor of 10) from 1. On large
scales, we then should consider how the remaining CFT degrees of
freedom behave. The operators $\psi$ are still attached to the
charges in the plasma. If this was ignored, we would integrate
over the positions of the $\psi$s as if the charge sector were
absent. Then the scaling dimension of $\bar{\psi}\psi$ would come
into play. Because the insertions of $\bar\psi\psi$ are in fact on
the charges in the plasma, fluctuations in the density of the
$\bar\psi\psi$s, which would be Poissonian if there were no
plasma, are in fact suppressed if one looks at a length scale
larger than the screening length. In the presence of the charge
sector, the screening length is the scale from which the RG
analysis must be started, rather than the scale $\varepsilon$
which tended to zero. As this scale is similar to the spacing of
particles, and thus of $\bar\psi\psi$ insertions, this corresponds
to a fugacity for the $\bar\psi\psi$ insertions of order $1$. In
this situation, it is hardly surprising that the fluctuations in
the number density of insertions are not Poissonian. This is
somewhat like a lattice model with $\bar\psi\psi$ on each lattice
site. In effect, the strength of the perturbation in the CFT (with
the charge sector removed) is here of order $1$.

We are then forced to consider the effect of the perturbation by
$\bar\psi\psi$ when its strength is of order one, rather than
small. When the perturbation is weak, the RG flow will carry the
system back to original CFT if the perturbation is irrelevant or
marginally irrelevant. If the perturbation is relevant, the flow
will usually carry the system to a different RG fixed point
(possibly passing close to other fixed points along the way). If
the perturbation is strong, the RG flow might be to a different
fixed point (or phase of the 2D theory) than when it is weak. In
particular, a perturbation that is irrelevant when weak might
still lead to a flow to different phase when strong. An example is
the KT theory again, according to Kosterlitz's RG flow diagram for
the two-component plasma \cite{kt}. On the other hand, a
perturbation that is relevant when weak seems unlikely to lead to
a flow back to the original CFT when it is strong. RG flows that
travel in a cycle back to the original fixed point are very
unusual, and impossible in unitary theories according to the
$c$-theorem \cite{cthm}. Thus a perturbation that is relevant for
small $\lambda$ should lead to a flow away from the original CFT
towards another fixed point also when it is strong, possibly to
the same one as starting from weak $\lambda$. If a perturbation
that is irrelevant when weak does flow to a different fixed point
when strong, it may be difficult to identify what fixed point that
is.

Thus the simplest assumption to make for the QH systems would be
that the behavior of the system viewed as a CFT without the charge
sector, perturbed by $\bar\psi\psi$ with coefficient $\lambda$ of
order 1 is the same as for the same perturbation with $\lambda$
small. In that case, if $\lambda$ is irrelevant, the system flows
back to the original CFT, and if $\lambda$ is relevant it flows to
the same fixed point different from the starting CFT, even though
the perturbation is not weak. We note that this point of view is
essentially a generalization of what has been used successfully in
studying composite particle approaches to the QH states, in which
the charge sector is factored off, and results for composite
particles in zero magnetic field (to correspond to the present
case) are used as a ``mean field'' approximation \cite{comppart}. But we will
learn from simple examples later that this plausible assumption
can fail in practice, in that even for cases in which $\lambda$ is
irrelevant (when weak), there can be flows to some other fixed
point. So we will not proceed by using this assumption. On the
other hand, if there are systems in which we are willing to use
the conformal blocks without the charge sector as trial
wavefunctions (with cutoff $\varepsilon$ much less than the
particle spacing), then the relevance or irrelevance of $\lambda$
comes directly into play. (In real non-QH systems, there could be
similar problems also, for example for paired states in the
weak-pairing phases. But for the paired states as studied in Sec.\
\ref{genpair}, this difficulty does not in fact seem to occur.)


\subsection{Long-distance fixed point}

Whether or not the simplest assumption for the effect of a
perturbation holds when its strength is of order one, we can
analyze the possible results for the holonomy at large length
scales. If we refer to the starting CFT as the short distance or
ultraviolet (UV) fixed point, then we will now be considering the
possible long-distance or infrared (IR) fixed point reached at
large length scales (possibly after the RG flow passes near other
fixed points at intermediate scale). There are generally two
classes of IR fixed points to consider. One is a ``massive'' fixed
point, in which all correlations go to constants with
exponentially decaying corrections. The other is a fixed point
that contains ``massless'' degrees of freedom, corresponding to a
critical theory, possibly a conformal field theory. In this there
are power-law correlations, though we should point out that the
power could be zero, and in all cases there will be power-law
subleading corrections also. We emphasize that these terms are
applied here to the behavior of {\em two-dimensional} (2D) field
theories. A further possibility is that of exponentially {\em
growing} correlations. This cannot happen in a unitary theory, but
may be allowed in non-unitary theories; in non-unitary CFTs, one
finds power-law correlations that increase with distance (negative
scaling dimensions), so it is not clear that exponentially growing
behavior is ruled out in the ``massive'' cases. This would
probably indicate an instability of the vacuum of the 2D theory.
We will not have much to say about this possibility.


\subsubsection{Massive IR fixed point}

We can consider the cases with or without the charge sector
together. In order to obtain holonomy equal to monodromy for the
trial functions given by conformal blocks (possibly including the
charge sector, on both sides of this relation), we need the
overlap integrals ${\cal Z}_{ab}$ to approach $\delta_{ab}$ at
large separations of the $w$s (up to a $w$-independent factor). We
now interpret this in terms of the RG flow of the underlying CFT.
The behavior is very much like correlation functions approaching a
constant in the perturbed phase, where the correlation functions
would however be $\sum_a{\cal Z}_{aa}$ (as noted in Ref.\
\cite{nw}, where the role of more than one conformal block is
neglected at this point). Thus we will suppose that the RG flow
goes to an IR fixed point that is ``massive'', that is one in
which correlations go to constants or to zero, with exponentially
small corrections at finite separation. (No power-law corrections
are possible in the absence of massless degrees of freedom in the
IR fixed point theory.)

We wish to consider then the behavior of the overlap {\em
matrices} ${\cal Z}_{ab}$ in such a phase. Important conditions
are placed on the possible behavior of these by the monodromy
properties they possess. It is easy to see that they inherit the
monodromy properties that they had in the starting CFT (in which
in the conformal blocks, the fields $\psi_{\rm e}$ or $\psi$ were
either omitted entirely, or located at fixed positions). That is,
under a braiding operation, either exchange of identical
quasiholes, or a circuit of one type of quasihole around another
one, the overlap changes by %
\be%
{\cal Z}_{ab}\to \sum_{c,d}(M^\dagger)_{ac}{\cal Z}_{cd}M_{db},
\label{Zmon}\ee%
where $M_{ab}$ is the monodromy matrix, describing the
corresponding effect on the conformal block. Now as noted above,
$\sum_a{\cal Z}_{aa}$ is a correlation function of local
operators, which goes to some constant. As the braiding matrices
are unitary, it is invariant under monodromy. (If the braiding
matrices in the monodromy were not unitary, it would be impossible
for the holonomy in the quantum mechanical theory to equal the
monodromy!) In a pure phase (in the sense here of 2D statistical
mechanics), this correlation approaches
$\prod_l\langle\bar\tau(\bar{w}_l)\tau(w_l)\rangle$, a product of
the expectation values of the local operator. If the perturbed
phase is not pure, this factorization may not hold, but the value
of the correlation approaches the product of the expectations that
would hold in one of the pure components of the phase, when the
phases are related by symmetry (for example, the low-temperature
phase of the Ising model). More generally, {\em all}\/ the
components of ${\cal Z}_{ab}$ can be considered as (linear
combinations of) correlation functions of some operators. The
non-trivial monodromy (\ref{Zmon}) implies that the operators are
not all mutually local. (Notice that Abelian factors in the
braiding drop out of this argument). In this situation, it is
well-known that the correlator cannot go to a constant. Thus, if
the phase were pure, it should reduce to a product of expectation
values of the various operators. As each expectation value is
independent of the locations of the other operators, the product
cannot have the required monodromy. Consequently, the overlaps
must go to zero at large distances (exponentially), except for
combinations that are invariant under all allowed braiding
operations, which can survive as constants.

The simplest example of this is the Ising model again, in which
$\bar\psi\psi$ (for $\psi$ the Majorana fermion) is the
perturbation. The Ising order operator is $\sigma(w,\bar{w})$, and
there is also the disorder operator $\mu(w,\bar{w})$; these two
fields are dual to each other. Under a circuit of one around the
other (along a path containing no others of either type) in a
correlator containing both, the monodromy is a factor $-1$. The
fields are mutually non-local, and cannot both have expectation
values in the same phase. $\sigma$ has a non-zero expectation in the
low-temperature phase, while $\mu$ has one in the high-temperature
phase. On the sphere, there are $2^{n/2-1}$ conformal blocks for
the chiral $\sigma$, so $2^{n-2}$ components ${\cal Z}_{ab}$ in
the overlap integral matrix. In fact there are this many blocks
both for even and for odd fermion numbers in the blocks (for
$n>0$), while $n$ must be even. In correlations of $\sigma$s and
$\mu$s in the {\em critical}\/ theory ($\lambda=0$), there are
$2^n$ possible choices (either $\sigma$ or $\mu$) of the field at
each $w_l$. However, interchanging the types $\sigma$ and $\mu$
leads to the same functions by duality, so there are only
$2^{n-1}$ distinct functions. Further, in the absence of any
fermions ($N=0$), the number of $\sigma$s and of $\mu$s must both
be even (this is because fusion of the nonchiral $\sigma$ and
$\mu$ produces either a $\psi$ or a $\bar\psi$ \cite{bpz}). More
generally, the number of $\mu$s plus $N$ must be even, and (hence)
also the same for $\sigma$s. Hence there are only $2^{n-2}$
distinct non-zero functions. This essentially shows that the terms
in ${\cal Z}_{ab}$ correspond to the various ways to choose order
and disorder fields in the correlator. In more detail, in the
grand-canonical version of the perturbed partition function, both
even and odd $N$ occur. The odd values of $N$ drop out if we sum
the cases $\lambda$, $-\lambda$. The counting of functions still
works, but one cannot say if it is the $\sigma$ or the $\mu$ that
has an expectation value. The only non-zero limit of large
separation corresponds to the correlator of all $\sigma$s (or all
$\mu$s), and this is equal to the trace $\sum_a{\cal Z}_{aa}$. If
one does not sum over the two signs of $\lambda$, but instead
fixes this sign, then it turns out that the two functions ($N$
even or odd, but close in value) are nearly equal, and so for one
sign of $\lambda$ they cancel, leaving a correlation function that
decays with distance (exponentially, with the correlation scale of
order the separation of $\overline\psi\psi$ insertions). For the
other sign of $\lambda$ they add. This is how the distinct
behavior of, for example, $\sigma\sigma$ and $\mu\mu$ correlations
in either the high- or low-temperature phase arises. Duality acts
by turning $\sigma$ to $\mu$ and $\mu$ to $\sigma$, and reversing
the sign of $\lambda$.

Returning to the general argument, if the overlap matrix
approaches a constant at large distances, then this must be
invariant under monodromy. Now suppose that the unitary monodromy
representation of the braid group on the conformal blocks is
irreducible (as it is in many CFTs). It now follows from Schur's
lemma that the only possible invariant form for the overlap matrix
is a constant times $\delta_{ab}$ (here we view the matrix ${\cal
Z}_{ab}$ as a map of the vector space of conformal blocks into
itself, defined in the given components). If the representation is
not irreducible, then we can find a (unitary) change of basis of
the space of conformal blocks so that the representation is block
diagonal with irreducible diagonal blocks. If the blocks are
mutually non-isomorphic as representations of the braid group,
then the overlaps of vectors in distinct blocks vanish, and the
same argument then applies to the overlap matrix of vectors from
the same block. It is not obvious that the constant multiplying
each of these is the same. If some of the distinct braid group
irreducibles occur more than once, then the overlap matrix can
only be reduced to a direct sum of tensor products of the identity
matrix on each irreducible, times a Hermitian matrix of overlaps
of size given by the multiplicity of that irreducible. By a choice
of basis that preserves the braid group action, these matrices
(and hence the full overlap matrix) can be made diagonal. The
diagonal $w$-independent constants in this matrix are expected
generically to be non-zero, so that they can be absorbed into
rescalings of the conformal blocks, though this may be forbidden
in conformal field theory, and it is possible that there are
deeper reasons why the constants must be equal. {}From the point of
view of the blocks as trial wavefunctions, rescaling them so that
they are all normalized is harmless, and by hypothesis can be done
independent of $w$.

This completes the argument, because when the overlap matrix
${\cal Z}_{ab}$ is proportional to the identity, so the basis is
orthonormal, as well as holomorphic in $w$, the Berry connection
vanishes, showing that {\em holonomy equals monodromy}. In
particular, for the Ising (or MR) example, this shows that the
expected non-Abelian statistics occurs, using only known
properties of the massive Ising (Majorana) field theory.
Generally, we have shown that the condition that the IR fixed
point of the two-dimensional theory is massive (or that
correlation functions of the $\tau$ insertions go to non-zero
constants with exponentially decaying corrections) is a sufficient
condition for holonomy to equal monodromy. It is also a necessary
condition, as a non-vanishing Berry connection is unlikely to have
no effect on the holonomy.

The question of whether or not the braid group representation is
irreducible is not crucial for the argument, as we have seen, but
is nonetheless worth a comment. In many familiar CFTs, it is
irreducible, including in those for the entire RR series. More
generally, we may make a connection with whether a theory, or QH
state, is universal for topological quantum computation. The
latter reduces to the question of whether the braid group
representations (for increasing numbers of quasiholes) are dense
in U($M$), where $M$ is the dimension of the space of blocks (more
accurately, as the U($1$) factors cannot be dense, it is whether
it is dense in the group PU$(M)$ of unitary matrices modulo phase
factors). If the representation is dense, then it is certainly
irreducible. Thus in all cases of interest for universal quantum
computation, the representation is irreducible.

We should add some further remarks on the case in which the monodromy matrices $M$ are not unitary. This occurs when the correlation functions in the CFT are not the sum of modulus-squares of the blocks (i.e.\ a positive-definite sesquilinear form is used to form the norm square of the vector of blocks, and the basis choice brings this to standard form) as assumed above, but some terms appear with a minus sign (this may occur in non-unitary CFTs). In this case, the monodromy matrices are pseudo-unitary. If the blocks form an irreducible representation of the braid group, then the invariance of the large-separation limit of the overlap matrix ${\cal Z}$ under monodromy forces it to be a non-positive sesquilinear form also (proportional to the same one as in the correlation functions). But the overlap matrix is a positive-definite form by construction, because of the positive-definite inner product in quantum mechanics, and so no such long-distance form is possible. This contradiction means that the hypothesis of a flow to the massive fixed point is untenable in such a case. Of course, as mentioned earlier, it is also impossible for the necessarily unitary holonomy to be equal to any non-unitary monodromy. 

We also wish to make a comment regarding the mapping class (or
modular) group, and the so-called twist operation on quasiholes.
This latter is an essential part of the structure of a modular
tensor category (or even just a ribbon category), or of a
topological phase or topological quantum field theory (TQFT) in
$2+1$ dimensions. It is effectively a counter-clockwise rotation
of a quasiparticle about its center. For a quasiparticle of a
definite type $\alpha$, the twist $\theta_\alpha$ is expected to be given by
$e^{2\pi i h_\alpha}$, where $h_\alpha$ is the conformal weight of the
corresponding field in the CFT (including the charge sector). This
cannot be produced using braiding only, which translates
quasiparticles without rotating them (we generally consider doing
so in the plane, which is flat). It is nonetheless a physical
operation on quasiparticles that can be obtained adiabatically, as
we will now explain.

First, we point out that for any non-negative number of
quasiparticles at marked points on a surface of genus ${\cal
G}\geq0$, there is a more general mapping class group of
diffeomorphisms that preserve the complex structure. It is
generated by the ``Dehn twists'', which are obtained by cutting
the surface along a closed simple curve, making a $2\pi$ rotation,
and rejoining; the closed curves can be non-contractible cycles,
either wrapping around a handle, or enclosing some number of
marked points. Thus the twist of a single quasiparticle is one of
these. In the absence of any marked points, this mapping class
group reduces to that mentioned in Sec.\ \ref{mapcl}; with marked
points, it includes the braid group for the plane as a subgroup.
Further, in the presence of marked points, their coordinates
(modulo complex analytic transformations) can be viewed as moduli.

Now the idea is to calculate the effect of any Dehn twist
adiabatically by deforming the metric of the surface until one
reaches a metric equivalent to the one we started with. This involves a generalization of what was done when we varied the aspect ratio of the torus. While we will not attempt to write down explicit expressions, we expect that the conformal blocks depend holomorphically on the moduli, except for simple
contributions in the charge part and some that relate to the conformal weight of each insertion, when written with the infinitesimal area element of the metric held fixed (these will produce effects involving the Hall viscosity). Moreover, the monodromy of the functions will generate a unitary ``monodromy representation'' of the mapping class group (strictly, it will be projective, so we should pass to a covering group). This can be used to show that the conformal blocks are orthonormal for each deformed metric, by a similar argument as for the braid group. Then adiabatic transport around a closed loop in the moduli space, which corresponds to implementing an element of the mapping
class group, should produce holonomy equal to the monodromy, up to a phase related to Hall viscosity that depends on the path taken in moduli space.
Applying this to the twist, we expect to show that after removing the Hall viscosity part (much like removing the phase related to the Magnus force from the statistics calculation), the twist $\theta$ for each quasiparticle type will be the same as in the underlying CFT (including the charge sector). This seems very likely to succeed in view of the analogy with the argument for the statistics given here, and would
essentially complete the derivation of the TQFT from the trial
wavefunctions given by conformal blocks, when the 2D theory flows
to a massive IR fixed point, up to some remaining details that
will be discussed in App.\ \ref{mtc}. Unfortunately, the explicit calculation must wait for another occasion.

There is an easier alternative approach, but it yields incomplete information. We can follow the techniques of sections \ref{halllaugh} and \ref{oddvis}, and consider modular transformations of the ground states on the torus, implemented adiabatically. When the hypothesis of a massive fixed point holds, we then find that (as in the CFT) the modular transformations obey $(ST)^3=S^2=C$, where $C$, which obeys $C^2={\rm 1}$, is the ``charge conjugation'' permutation matrix that exchanges type $\alpha$ with its antiparticle $\alpha^*$. The eigenvalues of the $T$ matrix are given by $\theta_\alpha e^{-2\pi i c/24}$, where $c$ is the central charge. Unfortunately, we do not have an adiabatic calculation of the central charge at present. If we know which eigenvalue corresponds to the identity object, which has twist $\theta=1$, then the problem is solved. In general we can still check the values of the {\em ratios} of the twists for distinct quasiparticles, and these clearly agree with the CFT (including the charge sector).


\subsubsection{Massless IR fixed point}

Now we turn to the case in which the IR fixed point is massless,
and thus scale invariant. We note that such a theory must in many
cases be conformally invariant as well, for example, in unitary
cases \cite{polch}. Then for a CFT we can again apply the
machinery of conformal blocks, etc. We begin here by considering
the case in which the perturbation is irrelevant (or marginally
irrelevant), so that the IR CFT is the same as the one with which
we began (with the charge part removed). It will be convenient
throughout to ignore the charge sector, which behaves exactly as
in the other cases.

The irrelevance of the perturbation implies that the overlap
matrix ${\cal Z}_{ab}$ of interest approaches at large separations
of the quasiholes the product of conformal blocks for the $\tau$
fields ({\em without} $\psi$ insertions):
\be%
{\cal Z}_{ab}\to \overline{{\cal F}_a(w_1,\ldots)}
{\cal F}_b(w_1,\ldots). \ee%
Thus it certainly does not approach a constant.

In this situation, we need more general expressions for the Berry
connection than those we used when we assumed that the states
$\Psi_a$ were orthonormal. Let us write%
\be%
{\cal Z}={\cal N}^{-2}\ee%
as matrices ($\cal N$ is Hermitian, and we choose signs so that it
is positive), so that $\sum_a|\Psi_a(w)\rangle{\cal N}_{ab}$ are
orthonormal. Then using the holomorphy of
$|\Psi_a(w)\rangle$, we find the matrix equations%
\bea%
A_w&=&-i\partial{\cal N}{\cal N}^{-1},\\
A_{\overline{w}}&=&i{\cal N}^{-1}\overline{\partial}{\cal N},\eea%
in which we suppress the index $l$ on $A$ and on
$\partial/\partial w_l$. These generalize expressions for the case
of a single block in Ref.\ \cite{stonebook}, in which there
appears to be a sign error.

On applying this to the limiting form of $\cal Z$ in the present
case, we see immediately that when there is more than one block,
there is a problem because, while ${\cal N}^{-1}={\cal Z}^{1/2}$
exists, its inverse ${\cal N}$ does not. Let us ignore this for a
moment by considering the two-quasihole case, for which there is
always only a single block, which has the form ${\cal
F}_1(w_1,w_2)=(w_1-w_2)^{-2h_\tau}$. Then a short calculation
shows that the Berry connection is non-zero, and in the holonomy
the path-ordered exponential exactly cancels the monodromy of the
function. (One can normalize the blocks, and then the cancellation
is exactly the same as occurs if one starts with a trivial example
and makes a gauge transformation to obtain both the connection and
the monodromy.)

Hence the adiabatic exchange of any quasiparticle type with itself
is trivial. If the trial wavefunctions represent a topological
phase of matter, then the quasiparticle properties are described
by a modular tensor category (when the underlying particles are
bosons; the fermion case should be similar). It is possible to
show that the only (unitary) MTC in which the exchange of two
identical quasiparticles is trivial for all quasiparticle types is
the trivial MTC containing only the identity or trivial
quasiparticle \cite{wangpc}. This shows that if the wavefunctions
in one of the present cases describe a topological phase, then it
is trivial. This is the opposite extreme from holonomy equal to
monodromy.

This however is not the whole story. As the IR fixed point CFT is
massless, one in general expects that there will be irrelevant
(and possibly also marginally irrelevant) operators that decay to
zero at large length scales. In the situation at present, in which
the perturbation of the UV theory is irrelevant, these can include
the same perturbation we began with. But there will be others
also, especially because our perturbation is not small. In
general, there is always an irrelevant operator $T\overline{T}$,
where $T$ and $\overline T$ are the components of the stress
tensor; this operator has dimension $4$. Thus there will in fact
generally be non-zero corrections of the form%
\be%
{\cal Z}_{ab}= \overline{{\cal F}_a(w_1,\ldots)}{\cal
F}_b(w_1,\ldots)
\left(1+O\left(\frac{1}{r^{2}}\right)\right), \ee%
as $r\to\infty$, where $r$ stands for $|w_k-w_l|$ for all pairs
$k$, $l$, and if there are other irrelevant operators more
relevant than $T\overline{T}$, then the correction factor will
decay as an even slower power. (The correction term will not
exhibit the factorized form of the first term.) When this is
included in a calculation of the Berry connection, the matrix
$\cal N$ may be well-defined for any number of conformal blocks,
and the resulting Berry connection will take the form $|A|\sim
O(1/r)+O(1/r^3)$. In particular, for two quasiholes, there will be
corrections to the trivial holonomy which decay as $1/r^2$.

More generally, even if the perturbation is not irrelevant, the IR
fixed point might be a non-trivial massless theory different from
the underlying CFT. If so, then the overlap matrices will possess
power-law corrections (even if the leading behavior is going to a
constant). If the leading behavior is a non-constant power-law
behavior, then the holonomy will disagree with the monodromy of
the blocks (which is always that of the underlying CFT with which
we started). If the massless theory is conformal, then there are
conformal blocks. Note that when the leading part of $\cal N$
exists, the integral of the leading part of the Berry connection
contributes just (minus) the logarithm of the monodromy of the
blocks of the IR CFT to the total holonomy in all cases. It is not
clear whether there will be any MTC that would contain the
resulting behavior (although the subtraction is reminiscent of a
coset construction \cite{dfms}). In all these cases there will be
power corrections to the leading results for the holonomy.

When the overlap matrix is viewed as that of trial states in a
quantum system in $2+1$ dimensions, the presence of power law
corrections to the holonomy signals that the behavior is not that
of a topological phase. Such a phase is supposed to be fully
gapped, and so the holonomy should possess only
exponentially-decaying corrections. The power-law corrections are
a clear break-down of the screening properties we were seeking. We
conclude then that the system is gapless. It may be in a gapless
phase, or it may be at a phase transition (a subspace of
codimension at least one) in the space of Hamiltonians for $2+1$
systems (that is, if there is a Hamiltonian for which the trial
wavefunctions are exact eigenstates). In such cases, adiabatic
transport presumably does not mean much anyway. Possibly, the
trial functions simply do not represent any respectable phase of
matter in $2+1$ dimensions, even one at a critical point.

\subsubsection{Summary}

Thus we have found that for the two main classes of possible IR
fixed points, massive and massless, there is corresponding
behavior of the holonomy: either (i) that consistent with a
topological phase of matter, and agreeing with the monodromy of
the underlying conformal field theory, or else (ii) none of these
statements hold and the $2+1$ system is gapless, respectively. In
general, although this is a little crude in case (ii), the
holonomy in the adiabatic transport can be thought of as the
monodromy of the blocks, which comes from the underlying (or UV)
CFT, divided by the monodromy of the corresponding blocks of the
IR CFT (the trivial CFT in the case of a massive phase). This is
also reminiscent of the normalization factors for the paired
states in the Sec.\ \ref{genpair}, which is a ratio referring to
the underlying CFT and the large distance behavior. Notice that we
do not find, for example, behavior like that of a strong-pairing
phase which is topological but trivial (apart from the charge
part). Trial wavefunctions that are conformal blocks are not known
for all the possible QH phases, so this may indicate that none can
exist for the strong-pairing type of phases. For the latter, the
closest one can get may be to use the trivial CFT (i.e.\ the
charge part only) in the UV, treating pairs or larger clusters of
particles as the basic particles represented by $\psi_e$. This
could be generalized to less trivial CFTs, but still applied to
clusters of particles. Finally, the third possibility, overlap
matrices that grow exponentially with distance, may mean a more
gross instability in the $2+1$ particle system, so that the trial
wavefunctions do not represent a true phase of matter at all, in
contrast to the preceding two possibilities which are well defined
phases, even when gapless.


\subsection{Examples}

\subsubsection{Abelian states}

We begin with trial wavefunctions for some Abelian QH phases. We
focus on the simplest ones (other than the Laughlin states), which
are the Halperin $mm'n$ states for particles that come in two
species or ``pseudospins'', which we will denote $\uparrow$ and
$\downarrow$. The states can be applied to bilayer QH systems,
with the pseudospin being the layer index. The trial states exist
for non-negative integers $m$, $m'$, $n$, but we will concentrate
on the cases $m'=m$ for which there is a symmetry under the
operation $\uparrow\to\downarrow$, $\downarrow\to\uparrow$ for all
particles. The trial wavefunctions
for the ground state (no quasiholes) are%
\bea%
\lefteqn{\Psi(z_1^\uparrow,\ldots;z_1^\downarrow,\ldots)=}\non\\
&&\prod_{i<j} (z_i^\uparrow-z_j^\uparrow)^m\cdot\prod_{i<j}
(z_i^\downarrow-z_j^\downarrow)^m\non\\
&&\cdot\prod_{i,j}
(z_i^\uparrow-z_j^\downarrow)^ne^{-\frac{1}{4}\sum_{i;\sigma=
\uparrow,\downarrow}
|z_i^\sigma|^2}\eea%
The filling factor is $2/(m+n)$, and the $331$ case, which means
the cases $m-n=2$, has been discussed above in Sec.\ \ref{statcalc}. There are
two-body local Hamiltonians for which the $mmn$ states are exact
zero-energy eigenstates (see e.g.\ Ref.\ \cite{rr1}). On removing
the charge sector, one has the fields
$\psi_{\uparrow,\downarrow}=e^{\pm i\sqrt{m-n}\varphi'/2}$, where
$\varphi'$ is a chiral scalar field, distinct from, but obeying
the same properties as, that for the charge sector. The conformal
weights for these are both $(m-n)/4$. Thus the perturbation of the
CFT is relevant when $m-n<4$, and irrelevant when $m-n>4$. We have
assumed here that $m-n>0$. If $m-n<0$, the plasma is unstable
\cite{dgrg}. At the same time, the scaling dimension of the fields
in the CFT becomes negative, and this corresponds to a non-unitary
CFT. Also, the case $m=n$ represents a ferromagnetic state. We
assume $m-n>0$ from here on.

When the charge part is removed, this problem is the two-component
Coulomb gas in two dimensions, which was considered by Kosterlitz
and Thouless (KT) \cite{kt}. Then if the perturbation of the CFT was weak,
the KT transition occurs at $m-n=4$. We would then expect flow to
the original CFT in the IR if $m-n>4$, resulting in no screening,
and gapless behavior, but flow to a screening phase if $m-n<4$.
However, for larger values of the perturbation (in this context,
called fugacity of the Coulomb charges) the system can flow to the
screening phase even for $m-n>4$, according to the RG flow diagram
of Kosterlitz, illustrating a point made above.

On the other hand, if we consider the cases $m>0$, $n=0$,
including the charge part, then we see that the two components of
pseudospin decouple, and we have a product of Laughlin states.
Hence in this case, screening does occur, even when $m-n=m$ is
large (for $m<70$ anyway). So in these cases it appears that the
perturbation is effectively so large that the perturbed CFT is in
the screening phase. For $n>0$ the plasma is again stable
\cite{dgrg}, and the widely-held belief that this plasma is in a
screening phase for both charge and pseudospin is consistent, at
least. Consequently, these examples form a cautionary note that
the irrelevance of the perturbation does not necessarily mean a
flow to a massless phase.

We note that {\em if} the flow had gone back to the CFT, it would
have been easy (using standard Coulomb gas arguments) to see that
screening fails in the $\varphi'$ sector, and that the pseudospin
associated with quasiholes is not localized near the locations
$w$, but spread over the region in between, consistent with a
gapless (possibly critical) phase.


\subsubsection{Read-Rezayi series}

We have already dealt at length with the Laughlin and MR states.
Now we address the RR series of states \cite{rr2}, labeled by an
integer $k>0$ (the other label $M$ corresponds to additional
Laughlin factors, and does not enter these considerations). For
all $k$, the CFT other than the charge part is the ${\bf Z}_k$
parafermion theory. The field $\psi$ involved in representing the
particles in the conformal blocks is the first parafermion
current, $\psi=\psi_1$, of conformal weight $h_\psi=1-1/k$. These
underlying CFTs are unitary. Moreover, the perturbation by
$\overline{\psi}_1\psi_1$ is relevant for all $k$. However, we
note that for $k>2$, the field $\psi$ is not equal to its own dual
(or ``antiparticle'') $\psi_1^\dagger$ (the adjoint taken in the
two-dimensional quantum field theory), which instead is
$\psi_1^\dagger=\psi_{-1}=\psi_{k-1}$. If the perturbation were
$\overline{\psi}_1\psi_1+\overline{\psi}_1^\dagger\psi_1^\dagger$,
then unitarity of the theory would be maintained by the
perturbation (as it is in the Abelian examples just discussed).
But for the perturbation we have, the perturbed theory is not
unitary, except for the cases $k=1$ (which is trivial), and $k=2$
(the MR state).

For the unitary perturbation $\overline{\psi}_1\psi_1+\overline{\psi}_1^\dagger\psi_1^\dagger$ of the parafermion theory, it is
generally believed that the system flows to a massive phase, at
least when the perturbation is weak and positive \cite{fz}. If this were the case here, we would immediately conclude that the adiabatic statistics of
quasiholes is the same as the monodromy of the conformal blocks.
But for the non-unitary perturbation by $\overline{\psi}_1\psi_1$  of the unitary underlying CFT, we may have some concern (though the coefficient is positive in the convention of Ref.\ \cite{fz})). It is possible for the flow still to arrive at the same IR massive fixed point, with non-unitarity
only manifesting itself in the behavior of the exponentially-small
corrections, and if such is the case then all is well, and we have
trial functions for a topological phase. In support of this we can
only point out that the omission of
$\overline{\psi}_1^\dagger\psi_1^\dagger$ from the perturbation
may not make that much difference, as in any case $k$ insertions
of $\psi_1$ can fuse to the identity in an operator product, so
that the other terms are not needed for ``charge neutrality'',
unlike the example of the $mmn$ states above. In addition, we
point out that for the charge sector (and thus even for the
Laughlin states), the perturbation is not unitary, because the
positive charges are the point particles, and the negative ones
form a fixed uniform background. Nonetheless, for $\nu^{-1}<70$,
the flow is to a screening phase, which behaves the same way as
for the screening phase in the two-component plasma which is a
unitary perturbed CFT, even though the conditions for {\em
reaching}\/ the screening phase are different in the two cases. We
conclude then that the non-unitarity of the perturbation is not
necessarily a problem, and the system may well flow to the same
massive 2D theory as in the case of the unitary perturbation,
which would lead to the expected non-Abelian statistics for all of
the RR series.


\subsubsection{Blok-Wen series}

In the Blok-Wen series of states \cite{bw}, the particles (of spin
$k/2$) are represented in the conformal blocks by the operator
$\psi=\phi_{k/2}$, where $\phi_s$ is the primary field of spin
$s=0$, $1/2$, \ldots, $k/2$ in SU(2) level $k$ current algebra (or
WZW theory). (We have suppressed labels for the $S_z$ component on
these fields.) The perturbation in the CFT is the spin-singlet
part of $\overline{\phi}_{k/2}\phi_{k/2}$. Unlike the RR series
for $k>2$, in this case one has a unitary perturbation of a
unitary theory for all $k\geq 0$ (except for the charge part, as
usual). However, we also notice that the conformal weight of
$\psi$ is $h_{\phi_s}=s(s+1)/(k+2)$, which is $k/4$ for $s=k/2$.
Hence the perturbation is relevant for $k<4$, marginal for $k=4$,
and irrelevant for $k>4$. As the perturbation is of order one, we
cannot be certain of the fate of each case. But suppose that the
IR fixed point is the same as if the perturbation were weak, as
discussed above. Then the cases $k<4$ lead to good topological
phases. But for $k> 4$, the return to the underlying CFT in the IR
would imply that the models are not topological after all. We note
that Blok and Wen proved \cite{bw} that the adiabatic statistics
is given by the monodromy of the conformal blocks, on the
assumption that there was screening in the spin sector. This also
entails that there be a well-defined spin of $s$ within a ``spin
screening length'' of a quasihole of the type labeled with spin
$s$ (when the quasihole is far from any others). The present
scenario would imply that the screening assumption breaks down for
$k>4$, and the statistics calculation then fails. In effect, the
trial states would not possess spin $s$ localized near each
quasihole, instead this spin would be spread over the region
containing several quasiholes. This would mean that the states are
not topological in the spin sector, but behave as a gapless phase,
or possibly a critical point. For the case $k=4$, further
calculations are needed to establish whether the perturbation is
marginally relevant, marginally irrelevant, or exactly marginal
(when weak). The latter two possibilities would both lead to
behavior that is not that of a topological phase.

The notion that screening in the spin sector breaks down for $k>4$
is intriguing, but by no means certain. The Blok-Wen states are
something like an SU($2$)-invariant version of the $mmn$ states
above, with $k$ in place of $m-n$, and corresponding conformal
weights and filling factor (note that $k=1$ coincides with the
$n+1,n+1,n$ state). The relation can be made even closer by
writing the SU($2$) level $k$ theory in terms of a scalar field
$\varphi'$ and the ${\bf Z}_k$ parafermion CFT. In view of the
$mmn$ example which has screening in pseudospin except for very
large $m+n$, it may well be that the Blok-Wen states are in the
screening phase for spin, and possess non-Abelian adiabatic
statistics.

There are other series of trial wavefunctions that have been
proposed for particles with spin, and for which there is a special
Hamiltonian for which the trial states are the zero-energy
eigenstates \cite{rlsr,nass}. These behave similarly to either the
RR or Blok-Wen series, and we will not discuss them further.


\subsubsection{Non-unitary theories and  minimal models}
\label{nonun}

Now we turn to further examples, but start with general remarks
about non-unitary underlying CFTs. In such cases, the perturbation
by $\psi$ may still be relevant, irrelevant or marginal, when
weak. We first consider the possibility of a flow to a massive
phase. This would lead to adiabatic statistics given by the
monodromy of the conformal blocks of the non-unitary CFT, as we
have seen. However, there are difficulties with the interpretation
of this scenario. We have emphasized previously \cite{read08} that
for a gapped bulk phase, the boundary is a massless field theory
that is both unitary and conformal. If one has statistics in the
bulk that is given by the monodromy of the blocks, then the same
CFT would be expected at the edge \cite{mr}. For a system with
trial wavefunctions given by conformal blocks of a non-unitary
theory, it is not clear what the candidate for this edge theory
can be. However, there is certainly more than one CFT that
produces the same fusion rules, braiding in monodromy, etc, or in
other words the same MTC.

Here we wish to make a point that is more direct, and stronger. If
the IR fixed point is massive, and so the MTC for the topological
phase is that of a non-unitary UV CFT, then under some conditions this
is not compatible with {\em any}\/ unitary  MTC (again, the MTC formalism applies to QH systems of bosons, and must be modified for fermions, so the arguments here are for bosons, but the fermion case will be similar; from the wavefunction point of view in QH systems, the two are very similar, differing only in the number of Laughlin-Jastrow factors). First, we should explain
that there is a notion of a unitary MTC, which is precisely the
correct concept to describe topological phases in $2+1$ dimensions
that arise from quantum mechanical systems with a
positive-definite inner product on the Hilbert space, and a
Hermitian Hamiltonian (with respect to this inner product)
\cite{turaev}. Unitarity of the MTC has certain consequences. For
example, in any MTC each quasiparticle type has an associated
``quantum dimension''. In a unitary MTC, all these quantum
dimensions must be positive. Next, for topological phases that
originate from trial wavefunctions that are conformal blocks in
some underlying CFT, the quantum dimensions can be calculated from
the CFT. [Strictly speaking, we have only so far argued that when
the IR fixed point is a massive phase, we obtain the same fusion
rules, braiding, and (probably) twist as in the underlying CFT. There are
some remaining structures used in defining an MTC, which are
needed to calculate the quantum dimensions; the derivation of
these for trial functions given by conformal blocks is discussed
in Appendix \ref{mtc}.] In non-unitary CFTs, there are usually
some negative quantum dimensions. [It can be shown using modular
transformations \cite{dfms} that if any conformal weights in a
rational CFT are strictly negative ($h<0$), then some quantum
dimensions are also negative. As a partial converse, if there are
negative quantum dimensions, then there are fields other than the
identity with $h\leq0$. It does not seem to be known whether a
non-unitary RCFT (MTC) must contain some negative conformal
weights (resp., quantum dimensions).] Hence, when this occurs,
{\em no unitary MTC can describe the resulting behavior} (see App.\ \ref{mtc} for further discussion). We conclude that such trial wavefunctions cannot flow to a massive phase. They must flow to some massless field theory, either back
to the original CFT, or to some other non-trivial massless theory,
or possibly exhibit exponentially-growing overlaps, which may
indicate a more serious instability. We should note that for
non-unitary theories a massless (scale-invariant) theory does not
have to be conformal \cite{polch} (for an example, see Ref.\
\cite{rivcar}).

As an interesting example, we consider the so-called Gaffnian
trial wavefunctions that have been proposed recently \cite{srcb}.
These are associated with one of the BPZ minimal CFTs, which in
general are denoted $M(p,p')$, parametrized by a pair of coprime
positive integers $p$, $p'$ with $p>p'$ \cite{dfms}. The Gaffnian
functions are conformal blocks from the $M(5,3)$ minimal model,
which is non-unitary. The perturbing field $\psi$ has conformal
weight $3/4$, so the perturbation is relevant. The fusion rule for
$\psi$ is $\psi\times\psi=1$, so that these states may be
considered to be paired states. Due to the relevance of the
perturbation (when weak), we expect that the RG flow does not lead
to the original CFT, and one might expect that the IR fixed point
is massive (the trivial CFT), which would give the same MTC as
that of the underlying CFT. But the quantum dimensions $d_\alpha$
of the primary fields in a rational CFT are given by the first row
(or column) of the unnormalized version, denoted $\widetilde{S}$,
of the modular $S$ matrix; $S$ was discussed above for the
Majorana theory [which is $M(4,3)$], though in a different basis
than here. One has $\widetilde{S}=DS$, where $D$ is the global
dimension determined by $D^2=\sum_\alpha d_\alpha^2$, where
$\alpha$ labels the distinct quasiparticle types, and we take the
positive square root for $D$. For the minimal models, the primary
fields $\phi_{r,s}$ are labeled by ordered pairs of integers
$(r,s)$, with $1\leq r\leq p'-1$, $1\leq s \leq p-1$ (the
so-called minimal block in the Kac table). There is an equivalence
$r\to p'-r$, $s\to p-s$, and each inequivalent primary occurs just
once in the theory. The conformal weights and central
charge are given by \cite{bpz}%
\bea%
c&=&1-6\frac{(p-p')^2}{pp'},\\
h_{r,s}&=&\frac{(pr-p's)^2-(p-p')^2}{4pp'},\eea%
and the modular $S$ matrix is %
\be%
S_{r,s;\rho,\sigma}=2\sqrt{\frac{2}{pp'}}(-1)^{1+s\rho+r\sigma}
\sin(\pi r\rho p/p')\sin(\pi s\sigma p'/p).\label{modS}\ee %
For the first column, $\rho=\sigma=1$, and the quantum dimensions
are $d_{r,s}=DS_{r,s;1,1}$. Now for the case of $M(5,3)$, we may
put $r=1$ also, and for the Gaffnian trial states
$\psi=\phi_{1,4}$. One finds that both $d_{1,3}$ and $d_{1,4}$ are
negative, while the other two are positive ($d_{1,1}=1$ for the
identity always). This behavior is then impossible in a unitary
MTC, and thus in a unitary topological phase. We conclude that, if
the RG flow driven by the perturbation does not lead back to the
original $M(5,3)$ CFT, then it must go to some other non-trivial
massless theory, and either way the system is in a gapless phase
as a $2+1$ system (possibly, at a critical point, as discussed in
Ref.\ \cite{srcb})---or a worse instability takes place. We note
that negative conformal weights, and hence negative quantum
dimensions, occur in all the minimal models other than the unitary
ones, for which $p=p'+1$.

While on the subject of minimal models, we may consider other
examples along similar lines as the Gaffnian. For $p'>2$, the
field at position $(p'-1,1)$ [or equivalently, $(1,p-1)$] in the
Kac table always has Abelian statistics: the fusion rule for it is
$\phi_{p'-1,1}\times\phi_{p'-1,1}=1$, so that all these examples
are paired states in the same general sense as the Gaffnian. [For
$p'=2$, the field at $(1,1)$ or $(1,p-1)$ is the identity, so no
interesting functions arise; the simplest non-trivial example is
the Yang-Lee theory, $M(5,2)$ \cite{dfms}.] Using this field as
$\psi$, one obtains a set of trial wavefunctions from conformal
blocks for {\em any} minimal model. For many non-unitary cases,
the perturbation is relevant, but a similar argument applies to all these non-unitary theories as for the Gaffnian (see App.\ \ref{mtc}). However, for
the unitary minimal models $M(p'+1,p')$, the scaling dimension is
$h_{p'-1,1}=(p'^2-3p'+2)/4$, which is relevant ($<1$) only in the
case $p'=3$, which is the MR example once again. The filling
factors of these states are determined by
$\nu^{-1}=(p'^2-3p'+2)/2$ plus integers. The next simplest example
in the unitary sequence is $p'=4$, in which the minimal model
describes the tricritical Ising model, which has many interesting
features including superconformal symmetry. The perturbation is by
a subleading ``thermal'' operator $\varepsilon''$ \cite{dfms}.
Again, in general, although the perturbation is irrelevant when
weak for $p'>3$, it may be that these cases flow to a massive
phase, representing a gapped topological phase in $2+1$
dimensions. In general, we are not aware of Hamiltonians for which
these trial wavefunctions are zero-energy eigenstates, so it is
not clear whether this approach leads to the construction of the
corresponding topological phases. However, S. Simon has informed us that there appears to be a Hamiltonian that produces the ``tricritical Ising'' state \cite{simonpriv}, based in part on techniques from Ref.\ \cite{src}.

Recently, some large families of functions have been proposed as possible trial wavefunctions \cite{bernhald,wenwang}. It is not generally clear if these come from any CFT, and if they do not it is not clear whether they truly represent a topological phase. However, those in Ref.\ \cite{bernhald} include the Gaffnian and possibly some based on other non-unitary minimal models, and so it is likely that many of them can be ruled out (as far as gapped topological phases are concerned) by arguments similar to those given here and in App.\ \ref{mtc}.


\subsubsection{Irrational theories}

In the preceding section, we considered non-unitary CFTs, but only
in the context of {\em rational} CFTs, so that they still
correspond to MTCs. We recall that a rational CFT contains a
finite set of irreducible representations of a chiral algebra, and
all the representations are fully reducible to direct sums of
irreducibles \cite{ms}. In this section, we will consider some
known examples in which the CFT may or may not be unitary, but is
in any case not rational.

The examples we consider here have been studied earlier. All of
them are paired states. In each case there is a ``special
Hamiltonian'' for which the trial states given by conformal blocks
are the exact zero-energy eigenstates, and all were found to be at
a critical point in the $2+1$ point of view. They are (i) the
Haldane-Rezayi state, (ii) the permanent state, and (iii) the
Haffnian state. First we will briefly discuss the CFT whose
conformal blocks are the trial wavefunctions in each case.

The Haldane-Rezayi (HR) state \cite{hr2} is a spin-singlet trial wavefunction
for spin-$1/2$ particles. The CFT (other than the charge part) is
the symplectic fermion theory with $c=-2$, which has the required
sl$_2$ symmetry \cite{mr}. This central charge corresponds to the
values $p=2$, $p'=1$ in the BPZ analysis. However, for the
$M(2,1)$ minimal model, the minimal block of the Kac table is
empty, as it is for all cases in which $p'=1$. The fields used in
the HR state \cite{milr} are the symplectic fermion of conformal
weight $1$), and the spin field of conformal weight $-1/8$, which
do lie in the Kac table, but are outside the (empty) minimal
block. Consequently, in this non-unitary theory, the
representations produced in operator products are not fully
reducible. It was argued in Ref.\ \cite{rg} that the state
corresponds to the weak- to strong-pairing transition for $d-id$
spin-singlet paired states of fermions.

The permanent state \cite{mr} is also a
spin-singlet state. The CFT is the $\beta$-$\gamma$ system \cite{fms}.
On the torus, or in the presence of sufficiently many quasiholes, the
dimension of the space of zero-energy states grows without bound
as the system size increases (this is because unpaired bosons can
occupy zero modes). The CFT is non-unitary, and also irrational
because of the ``picture-changing operators''. This system has
been argued to correspond to the transition point between the
spin-polarized Laughlin state, and a spin-density wave state, with
the permanent ground state itself corresponding in fact to an
anti-skyrmion spin texture \cite{rr1,grr}.

Finally, the Haffnian state is a paired state for spinless (or
spin polarized) particles in which spinless composite bosons form
pairs of angular momentum $-2$ \cite{wenwu,grr}. In this case, the CFT is unitary, and consists once again of a chiral scalar $\varphi'$. The
particles contain $\psi=\partial\varphi'$ which is a U$(1)$
current of conformal weight $1$, and is a good conformal field,
unlike $\varphi'$ itself. Using Wick's theorem, the correlator of
$\psi$ in the case of the trial ground state is the Haffnian,%
\be%
\langle
\partial\varphi'(z_1)\cdots\partial\varphi'(z_N)\rangle=
{\rm Hf}\left(\frac{1}{(z_i-z_j)^2}\right),\ee%
where the Haffnian of a symmetric matrix $M$ is defined by%
\be%
{\rm Hf}\, M_{ij}={\cal S}\left( M_{12}M_{34}\cdots
M_{N-1,N}\right),\ee%
in which the symmetrizer ${\cal S}$ sums over all permutations
that produce distinct pairings $(i,j)$, similarly to the Pfaffian
(from which the name is derived), but without the sign of the
permutation. The quasihole functions contain the twist field
$\tau$, around which the $\psi$ field has a square root
behavior of ope:%
\be%
\partial\varphi'(z)\tau(0)\sim z^{-1/2}\tau(0)+\ldots.\ee%
This field $\tau$ features in orbifold theories based on
$\varphi'$, and has conformal weight $1/8$ \cite{dixon}. (There
are many other possible twist fields with other fractional-power
ope's \cite{dixon}, but only this one can be used to make
single-valued functions of the particle coordinates when the charge sector
is included.) Like the permanent case, this state has
highly-degenerate torus ground states and quasihole states, and is
believed to be at a transition point, to one side of which is the
Laughlin state \cite{grr}.

We will attempt to explain what is happening in these examples,
focusing primarily on the Haffnian, as it uses a more familiar
unitary CFT. To obtain unambiguously-defined systems, we will
invoke here the special Hamiltonians mentioned above. Such a
Hamiltonian is an operator on the particles. The detailed form for
each case will not be important here. The point is that in seeking
zero-energy eigenstates of such a Hamiltonian, we are forced to
fix the ope's of the fields $\psi_e$ that correspond to the
particles, in a form that of course depends on the Hamiltonian.

For the Haffnian state, the special Hamiltonian implies that the
CFT should have the ope's of the U$(1)$ current $\psi=\partial\varphi'$:%
\be%
\psi(z)\psi(0)\sim 1/z^2+\ldots.\ee%
If the scalar field $\varphi'$ were compactified with radius $R$,
then the scalar theory would contain such fields as $\exp
i\alpha\varphi'$ for a discrete set of real values of $\alpha$
(integer multiples of one value, related to $R$ \cite{dfms}). For
the orbifold of this rational theory, there are similar fields,
and the ope's among $\psi$ and $\tau$ are {\em independent} of the
radius $R$. Because the Hamiltonian can only determine the ope for
$\psi$, in counting states we must consider all possible values of
$R$ and $\alpha$. As $R$ goes to infinity, the exponentials of
$\varphi'$ can be expanded in powers of $\varphi'$, and insertions
of these appear to produce the conformal blocks that we described
as bosons occupying zero modes (on the torus or in the presence of
quasiholes). In the limit, the CFT is clearly not rational.
However, one also expects that only a rational theory can
correspond to a topological phase; in particular, the number of
quasiparticle types should be finite, otherwise one suspects that
the system will be gapless.

Similar arguments apply to the symplectic fermion and
$\beta$-$\gamma$ systems. For symplectic fermions, being fermions,
the use of different radii does not directly apply. But there is
still an infinite set of ``rational'' versions, in the sense of
possessing an extended chiral algebra that has only a finite
number of representations (though they still fail to be
semi-simple) \cite{kausch}.

It is interesting also to pursue some of the other arguments above
in the present cases. We notice that for the HR and Haffnian
cases, the perturbing field $\psi$ is weight one, and so is
marginal when weak (as a perturbation of the CFT without the
charge part). The same scaling is found in the pairing function
$g$ of the p+ip paired states at the critical point \cite{rg}
(though then $g$ is not meromorphic, and does not correspond to a
conformal field theory). If we consider this perturbation as weak,
and write
$\varphi'(z,\overline{z})=\varphi'(z)+\overline{\varphi}'
(\overline{z})$
in the non-chiral theory, then in the action we have
$\lambda\overline\partial\varphi'\partial\varphi'$. This is the
same as the unperturbed action, and can be absorbed by rescaling
$\varphi'$. Thus there is no effect on the theory of the
perturbation (though there would be if the field were
compactified). The perturbation is not only marginal, it is
redundant. The same argument goes through for the HR theory also.
We should note however that since the perturbation is actually of
order one, there may be concerns about the correctness of the
argument. At any rate, other irrelevant operators will be
generated, whereas the rescaling argument would suggest that there
are none. We cannot enter into this further here. If we accept the
argument, then the IR fixed point is the same CFT with which we
started the RG flow, and we argued above that this must correspond
to a gapless phase. This is consistent with the quite different
arguments that these states correspond to critical points, given
previously.

As a related side remark, we return to examples such as the
Yang-Lee CFT mentioned previously. The ground state wavefunctions
(which contain $\psi=1$ for the CFT) can be produced by the usual
pseudopotential Hamiltonian that produces the Laughlin states,
which of course are the same functions. One is free to write down
the functions derived from the Yang-Lee theory, including the only
non-trivial field $\tau=\phi_{1,2}$. But as functions of the
particle coordinates, the trial functions are independent of the
positions $w$ of these insertions, which consequently can carry
zero charge in the charge part. Thus these trial functions are not
linearly independent of the usual Laughlin ground and quasihole
states, and are in fact the Laughlin functions times functions of
the $\tau$ locations alone (thus the theory is a direct product).
This is why the Yang-Lee theory leads to nothing new. We notice
that the field $\psi_e$ (with its dual) generates the chiral
algebra of the Laughlin state \cite{mr}, which never includes the
stress tensor of the Yang-Lee theory. In general, we expect that
the trial functions based on conformal blocks of a CFT are
linearly independent as functions of the particle coordinates only
if the particle field generates the full chiral algebra of the CFT
(including the charge sector).

Put another way, a special Hamiltonian determines the field
$\psi_e$ which is part of some CFT. The CFT (including the charge
part) that one should use to analyze the problem is the
``minimal'' one, that is the one with the smallest possible chiral
algebra, and that means that $\psi_e$ and its dual should generate
the chiral algebra. For the Haffnian and other examples above, the
resulting chiral algebra is however too small to define a rational
theory (but extensions exist that correspond to a rational theory,
at least in the sense of a finite number of types).


\section{Conclusion}

The main message to emerge from this work is that when conformal blocks are used as trial wavefunctions, whether in QH or other systems, the properties of the topological phase (if any) that they represent are directly determined by, and actually equal to, those of the underlying CFT. (This has frequently been assumed, rather than demonstrated, in the literature.) That is, the holonomy (such as adiabatic statistics) is equal to the monodromy (analytic continuation of blocks) if and only if the 2D perturbed CFT is in the massive phase. If not, then there are signs that the system is gapless as a $2+1$ phase of matter (though we did not address Hamiltonians for our states).

When the trial wavefunctions represent a topological phase, our results almost completely constrain it. The loose end is that the twist (the effect of rotating a quasiparticle about its center) should also be calculated by adiabatic transport, and this has not been done. However, it should be possible by adiabatically varying the metric of the 2D surface. A first step in this direction is the determination of the Hall viscosity, which determines an Abelian contribution to this transport that is present in all cases. The fact that this Hall viscosity is itself related to the density of conformal weight in the trial ground states strongly suggests that the conformal weight of the quasihole field will also emerge in the twist, as expected.

If the latter calculation is completed, it will remove the final possible loophole in our argument that non-unitary RCFTs that contain negative quantum dimensions (or conformal weights) do not give rise to a topological phase (or unitary modular tensor category) when their blocks are used as trial wavefunctions; instead they should apparently be gapless, or suffer a worse instability.

Another loose end of this work is that the BCS trial wavefunctions for states containing vortices have not been shown to reduce to the conformal block form at large length scales. Instead, the blocks were used directly (see Sec.\ \ref{statcalc}). It would be desirable to show this, so as to broaden the domain of explicit calculations to the wider class of BCS functions.

Finally, we note that the rather simple form of all these results seems to be related to the fact that the underlying particles are Abelian, as is the perturbing field in the CFT point of view. It would be of interest to consider states formed of a non-zero density of non-Abelian particles (which could be useful in hierarchical constructions of QH or anyon superfluid states), but this is clearly much more difficult to analyze.

{\it Note added:} Recent work has tied up the loose end mentioned above, by deriving the twist for quasiholes in the trial states from their spin, which can be obtained by adiabatic transport in the presence of curved space, such as on a sphere \cite{read08c}. This completes the argument that conformal blocks from {\em any} non-unitary CFT containing negative quantum dimensions cannot correspond to a topological phase.

\acknowledgments

We are grateful for discussions with M. Freedman, A. Ludwig, S. Simon, I. Tokatly, G. Vignale, and especially with N. Cooper, P. Fendley, and Z. Wang. This work was supported by NSF grant no.\ DMR-0706195.

\begin{appendix}

\section{Remaining data of an MTC, and non-unitary cases}
\label{mtc}

In this Appendix we discuss the remaining structures needed for
the definition of an MTC, other than the fusion rules, braiding,
and twist, which have been discussed in the main text. As we do
not intend to give a full exposition of MTCs, we cannot be
entirely self-contained, and must make use of the references for
full technical details. Accordingly, this Appendix is intended for
readers more expert about such matters. However, its contents are
mostly general definitions, which many readers might be willing to
take on trust. We do address the Gaffnian example again at the
end.

The main goal for the definitions is to be able to calculate the
quantum dimension associated with each quasiparticle
type, from trial wavefunctions given by conformal blocks of some
CFT. This requires that the full structure of a ribbon category
\cite{turaev,bk,wang} be available. (A MTC is a ribbon category in
which the $\widetilde{S}$ matrix, which can be defined in a ribbon
category, is invertible.)

Informally, the quantum dimension for a type of quasiparticle is defined as
the amplitude for a process in which the quasiparticle, of type
$\alpha$ say, and its antiparticle are created from the vacuum,
separated, and then annihilated again. The motion of
quasiparticles can be assumed to be adiabatic. The definition is
not vacuous because the creation and annihilation processes used
have to be normalized to satisfy other topological criteria. We
note immediately that in a unitary (i.e.\ quantum-mechanical)
theory (in $2+1$ dimensions), the creation and annihilation
processes are adjoints to each other, and the quantum dimension is
then the modulus-square of the creation amplitude. Hence it must
be positive in the ribbon category describing a quantum-mechanical
topological phase (or unitary TQFT). The latter corresponds to a
unitary ribbon (and modular tensor) category in the sense of
Turaev \cite{turaev}.

In the language of tensor categories, the distinct quasiparticle
types $\alpha$ correspond to isomorphism classes of simple objects
$V_\alpha$. The trivial or identity quasiparticle type $0$
corresponds to $V_0={\bf 1}$. More general objects can be formed
by direct sums and by tensor products, written $V\oplus W$ and
$V\otimes W$ respectively (the notation is based on that for
vector spaces, but for us these objects do not have to be thought
of as such). The fusion rules correspond to the decomposition of a
tensor product
of simples as (isomorphic to) a direct sum:%
\be%
V_\alpha\otimes V_\beta\cong\bigoplus_\gamma N_{\alpha\beta}^\gamma V_\gamma,\ee%
where we write $NV$ for $V\oplus V\oplus\cdots$ with $N$ terms $V$
in the sum. There are isomorphisms ${\bf 1}\otimes V\cong V$,
$V\otimes {\bf 1}\cong V$, for all $V$. Although the objects may
not be vector spaces, the morphisms (or maps) between objects, say
from $V$ to $W$, form a complex vector space ${\rm Hom}(V,W)$.

The tensor product is in general not strictly associative, but
there are ``natural'' isomorphisms $F$ between different orders of
brackets:%
\be%
F_{U,V,W}:(U\otimes V)\otimes W\to
U\otimes (V\otimes W).\ee%
These may be built from the $F$ ``matrices'' for products of
simple objects (replacing $U$, $V$, $W$, by $V_\alpha$, $V_\beta$,
$V_\gamma$), $F_{\alpha\beta\gamma}$. The tensor product on either
side decomposes as the same sum of simple terms, and after
choosing bases, $F_{\alpha\beta\gamma}$ becomes a matrix in this
space \cite{wang}, which has dimension $\sum_{\epsilon,\delta}
N_{\alpha\beta}^\epsilon
N_{\epsilon\gamma}^\delta=\sum_{\epsilon',\delta}
N_{\beta\gamma}^{\epsilon'} N_{\alpha\epsilon'}^\delta$. As there
are no morphisms between distinct $V_\delta$s, the $F$ matrices
break into block diagonal matrices $F_{\alpha\beta\gamma}^\delta$
of size $\sum_{\epsilon} N_{\alpha\beta}^\epsilon
N_{\epsilon\gamma}^\delta=\sum_{\epsilon'}
N_{\beta\gamma}^{\epsilon'} N_{\alpha\epsilon'}^\delta$:
$F_{\alpha\beta\gamma}=\sum_\delta F_{\alpha\beta\gamma}^\delta$.
As a tensor product is isomorphic to a direct sum, say
$V_\alpha\otimes V_\beta\cong \oplus_\epsilon
N_{\alpha\beta}^\epsilon V_\epsilon$, the
$F_{\alpha\beta\gamma}^\delta$ have blocks that can be labeled
$F_{\alpha\beta\gamma,\epsilon\epsilon'}^\delta$. The
decomposition to blocks
$F_{\alpha\beta\gamma,\epsilon\epsilon'}^\delta$ is well-defined
up to isomorphisms, and these maps become actual matrices (or
simply numbers if $N_{\alpha\beta}^\epsilon
N_{\epsilon\gamma}^\delta$ and $N_{\beta\gamma}^{\epsilon'}
N_{\alpha\epsilon'}^\delta$ equal $1$) only after fully specifying
a basis. The $F$ matrices must satisfy some consistency conditions
(pentagon equations) \cite{ms,turaev,bk}.

Braiding is strictly defined only for two objects not separated by
any parentheses: $c_{V,W}:V\otimes W\to W\otimes V$ for any $V$,
$W$. The braiding isomorphisms $c$ may be built up from simple
components, and (choosing bases) give rise to matrices
$c_{\alpha\beta}^\gamma$ on the sum of terms isomorphic to
$V_\gamma$ in $V_\alpha\otimes V_\beta$. In the case of a RCFT,
the eigenvalues of these can be found from the ope of
$\phi_\alpha$ with $\phi_\beta$. The braiding and $F$ matrices
must satisfy the hexagon equations. For a ribbon category arising
from a CFT, the pentagon and hexagon equations are satisfied
automatically.

The formula for the number of simple terms, which when the ribbon
category comes from a RCFT is the dimension of the space of
conformal blocks for correlators in the plane of corresponding
fields, illustrates how in an iterated tensor product of simples,
of any length, each lexicographically valid way of inserting
brackets corresponds to a labeling of states (the labeling is only
a partial labeling if some $N_{\alpha\beta}^\gamma >1$). For each
such choice there is a basis of the space of conformal blocks. The
$F$ matrices then describe the change between two such bases. {}From
the physical point of view, these basis changes are ``passive''
transformations, whereas the braid and twist operations are
``active''. For a CFT, the $F$ matrices can be read off from the
behavior of the conformal blocks. To consider this, it is useful
to place all fields (or quasiholes) $\phi_{\alpha_l}$ on the $x$
axis at positions $w_l=x_l$. If three adjacent such fields are
types $\alpha$, $\beta$, $\gamma$ as above, with $\alpha$ at
$x=0$, $\gamma$ at $x=1$, and $\beta$ at $x$, $0<x<1$, then as
$x\to0$ one would use the ope and analyze the block as on the
left-hand side of the definition of $F$. As $x\to1$, one analyzes
as on the right-hand side. The conformal blocks then determine the
$F$ matrices. The ($w$-independent) linear transformation between
bases remain valid even when the $w_l$s are not asymptotically
close together, which is useful as we want them further apart than
the screening length. In addition, they remain unchanged in the
presence of added fields $\psi_e$, which are in the chiral algebra
of the full CFT (including the charge sector) \cite{mr,milr}. For
us, the conformal blocks are the trial states, and we use one of
these bases, as convenient, and transform between them using the
$F$ matrices which are determined by our choice of the trial
functions themselves (elsewhere in this paper, we did not need to
specify our bases explicitly). For the braid group representation,
all matrices must be expressed in a single basis, and $F$ moves
are used to transform this to one appropriate for an application
of any desired braid generator $c$ or $c^{-1}$.

The other structure needed in a ribbon category is the set of
duality maps. These correspond to the creation and annihilation
processes mentioned above. For each object $V$ there is (an
isomorphism class of) dual object(s) $V^*$, and morphisms
$e_V:V^*\otimes V\to {\bf 1}$ (evaluation), $i_V:{\bf 1}\to
V\otimes V^*$ (co-evaluation). The order of factors in these
tensor products should be carefully noted. These maps are required
to satisfy relations (as maps $V\to V$):%
\be%
{\rm id}_V\otimes e_V\circ F_{V,V^*,V}\circ i_V\otimes {\rm
id}_V={\rm id}_V,\label{bdrel}\ee%
where ${\rm id}_V$ is the identity morphism on $V$, and $\circ$ is
composition of maps, which act from the left. A similar relation
with $V$ and $V^*$ interchanged, and the order of tensor factors
reversed, is also a requirement. When viewed as creation and
annihilation processes in spacetime, these relations allow us to
straighten out worldlines that sometimes reverse course and run
backwards (by a creation or annihilation event). Thus for
worldlines with a zigzag of this form, the zig and the zag cancel
(however, the similar identity for the mirror image zigzag is not
a consequence of these identities). For simple objects, we will
sometimes write $V_\alpha^*$ as $V_{\alpha^*}$. (Here and below,
for simplicity we do not explicitly write the maps $l_V:{\bf
1}\otimes V\to V$ and $r_V:V\otimes {\bf 1}\to V$ \cite{bk}, which
strictly should appear together with their inverses on the left
hand side of this relation.)

Finally, we recall the twist $\theta$ which is a map
$\theta_V:V\to V$ for each object $V$. For simple objects
$V_\alpha$, we can write $\theta_{V_\alpha}$ as a number, which is
$e^{2\pi ih_\alpha}$ when a ribbon category is obtained from a
RCFT. The twist must obey%
\be%
\theta_{V\otimes W}=c_{W,V}c_{V,W}(\theta_V\otimes \theta_W),
\label{thetarel}\ee%
and also $\theta_{\bf 1}={\rm id}_{\bf 1}=1$,
$\theta_{V^*}=(\theta_V)^*$ (the latter uses the dual map, which
is the transpose matrix---not complex conjugation!).

The quantum dimension for any object $V$ can now
be formally defined as a composite map from $\bf 1$ to $\bf 1$, and hence
can be represented as a complex number:%
\be%
{\rm qdim}_V=e_V c_{V,V^*}(\theta_V\otimes{\rm id}_{V^*})i_V,\ee%
which obeys ${\rm qdim}_{V\oplus W}={\rm qdim}_V+{\rm qdim}_W$ and
${\rm qdim}_{V\otimes W}=({\rm qdim}_V)({\rm qdim}_W)$. For simple
objects $V_\alpha$ we define $d_\alpha={\rm qdim}_{V_\alpha}$ (and
note that $d_0=1$). The quantum dimension can be defined more
simply, using maps $e_V'$, defined by $e_{V_\alpha}'= e_{V_\alpha}
c_{\alpha,\alpha^*}(\theta_{V_\alpha}\otimes{\rm
id}_{V^*_\alpha}):V_\alpha\otimes V_\alpha^*\to{\bf 1}$, and
similarly $i_V'$. Then one has ${\rm qdim}_V=e_V'i_V=e_Vi_V'$. As
mentioned above, in a unitary ribbon category, $e_V'$ ($e_V$) is
the adjoint of $i_V$ (resp., $i_V'$), and so $d_\alpha>0$ for all
$\alpha$.

Now we address these structures for our conformal blocks viewed as
trial wavefunctions, in the case when the flow in the 2D theory is
to a massive IR fixed point. Then we have argued in the main text
that the overlap matrix is proportional to a sum of identity
matrices, in a basis in which the braid group acts by unitary
transformations (for this, any one of the bases above is
suitable), with proportionality constants that are independent of
quasihole positions $w$, provided they are well separated. Again,
we will consider here $w_l$ on the real axis. The $F$ matrices
allow us to transform the basis in this statement. We can assume
that a basis exists in which the $F$ matrices are unitary, as
otherwise there will be inconsistencies. For the $e_V$ maps, we
will adiabatically transport quasiholes corresponding to
$V_\alpha$ and $V_\alpha^*$ together, but not closer than the
screening length. By definition, from far away they can be viewed
as a single quasihole of types occurring in the fusion rule for
$V_\alpha$ with $V_\alpha^*$, and this must include the identity
$\rm 1$, just once. (We must digress to point out that the fusion
of the negatively-charged quasiholes in the QH effect will produce
a negatively-charged result. But the destruction of the underlying
particle produces a quasihole of exactly this type, and is viewed
as equivalent to the identity as it has trivial braiding with all
quasiparticle types.) There is also a trial state with the pair
replaced by the quasiparticle type $\bf 1$, which can also be
normalized. The map $e_{V_\alpha}$ is a map between these, and can
be defined without further consideration of any short-scale
physics of the states. It is tempting to say that the map
$e_{V_\alpha}$ is the map between these two normalized states,
times some phase factor. But this is not generally correct, and
the reason is that the $i$ and $e$ maps must be normalized by
relation (\ref{bdrel}) above. As we will see, the $F$ matrix
element is something like $1/d_\alpha$, while $e$ and $i$ are like
$\sqrt{d_\alpha}$.

In terms of conformal blocks viewed as trial wavefunctions, if a
quasiparticle type $\alpha^*$ that is suitable to be the dual of
type $\alpha$ exists for all $\alpha$, then it can be shown that
${\rm Hom}({\bf 1},V_\alpha\otimes V_\alpha^*)$ and ${\rm
Hom}(V_\alpha^*\otimes V_\alpha,{\bf 1})$, to which $i_{V_\alpha}$
and $e_{V_\alpha}$ belong, are one-dimensional. Then the
normalization condition allows the definition of $i_{V_\alpha}$
and $e_{V_\alpha}$ maps to be varied only by multiplication by
scalars that are inverses of each other for each $\alpha$, which
cannot change the quantum dimensions. This is true for the CFT
itself, also. But the quantum dimensions can be
evaluated in the CFT, and if any is negative, then there is no way to avoid a
conflict with the quantum mechanical definition which leads to
positive quantum dimensions.

Here all structures other than the duality maps were already
fixed, either from properties of the conformal blocks used, or by
adiabatic transport that agrees with the monodromy of the blocks.
However, there was some uncertainty about $\theta$ (which is
independent of the choice of $F$s, $c$, $i$, $e$), because the adiabatic calculation of the twist by varying the metric has not been performed.
Perhaps the system can be rendered consistent by the presence of
different values $\theta'_{V_\alpha}$, with the other
structures which we have either defined or calculated ($c$, $F$,
$i$, $e$) unchanged. A different choice for the values of $\theta$
on simple objects can sometimes change the values of the quantum
dimensions, making some positive, as we will see in a moment. But
this cannot change the value of the $F$ matrices, or the
normalization condition for $i$ and $e$. We will see that, in many cases, no consistent choice exists for the twist that makes all quantum dimensions positive (for the given $F$s, $c$, $i$, $e$).

We already know one set of values of $\theta_{V_\alpha}$ that is
consistent, which come from the CFT. A different choice $\theta'$ must
still satisfy relation (\ref{thetarel}), so that%
\be%
\theta_{V_\alpha\otimes V_\beta}'/\theta_{V_\alpha\otimes V_\beta}
=(\theta_{V_\alpha}'/\theta_{V_\alpha})(\theta_{V_\beta}'/
\theta_{V_\beta}).\ee
Moreover, $\theta_{\bf 1}'/\theta_{\bf 1}=1$. This implies that
the map
$\phi_\alpha\to(\theta_{V_\alpha}'/\theta_{V_\alpha})\phi_\alpha$
is an automorphism of the ring defined by the fusion rules (the
Grothendieck ring). This and
$\theta'_{V_\alpha^*}=\theta'_{V_\alpha}$ then imply further that
$(\theta_{V_\alpha}'/\theta_{V_\alpha})^2=1$, so
$\theta_{V_\alpha}'/\theta_{V_\alpha}=\pm 1$ for all $\alpha$. Now
as the definition of quantum dimension used $\theta$, if we keep
$c$, $i$, $e$ fixed we may be able to reverse the signs of some of
the quantum dimensions.

As an example, for the Gaffnian state, where the CFT apart from
the charge sector is the $M(5,3)$ minimal model, the fusion rules
are the same as for SU($2$) level $3$, and [using the notation of
SU(2) spin $s$,  $\alpha=s=0$, $1/2$, \ldots, $3/2$ for the
fields], the only automorphism is given by multiplication by
$(-1)^{2s}$. This leaves $d_{1}$ ($d_{1,3}$ in the notation in the
main text) still negative. More generally, for the class of states mentioned in Sec.\ \ref{nonun}, in which the CFT is a BPZ minimal model, the twist of the field $\phi_{r,s}$ can be changed by multiplication by $(-1)^{r-1}$, or by $(-1)^{s-1}$, or both. Using the formula (\ref{modS}) for the modular $S$ matrix of the minimal models, for the non-unitary cases $p\neq p'+1$ there are some negative quantum dimensions, and one can show that not all of the negative ones become positive under any of these changes in the twist (indeed, additional ones are generated). Thus the family of trial states constructed from the minimal models, mentioned in Sec.\ \ref{nonun}, fails to produce valid (i.e.\ unitary)
topological phases except in the case of the unitary minimal models $M(p'+1,p')$.

We also record here a basis-independent relation involving the
part of the $F$ matrix that enters the normalization condition,
together with braiding and twist matrix elements, and the quantum
dimensions. The result, which is close to Moore and Seiberg
\cite{ms} (and references therein), eqs.\ (7.12) and (C.15), is
\be%
(c_{0,\alpha}^\alpha)^{-1} (c_{\alpha,\alpha^*}^0)^{-1}
F_{\alpha\alpha^*\alpha,00}^\alpha=\theta_\alpha/d_\alpha.\ee%
All the factors in this expression are simply numbers. (We note
that $c_{\alpha^*\alpha}^0c_{\alpha\alpha^*}^0=\theta_\alpha^{-2}$
and $c_{\alpha,0}^\alpha c_{0,\alpha}^\alpha=1$ in any basis, so
up to some choices of basis the result says that
$F_{\alpha\alpha^*\alpha,00}^\alpha=1/d_\alpha$.) If we can change
the sign of $\theta_\alpha$ as discussed above, then $d_\alpha$
also changes sign, while the left hand side is invariant. The
relation is obtained by evaluating
\be%
c_{\alpha,0}\circ ({\rm id}_{V_\alpha}\otimes
e'_{V_\alpha})\circ({\rm id}_{V_\alpha}\otimes
c_{\alpha,\alpha^*}^{-1}) \circ
F_{\alpha\alpha^*\alpha}^\alpha\circ (i_{V_\alpha}\otimes {\rm
id}_{V_\alpha}),\ee%
which is a map from ${\bf 1}\otimes V_\alpha$ to itself, in two
ways, using the relation (\ref{bdrel}).

\end{appendix}

\end{document}